\documentclass[acmsmall,screen]{acmart}

\setcopyright{none}

\AtBeginDocument{%
  \providecommand\BibTeX{{%
    \normalfont B\kern-0.5em{\scshape i\kern-0.25em b}\kern-0.8em\TeX}}}

\setcopyright{acmcopyright}
\copyrightyear{2018}
\acmYear{2018}
\acmDOI{XXXXXXX.XXXXXXX}

\usepackage{xcolor}
\usepackage{wrapfig, amsthm, pict2e, booktabs, subcaption, subfiles,
  rotating, pifont, indentfirst, amsmath, mathbbol, amsfonts,
  mathrsfs, adjustbox, mathtools,booktabs, microtype, ebproof,
  enumitem, multirow, multicol, setspace, hyperref, lineno, listings,
  stmaryrd, xspace, stackengine}
\usepackage{empheq}

\usepackage[tableposition=top]{caption}
\usepackage{tikz-cd}
\usetikzlibrary{decorations.pathmorphing}
\usepackage[ruled,linesnumbered,vlined]{algorithm2e}
\usepackage{graphicx}
\usepackage[frozencache=true,cachedir=minted-cache]{minted}
\usepackage{tikz}
\newcounter{markeq}
\usepackage{spacingtricks}

\usepackage[T1]{fontenc}

\usepackage[xcolor]{changebar}
\cbcolor{blue}

\usepackage{spacingtricks, commands, refinementtydef}

\begin{document}

\title{A HAT Trick: Automatically Verifying Representation Invariants Using Symbolic Finite Automata}

\author{Zhe Zhou}
\affiliation{
  \institution{Purdue University}            
  \country{USA}                    
}

\author{Qianchuan Ye}
\affiliation{
  \institution{Purdue University}            
  \country{USA}                    
}

\author{Benjamin Delaware}
\affiliation{
  \institution{Purdue University}            
  \country{USA}                    
}

\author{Suresh Jagannathan}
\affiliation{
  \institution{Purdue University}            
  \country{USA}                    
}


\begin{abstract}
  Functional programs typically interact with stateful libraries that
  hide state behind typed abstractions. One particularly important
  class of applications are data structure implementations that rely
  on such libraries to provide a level of efficiency and scalability
  that may be otherwise difficult to achieve. However, because the
  specifications of the methods provided by these libraries are
  necessarily general and rarely specialized to the needs of any
  specific client, any required application-level invariants must
  often be expressed in terms of additional constraints on the (often)
  opaque state maintained by the library.

  In this paper, we consider the specification and verification of
  such \emph{representation invariants} using \emph{symbolic finite
    automata} (SFA).  We show that SFAs can be used to succinctly and
  precisely capture fine-grained temporal and data-dependent histories
  of interactions between functional clients and stateful libraries.
  To facilitate modular and compositional reasoning, we integrate SFAs
  into a refinement type system to qualify stateful computations
  resulting from such interactions.  The particular instantiation we
  consider, \emph{Hoare Automata Types} (HATs), allows us to both
  specify and automatically type-check the representation invariants
  of a datatype, even when its implementation depends on stateful
  library methods that operate over hidden state.

  We also develop a new bidirectional type checking algorithm that
  implements an efficient subtyping inclusion check over HATs,
  enabling their translation into a form amenable for SMT-based
  automated verification.  We present extensive experimental results
  on an implementation of this algorithm that demonstrates the
  feasibility of type-checking complex and sophisticated HAT-specified
  OCaml data structure implementations layered on top of stateful
  library APIs.

\end{abstract}

\ccsdesc[500]{Theory of computation~Automated reasoning}
\ccsdesc[500]{Software and its engineering~Language types}
\ccsdesc[500]{Theory of computation~Modal and temporal logics}
\ccsdesc[500]{Software and its engineering~Software verification and validation}

\keywords{refinement types, symbolic finite automata, representation invariants}

\maketitle
\section{Introduction}
\label{sec:intro}

Functional programs often interact with stateful libraries that hide
internal state behind their APIs.  Such libraries are a vital
component of a programming language's ecosystem, providing well-tested
implementations of databases, key-value stores, logging and system
services, and other stateful abstractions.  A common use case for how
these libraries are used is for programmers to build higher-level
data structures on top of the abstractions they provide~\cite{vocal}.
A set abstract data type (ADT) equipped with intersection and
cardinality operations, for example, might use the aforementioned
key-value library to maintain a persistent collection of its elements
by associating those elements (represented as values) with unique ids
(their associated keys). Since the internal state of the underlying
stateful library is opaque to its clients, the implementations of
these higher-level modules will necessarily be expressed in terms of
the operations provided by the library: the methods of the set ADT
might use methods like \zput{} and \zget{} to interact with the
backing key-value store, for example. To maximize reusability, these
low-level libraries are typically quite permissive in how client may
use them, resulting in weak specifications that only describe simple
algebraic or equational constraints. For example, the specification of
the key-value store might state that two \zput{} operations over
distinct keys commute or that a \zget{} operation on a key
always returns the last value \zput{} in that key.

In contrast, the higher-level abstractions provided by clients built
on top of these libraries are equipped with stronger guarantees: for example, the
operations of our set ADT need to respect the semantics of
mathematical sets, for example, so that
$\forall A\; B : \Code{Set.t}.\; |A \cup B| + |A \cap B| = |A| +
|B|$. Verifying such semantically rich properties typically requires a
strong guarantee that the internal state of the set library is
consistent with the abstract states of the datatype, i.e., the ADT implementation should be equipped with a 
\emph{representation invariant}~\cite{MP+07, MP+20}. Ensuring that
distinct keys are always associated with distinct values allows us to
prove the above relationship between the intersection and cardinality
operations of the set ADT, for example. Establishing such an invariant
is complicated by the fact that our set ADT is built on top of another
library with an opaque internal state, and whose weak specifications do
not directly address the notion of uniqueness central to the set
ADT. This paper presents an automated type-based verification
procedure that addresses these complications.

Our proposed solution augments a refinement type system for a
call-by-value functional core language that supports calls to stateful
libraries, with \emph{Hoare Automata Types} (HATs). HATs are a new
type abstraction that qualifies basic types with pre- and
post-conditions expressed as \emph{symbolic finite-state
  automata}~\cite{VBM10,Vea13}, a generalization of finite-state
automata equipped with an unbounded alphabet, and whose transitions
are augmented with predicates over the elements of this alphabet.
HATs serve as a compact representation of the trace of stateful API
calls and returns made prior to a function invocation and those
performed during its execution.
As a result, HATs serve as an
expressive basis for a fine-grained effect system within a general
refinement type framework, enabling compositional reasoning about
invocations of stateful APIs in a form amenable to automated
verification of a client module's representation invariant.  For example, he
representation invariant from our set ADT is captured by
the following symbolic automaton, written using a variant of linear
temporal logic on finite sequences~\cite{LTLf}. The base predicates of
this formula describe \emph{events}, or invocations of the methods of
a key-value store whose values are drawn from a
{\small$\Code{setElem}$} type:
{\small
\setlength{\abovedisplayskip}{3pt}
\setlength{\belowdisplayskip}{3pt}
\setlength{\abovedisplayshortskip}{3pt}
\setlength{\belowdisplayshortskip}{3pt}
  \begin{align*}
    &\riA{Set}(\Code{el}) \doteq \globalA
    (\evop{put}{key\,val}{\I{val} = \Code{el}} \impl \nextA \neg
    \finalA \evop{put}{key\,val}{\I{val} = \Code{el}})
  \end{align*}
}\noindent
This representation invariant ({\small$\riA{Set}$}) is expressed using  standard
temporal logic modalities (e.g., {\small$\globalA$} (always),
{\small$\nextA$} (next), and {\small$\finalA$} (eventually)) to
express the history of valid interactions between the set
implementation and the underlying key-value store that ensure the
expected element uniqueness property holds. Informally, the invariant
establishes that, at every step of the execution
({\small$\globalA(...)$}), once a value {\small$\Code{el}$} has been
inserted into the set
({\small$\evop{put}{key\,val}{\I{val} = \Code{el}}$}), it will never
be reinserted
({\small$\nextA \neg \finalA \evop{put}{key\,val}{\I{val} =
    \Code{el}}$}).  An implementation that preserves this invariant
guarantees that every value is always associated with at most one key.

To verify it, we can check the {\small$\Code{insert}$} method of the set
ADT against the following type
{\small
  \setlength{\abovedisplayskip}{3pt}
  \setlength{\belowdisplayskip}{3pt}
  \setlength{\abovedisplayshortskip}{3pt}
  \setlength{\belowdisplayshortskip}{3pt}
  \begin{align*}
    &\tau_{\Code{insert}} \doteq \gvn{el}{setElem}
    \Code{x}{:}\nuot{setElem}{\top} \sarr
    \htriple{\riA{Set}(\Code{el})}{\nuot{unit}{\top}}{\riA{Set}(\Code{el})}
  \end{align*}
}\noindent
The output type of $\tau_{\Code{insert}}$
({\small$\htriple{\riA{Set}(\Code{el})}{\nuot{unit}{\top}}{\riA{Set}(\Code{el})}$})
is a Hoare Automata Type (HAT) over the representation invariant
{\small$\riA{Set}$}.  This type captures the behavior required of any
correct implementation of {\small$\Code{insert}$}: assuming the
invariant holds for any value {\small$\Code{el}$} (encoded here as a
ghost variable scoped using {\small$\garr$}) prior to the insertion of
a new element {\small$\Code{x}$}, it must also hold after the
insertion as well.  The precondition of the HAT captures the set of
interactions that lead to a valid underlying representation state
(namely, one in which every value in the store is unique), and the
postcondition of the HAT specifies that this condition remains valid
even after executing {\small$\Code{insert}$}.

Our use of HATs differs in important ways from other type-based
verification approaches for functional programs that allow
computations to perform fine-grained effects, such as F*~\cite{FStar}
or Ynot~\cite{NM+08}.  These systems use the power of rich dependent
types to specify fine-grained changes to the heap induced by an
stateful operation.  The Dijkstra monad~\cite{MAA+19,SW+13} used by
F*, for example, extends ideas first proposed by Nanevski
\emph{et. al}~\cite{NMB08} in their development of Hoare Type Theory
to embed a weakest precondition predicate transformer semantics into a
type system that can be used to generate verification conditions
consistent with a provided post-condition on the program heap. %
\cbnewadding{%
  In contrast, our setting assumes that such operations are executed
  by a library with a \emph{hidden} state, requiring the
  client's representation invariant to be
  expressed solely in terms of observations on calls (and returns) to
  (and from) the methods of the underlying library. %
}%

Our approach also bears some similarity to recent (refinement) type
and effect systems for verifying temporal properties of event traces
\emph{produced} during a computation~\cite{NUKT+18,
  Temporal-Verification}.  While the idea of using symbolic traces of
previous observations as a handle to reason about the hidden state of
a library is similar to these other efforts, incorporating return
values of operations as part of such traces is a significant point of
difference.  This capability, critical for the validation of useful
representation invariants, allows the precondition encoded in a HAT to
describe a fine-grained \emph{context} of events in which a program
can run, expressed in terms of both library method arguments
\emph{and} return values.  These important differences with prior work
position HATs as a novel middle-ground in the design space between
explicit state-based verification techniques and trace-based effect
systems.  Our formulation also leads to a number of significant
technical differences with other refinement type systems~\cite{JV21}
and their associated typing algorithms, since we must now contend with
type-level reasoning over symbolic automata qualifiers.

In summary, this paper makes the following contributions:
\begin{enumerate}
\item We formalize a new trace-based specification framework for
  expressing representation invariants of functional clients of
  stateful libraries that manage hidden state.

\item We show how this framework can be embedded within a
  compositional and fine-grained effect-tracking refinement type
  system, by encoding specifications as symbolic automata and
  instantiated as HATs in the type system.

\item We develop a bidirectional type-checking algorithm that
  translates the declarative type system into efficient
  subtype inclusion checks amenable to SMT-based automated
  verification.

\item Using a tool (\name{}) that implements these ideas, we present a
  detailed evaluation study over a diverse set of non-trivial OCaml
  datatype implementations that interact with stateful libraries.  To
  the best of our knowledge, \name{} is the first system capable of
  automated verification of sophisticated representation invariants
  for realistic OCaml programs.
\end{enumerate}

\noindent The remainder of this paper is organized as follows.  In the next
section, we provide further motivation for the problem by developing a
running example, which is used in the rest of the paper.
\autoref{sec:lang} introduces a core functional language equipped with
operators that model stateful library methods.  \autoref{sec:typing}
presents a refinement type system that augments basic type qualifiers
to also include HATs; a brief background on symbolic automata is also
given.  An efficient bidirectional typing algorithm is then presented
in ~\autoref{sec:algo}.  Implementation details, as well as experimental results on its effectiveness in automatically verifying the representation
invariants of a number of realistic data structure client programs
that interact with stateful libraries is provided
in~\autoref{sec:evaluation}.  The paper ends with a
discussion of related work and conclusions.\\[-.8em]

\section{Motivation}\label{sec:overview}

\begin{figure}[!t]
  \begin{tabular}{cc}
    \begin{minipage}{.38\textwidth}
\begin{minted}[xleftmargin=-3pt, numbersep=4pt, fontsize = \footnotesize, escapeinside=!!]{ocaml}
module type Bytes = sig
  type t
  ...
end

module type Path = sig
  type t = string
  val parent: t -> t
  val isRoot: t -> bool
  ...
end

module type File = sig
  val init: unit -> Bytes.t
  val isDir: Bytes.t -> bool
  val isFile: Bytes.t -> bool
  val isDel: Bytes.t -> bool
  val setDeleted: Bytes.t -> Bytes.t
  val delChild:
    Bytes.t -> Path.t -> Bytes.t
  val addChild:
    Bytes.t -> Path.t -> Bytes.t
  ...
end

module type KVStore = sig
  val !$\textbf{put}$!: Path.t -> Bytes.t -> unit
  val !$\textbf{exists}$!: Path.t -> bool
  val !$\textbf{get}$!: Path.t -> Bytes.t
end
\end{minted}
\end{minipage}
    &
\begin{minipage}{.60\textwidth}
\begin{minted}[xleftmargin=5pt, numbersep=4pt, linenos = true, fontsize = \footnotesize, escapeinside=!!]{ocaml}
module FileSystem = struct
  open KVStore
  let init () : unit = !$\textbf{put}$! "/" (File.init ())

  let add (path: Path.t) (bytes: Bytes.t): bool =
    if !$\textbf{exists}$! path then false
    else
      let parent_path = Path.parent path in
      if not (!$\textbf{exists}$! parent_path) then false
      else
        let bytes' = !$\textbf{get}$! parent_path in
        if File.isDir bytes' then (
          !$\textbf{put}$! path bytes;
          !$\textbf{put}$! parent_path (File.addChild bytes' path);
          true)
        else false

  let delete (path : Path.t): bool =
    if (isRoot path) or not (!$\textbf{exists}$! path) then false
    else (
      let bytes = !$\textbf{get}$! path in
      if File.isDir bytes then deleteChildren(path);
      let parent_path = Path.parent path in
      let bytes' = !$\textbf{get}$! parent_path in
      !$\textbf{put}$! path (File.setDeleted bytes);
      !$\textbf{put}$! parent_path (File.delChild bytes' path);
      true)
end
\end{minted}
\end{minipage}
  \end{tabular}
  \caption{A file system datatype based on underline key-value store
    library.}
\label{fig:ex-fs}
\end{figure}

To further motive our approach, consider the (highly-simplified)
Unix-like file-system directory abstraction (\motidt{}) shown in
\autoref{fig:ex-fs}.  This module is built on top of other libraries
that maintain byte arrays ({\small\Code{Bytes.t}}) which hold a file
or a directory's contents and metadata, and which manipulate
\emph{paths} ({\small\Code{Path.t}}), strings that represent
fully-elaborated file and directory names. \motidt's implementation
also uses an underlying key-value store library
({\small\Code{KVStore}}) to provide a persistent and scalable storage
abstraction, where keys represent file paths and values are byte
arrays.  The interface to this store is defined by three
\emph{stateful} library methods: \zput{} persistently stores a
key-value pair {\small$\Pair{k}{v}$}, adding the pair to the store if
{\small$k$} does not already exist, and overwriting its old value with
{\small$v$} if it does; \zexist{} returns true if the given key has
been previously \zput{} and false otherwise; and, \zget{} returns the
value associated with its key argument, and fails (i.e., raises an
exception) if the key does not exist in the store.

\autoref{fig:ex-fs} shows three of the methods provided by \motidt{}:
\motiinit{} initializes the file system by storing a root path
(``{\small\Code{/}}''); \motiadd{} allows users to add regular files or
directories to the filesystem; and, \motidel{} logically removes files
or directories from the filesystem.  The latter two operations return
a Boolean value to indicate whether they were successful. The {\small
  $\Code{add}$} operation succeeds (line {\small$15$}) only if (a) the
given path does not already exist in the file system (line
{\small$6$}) and (b) a path representing the parent directory for the
file does exist (line {\small$9$}). Besides adding the file (line
{\small$13$}), {\small $\Code{add}$} also updates the contents of the
parent directory to make it aware that a new child has been added
(line {\small$14$}). Similar to \motiadd{}, \motidel{} requires that
the supplied path exists and is not the root (line {\small$19$}).  If
the path corresponds to a directory, its children are marked deleted
(line {\small$22$}) using procedure {\small$\Code{deleteChildren}$}
(not shown).  Additionally, the directory's metadata, and that of its
parent, are updated to record the deletion (lines {\small$25$} and
{\small$26$}).  Implicit in \motidel{}'s implementation is the
assumption that the parent path of the file to be deleted exists in
the store: observe that there is no check to validate the existence of
this path before the \zget{} of the parent's contents is performed on
line {\small$24$}.  This assumption can be framed as a representation
invariant that must hold prior to and after every call and return of
\motiadd{} and \motidel{}.  Intuitively,
this invariant captures the following property: \\[-.8em]

\motiinv{}: \textit{Any path that is stored as a key in the key-value
  store (other than the root path) must also have its parent stored as a
  non-deleted directory in the store.}

\paragraph{Representation Invariants} In a language that guarantees
proper data abstraction, library developers can safely assume that any
guarantees about an API established using a representation invariant
are independent of a particular client's usage. However, as we noted
earlier, numerous challenges arise when validating a representation
invariant in our setting because we can only characterize the state of the underlying library
in terms of its admissible sequence of
interactions with the client. In
our example, this characterization must be able to capture the
structure of a valid filesystem, in terms of sequences of \zput{},
\zget{}, and \zexist{} calls and returns that are sufficient to ensure
that the invariant is preserved.  Verifying that \motiinv{} holds
requires us to prove that, except for the root, the path associated
with a file or directory can only be recorded in the store when its
parent path exists, i.e., its parent has previously been \zput{} and
has not yet been deleted.  Establishing that this property holds
involves reasoning over both data dependencies among filesystem calls
and temporal dependencies among store operations.



\paragraph{Events, Traces, and Symbolic Finite Automata} We address
these challenges by encoding desired sets of stateful interactions
between the client and library as \emph{symbolic finite automata}
(SFA).  Besides constants and program variables, the alphabet of an
automaton that captures these interactions includes a set of stateful
operators {\small $\effop$} (e.g., \zput{} and \zget{}).  The language
recognized by an automaton specifies the sequences of permitted library
method invocations.  Each such invocation (called an \emph{event})
{\small $\pevapp{\effop}{\overline{v_i}}{v}$} records the operator
invoked ({\small $\effop$}), a list of its arguments ({\small
  $\overline{v_i}$}), and a single result value {\small$v$}. It is critical
that we record \emph{both} inputs and outputs of an
operation, in order to enable our type system to distinguish among
different hidden library states. Consider, for example, a stateful
library initialization operator {\small $\S{initialize}$}
{\small $\Code{: unit\sarr bool}$}. The hidden state after observing
``{\small$\evapp{initialize}{()}{true}$}'', representing a successful
initialization action, is likely to be different from the hidden state
that exists after observing the event
``{\small$\evapp{initialize}{()}{false}$}'', which indicates an
initialization error. The language accepted by an SFA is a potentially
infinite set of finite sequences of events, or \emph{traces}.


\begin{example}[Traces]
  \label{ex:traces} In contrast to the correct implementation of
  \motiadd{} shown in \autoref{fig:ex-fs}, consider the following
  incorrect version:
\begin{minted}[xleftmargin=30pt, numbersep=4pt, fontsize = \small, escapeinside=!!]{ocaml}
    let add!$_\Code{bad}$! (path: Path.t) (bytes: Bytes.t) = !$\S{put}$! path bytes
\end{minted}
This implementation simply records a path without considering whether
its parent exists.  We assume the file system initially contains only
the root path, {\small$``/"$}, which is \zput{} when the file system
is created by the method {\small$\Code{init}$}.  The traces $\alpha_1$
and $\alpha_2$, shown below, represent executions of these two
implementations that reflect their differences:
{\footnotesize
    \begin{align*}
      {\small\text{under context trace}}\; \ \alpha_0 \doteq \ &[\evapp{put}{``/"\, bytesDir}{()}]\\
      \Code{add_{bad}\,``/a/b.txt"\, bytesFile}\ \ &{\small\text{yields}}\ \
        \alpha_1 \doteq [\evapp{put}{\Code{``/"}\,bytesDir}{()};\,
        \evapp{put}{``/a/b.txt"\, bytesFile}{()}] \\
      \Code{add\,``/a/b.txt"\,bytesFile}\ \ &{\small \text{yields}}\ \
        \alpha_2 \doteq [\evapp{put}{\Code{``/"}\,bytesDir}{()};\,
        \evapp{exists}{``/a/b.txt"}{false};\,
        \evapp{exists}{``/a"}{false}]
    \end{align*}
  }\noindent where {\small$\Code{bytesDir}$} and
  {\small$\Code{bytesFile}$} represent the contents of a directory and
  a regular file, resp.  The trace {\small$\alpha_1$} violates
  \motiinv{} since it registers the path ``{\small\Code{/a/b.txt}}''
  in the file system even though the parent path
  (``{\small\Code{/a}}'') does not exist.  Thus, performing the
  operation {\small\Code{delete}\ ``\Code{/a/b.txt}''} after executing
  the actions in this trace will cause a runtime error when it
  attempts to perform a \textbf{get} operation on the parent path
  ``{\small\Code{/a}}''.  On the other hand, executing this operation
  after executing the actions in trace {\small$\alpha_2$} terminates
  normally and preserves \motiinv{}.
\end{example}

\begin{example}[Symbolic Finite Automata]
  \label{ex:SFAs}
  We express symbolic automata as formulae in a symbolic version of
  linear temporal logic on finite sequences~\cite{LTLf}.  \motiinv{}
  can be precisely expressed in this language as the following formula
  (denoted {\small$\riA{FS}(\Code{p})$}) parameterized over a path
  {\small$\Code{p}$}: {\small\begin{align*} &\riA{FS}(\Code{p}) \doteq
      \globalA \evparenth{ \Code{isRoot(p)} } \lorA
      (\predA{isFile}{}(\Code{p}) \lor \predA{isDir}{}(\Code{p}) \impl
      \predA{isDir}{}(\Code{parent(p)}))
\end{align*}}\noindent
where
{\footnotesize
  \begin{align*}
    &\predA{isDir}{}(\Code{p}) \doteq \finalA(\evop{put}{key\,val}{\I{key} = \Code{p}
      \land \Code{isDir}(\I{val})} \land \nextA\globalA \neg
      \evop{put}{key\,val}{ \I{key} = \Code{p} \wedge (\Code{isDel}(\I{val}) \lor \Code{isFile}(\I{val})) })
    \\&\predA{isFile}{}(\Code{p}) \doteq \finalA(\evop{put}{key\,val}{\I{key} = \Code{p}
    \land \Code{isFile}(\I{val})} \land \nextA\globalA \neg
    \evop{put}{key\,val}{ \I{key} = \Code{p} \wedge (\Code{isDel}(\I{val}) \lor \Code{isDir}(\I{val})) })
  \end{align*}}\noindent
Here, {\small$\Code{isRoot}$}, {\small$\Code{isDir}$}, {\small$\Code{isFile}$}, and
{\small$\Code{parent}$} are pure functions on paths and bytes.
The invariant is comprised of two disjunctions involving $\small{\Code{p}}$:
\begin{enumerate}
\item {\small$\globalA \evparenth{ \Code{isRoot(p)} }$} asserts that
  {\small$\Code{p}$} is the filesystem root and is always a valid
  path, regardless of any other actions that have been performed on
  the underlying key-value store.

\item
  {\small$\predA{isFile}{}(\Code{p}) \lor \predA{isDir}{}(\Code{p})
    \impl \predA{isDir}{}(\Code{parent(p)})$} asserts that if
  {\small$\Code{p}$} corresponds to either a file or directory in the
  file system, then its parent path must also correspond to a
  directory in the file system. The predicate
  {\small$\predA{isDir}{}(\Code{p})$} captures contexts in which
  {\small$\Code{p}$} has been registered as a directory and then not
  subsequently deleted or replaced by a file (i.e., {\small$\nextA
    \globalA \neg \ltortoise \S{put}\,\I{key\,val} \eveq \vnu \,|\,
    \I{key} = \Code{p} \land (\Code{isDel}(\I{val})\, \lor\,
    \Code{isFile}(\I{val})) \rtortoise$}).  The predicate
  ({\small$\predA{isFile}{}(\Code{p})$}) is defined similarly except
  that it requires that an existing file not be deleted or modified to
  become a directory.  Observe that atomic predicates (e.g.,
  {\small$\ltortoise \S{put}\,\I{key\,val} \eveq \vnu \,|\, \I{key} =
    \Code{p} \land \Code{isFile}(\I{val}) \rtortoise$}) not only
  capture when an event has been induced by a library operator
  (e.g., $\zput{}$), but also \emph{qualify} the arguments to that
  operator and the result it produced (e.g., the argument
  {\small$\I{key}$} must be equal to the path
  {\small\Code{p}}).\footnote{Throughout this paper, we italicize
    event arguments.}


\end{enumerate}
This specification of the representation invariant
{\small$\riA{FS}(\Code{p})$} thus formalizes our informal
characterization of a correct file system, \motiinv.  The two
subformulas capture the property that any path recorded
as a key in the key-value store, other than the root, must have a
parent directory.  Notably, this specification is defined purely by
constraining the sequence of \zput{} operations allowed on the store
and does not require any knowledge of the store's underlying representation.

\end{example}

\paragraph{Hoare Automata Types} While SFAs provide a convenient
language in which to encode representation invariants, it is not
immediately apparent how they can help with verification since, by
themselves, they do not impose any safety constraints on
expressions. To connect SFAs to computations, we embed them as
refinements in a refinement type (and effect) system.  In a standard
refinement type system, the refinement type {\small$\rawnuot{b}{\phi}$} of an expression $e$
uses a qualifier {\small$\phi$} to refine the value of
{\small$e$}. Similarly, a Hoare Automata Type (HAT),
{\small$\htriple{A}{\rawnuot{b}{\phi}}{B}$}, additionally refines a
pure refinement type {\small$\rawnuot{b}{\phi}$} with two symbolic
automata: a precondition automaton {\small$A$} defining an effect
\emph{context} under which {\small$e$} is allowed to evaluate, and a
postcondition automata {\small$B$} that augments this context with the
effects induced by {\small$e$} as it executes.  A HAT thus allows us
to track the sequence of stateful interactions between higher-level
datatypes and the underlying stateful library, as well as to record the
consequences of this interaction in terms of the return values of
stateful calls.

Our type system leverages this information to verify the preservation of
stated representation invariants in terms of these interactions. For
example, \motiadd{} and \motidel{} can be ascribed the following
HAT-enriched types:
{\small
  \begin{align*}
    \Code{add} :&\ \gvn{p}{Path.t} \normalv{path}{\nuot{Path.t}{\top}}
                  \normalv{bytes}{\nuot{Bytes.t}{\top}} \tag{$\tauadd$}
                  \htriple{\riA{FS}(\Code{p})}{\nuot{bool}{\top}}{\riA{FS}(\Code{p})} \\
    \Code{delete} :&\ \gvn{p}{Path.t}
                     \normalv{path}{\nuot{Path.t}{\top}}
                     \htriple{\riA{FS}(\Code{p})}{\nuot{bool}{\top}}{\riA{FS}(\Code{p})}
                     \tag{$\taudel$}
  \end{align*}}\noindent
Here, the notation ``{\small$\gvn{p}{Path.t}$}'' indicates that
{\small$\Code{p}$} is a ghost variable representing an
arbitrary path; thus, the representation invariant must hold over
\emph{any} instantiation of {\small$\Code{p}$}.  While their
input arguments ({\small$\nuot{Path.t}{\top}$} and
({\small$\nuot{Bytes.t}{\top}$})) are unconstrained, the return types of these functions
are expressed as HATs in which both the precondition and postcondition
automata refer to the representation invariant for the file system,
{\small$\riA{FS}(\Code{p})$}.  Informally, this type reads ``if we
invoke the function in a context that is consistent with $\riA{FS}$,
then the state after the function returns should also be consistent
with $\riA{FS}$''.  The traces admitted by an SFA are used to
represent these contexts and states.

More concretely, observe that the trace {\small$\alpha_2$} from
Example~\ref{ex:traces} is derived by concatenating $\alpha_0$ with
the library calls generated during the (symbolic) execution of
\motiadd{}: %
{\small
  \begin{align*} {\alpha_{\Code{new}} \doteq
      [\evapp{exists}{``/a/b.txt"}{false};\,
      \evapp{exists}{``/a"}{false}]}
\end{align*}}\noindent
(i.e., {\small$\alpha_2 = \alpha_0 \listconcat
  \alpha_{\Code{new}} $}).  Note that both {\small$\alpha_0$} and
{\small$\alpha_2$} satisfy the representation invariant
{\small$\riA{FS}$}. To admit this trace, our type system
automatically infers a new automata {\small $A$} that
accepts the sequence of events in {\small$\alpha_{\Code{new}}$} by
type-checking \motiadd{}.  Verifying that the
representation invariant holds now reduces to an inclusion check between
two symbolic automata, requiring that the trace expected by the
\emph{concatenated} automata {\small$\riA{FS};A$} is also accepted by the
representation invariant {\small$\riA{FS}$}.

\section{Language}\label{sec:lang}

\begin{figure}[t!]
{\small
    \begin{alignat*}{2}
    \text{\textbf{Variables }}& \quad &\quad& x, y, z, f, \vnu, ... \\
    \text{\textbf{Pure Operations}}& \quad & \primop ::= \quad & {+} ~|~ {-} ~|~ {==} ~|~ {<} ~|~ {\leq} ~|~ \Code{mod} ~|~ \Code{parent} ~|~ ...\\
    \text{\textbf{Effectful Operations }}& \quad & \effop{} ::= \quad & \S{put} ~|~ \S{exists} ~|~ \S{get} ~|~ \S{insert} ~|~ \S{mem} ~|~ ...
    \\ \text{\textbf{Data constructors }}& \quad &d ::= \quad & () ~|~ \Code{true} ~|~ \Code{false} ~|~ \Code{O} ~|~ \Code{S} ~|~ ...
    \\ \text{\textbf{Constants }}& \quad &c ::= \quad & \mathbb{Z} ~|~ d(\overline{c})
    \\ \text{\textbf{Values }}& \quad  & v ::= \quad & c ~|~ x ~|~ d(\overline{v}) ~|~ \zlam{x}{t}{e} ~|~ \zfix{f}{t}{x}{t}{e}
    \\ \text{\textbf{Expressions (Computations)}}& \quad & e ::=\quad &
    v
    ~|~ \zlet{x}{\effop\ \overline{v}}{e}
    ~|~ \zlet{x}{\primop\ \overline{v}}{e}
    ~|~ \zlet{x}{v\ v}{e}
    \\ &\quad&\quad& ~|~ \zlet{x}{e}{e} ~|~ \match{v}{\overline{d\ \overline{y} \sarr {e}}}
  \end{alignat*}
}
\vspace{-.5cm}
  \caption{\langname{} term syntax.}
  \label{fig:term-syntax}
\end{figure}

To formalize our approach, we first introduce \langname{}, a core
calculus for effectful programs.\footnote{%
\cbnewadding{ %
    Although the focus of this paper is the verification of clients of
    stateful libraries, our proposed approach applies to effectful
    programs more generally. Our formalism reflects the more general
    setting of effectful programs. } %
}


\paragraph{Syntax}
\langname{} is a call-by-value lambda calculus with built-in inductive
datatypes, pattern matching, a set of pure and effectful operators,
and recursive functions. Its syntax is given in
\autoref{fig:term-syntax}. For simplicity, programs are expressed in
monadic normal-form (MNF)~\cite{JO94}, a variant of A-Normal Form
(ANF)~\cite{FSDF93} that permits nested let-bindings. The terms in
\langname{} are syntactically divided into two classes: values
{\small$v$} and expressions {\small$e$}, where the latter include both pure terms
and those that have computational effects.  Similar to the
simply-typed lambda calculus, function parameters have explicit type
annotations. 
\langname{} is parameterized over two sets of primitive operators:
pure operators ({\small$\primop$}) include basic arithmetic and
logical operators, while examples of effectful operators
({\small$\effop{}$}) include state manipulating operators (e.g.,
\zput{}/\zexist{}/\zget{} for interacting with a key-value store,
or \zinsert{}/\zmem{} for manipulating a stateful set library).  We
express sequencing of effectful computations {\small$e_1; e_2$} in
terms of $\Code{let}$-bindings: {\small$e_1;e_2 \doteq \zlet{x}{e_1}{e_2}$},
where {\small $x$} does not occur free in {\small$e_2$}. The
application $e_1\,e_2$ is syntactic sugar for
{\small$\zlet{x}{e_1}{\zlet{y}{e_2}{\zlet{z}{x\,y}{z}}}$}.

\begin{figure}[t!]
  \vspace{-.3cm}
{\footnotesize
{\small
  \begin{flalign*}
    \text{\textbf{Traces}}& \quad \alpha ::= \quad
    \emptr~|~ (\pevapp{\effop}{\overline{v}}{v}) {::} \alpha
  \end{flalign*}}
\vspace{-1.5em}
{\small
\begin{flalign*}
 &\text{\textbf{Small-Step Reduction Rules}} & \fbox{$\steptr{\alpha}{e}{\alpha}{e}$}
\end{flalign*}}
\begin{prooftree}
\hypo{\primop\ \overline{v} \Downarrow v_x }
\infer1[\textsc{StPureOp}]{
\steptr{\alpha}{\zlet{x}{\primop\ \overline{v}}{e}}{\emptr}{e[x\mapsto v_x]}
}
\end{prooftree}
\quad
\begin{prooftree}
\hypo{ \alpha \vDash \effop\, \overline{v} \Downarrow v_x }
\infer1[\textsc{StEffOp}]{
\steptr{\alpha}{\perform{x}{\effop{}}{\overline{v}}{e}}{[\pevapp{\effop{}}{\overline{v}\,}{\,v_x}]}{e[x\mapsto v_x]}
}
\end{prooftree}
}
\caption{Trace syntax and selected operational semantics. }
\label{fig:selected-semantics}
\end{figure}

\paragraph{Operational Semantics} Our programming model does not have
access to the underlying implementation of effectful operators, so the
semantics of \langname{} constrains the behavior of impure operations
in terms of the outputs they may produce.  To do so, we structure its
semantics in terms of \emph{traces} that record the history of
previous effectful operations and their results. The syntax of traces
adopts standard list operations (i.e., cons {\small${::}$} and
concatenation {\small$\listconcat{}$}), as shown in
\autoref{fig:selected-semantics}.  \emph{Traces} are lists of
\emph{effect events} {\small$\pevapp{\effop}{\overline{v}}{v}$} that
record the application of an effectful operator {\small$\effop$} to
its arguments {\small$\overline{v}$} as well as the resulting value
{\small$v$}. The traces {\small$\alpha_2$} and {\small$\alpha_1$} from Example~\ref{ex:traces}
record the events generated by correct and incorrect implementations
of the \motiadd{} function of the \motidt{} ADT, for example.

The operational semantics of \langname{} is defined via a small-step
reduction relation, {\small$\steptr{\alpha}{e}{~\alpha'}{e'}$}. This
relation is read as ``under an effect context (i.e., trace)
{\small$\alpha$}, the term {\small$e$} reduces to {\small$e'$} in one
step and performs the effect {\small$\alpha'$}'', where
{\small$\alpha'$} is either empty {\small$\emptr{}$} or holds a single
event. The semantics is parameterized over two auxiliary relations,
{\small$e \Downarrow v$} and {\small$\alpha \vDash e \Downarrow v$},
used to evaluate pure and impure operations,
respectively. \autoref{fig:selected-semantics} shows how the
corresponding reduction rules use these relations.\footnote{The
  remaining rules are completely standard and provided in the
  \techreport{}.}  
%
The reduction rule for pure operators (\textsc{STPureOp}) holds in an
arbitrary effect context (including the empty one), and produces
no additional events. The rule for effectful operators
(\textsc{STEffOp}), in contrast, can depend on the current context
{\small$\alpha$}, and records the result value of the effectful
operator in the event it emits. The multi-step reduction relation
{\small$\msteptr{\alpha}{e}{\alpha'}{e'}$} defines the reflexive,
transitive closure of the single-step relation; its trace {\small$\alpha$}
records all events generated during evaluation.


\begin{example}
  \label{ex:reduction}
  We can define the semantics for the \zput{}, \zexist{}, and \zget{}
  operators found in a key-value store library and used in \motidt{},
  via the following rules: %
  {\footnotesize
    \\ \ \\
    \begin{prooftree}
      \hypo{}
      \infer1[\textsc{Put}]{
        \alpha \vDash \S{put}\, k\, v \Downarrow ()
      }
    \end{prooftree}
    \ \
    \begin{prooftree}
      \hypo{ \pevapp{put}{k\,v}{()} \not\in \alpha}
      \infer1[\textsc{ExistsF}]{
        \alpha \vDash \S{exists}\,k \Downarrow \Code{false}
      }
    \end{prooftree}
    \ \
    \begin{prooftree}
      \hypo{ \pevapp{put}{k\,v}{()} \in \alpha}
      \infer1[\textsc{ExistsT}]{
        \alpha \vDash \S{exists}\,k \Downarrow \Code{true}
      }
    \end{prooftree}
    \ \
    \begin{prooftree}
      \hypo{ \pevapp{put}{k\,v}{()} \not\in \alpha'}
      \infer1[\textsc{Get}]{
        \alpha \listconcat [\pevapp{put}{k\,v}{()}] \listconcat \alpha' \vDash \S{get}\,k \Downarrow v
      }
    \end{prooftree}
    \label{ex:reduction}
    \\ \ \\ }
  \noindent The rule \textsc{Put} asserts that \zput{} operations
  always succeed. For a particular key, the \textsc{ExistsF} and
  \textsc{ExistsT} rules stipulate that \zexist{} returns
  {\small$\Code{false}$} or {\small$\Code{true}$}, based on the
  presence of a corresponding \zput{} event in the effect
  context. Finally, the rule \textsc{Get} specifies that the result of
  a \zget{} event is the \emph{last} value written by a \zput{} event
  to that key. Note that a program will get stuck if it attempts to
  \zget{} a key in a trace without an appropriate \zput{} event, since
  there is no rule covering this situation.

  All the traces appearing in \autoref{sec:overview} can be derived
  from these rules. Under the effect context $\alpha_{0}$, for
  example, the expression,
  {\small$\Code{add\ ``/a/b.txt"\ bytesFile}$}, induces the trace
  {\small$\alpha_2$} as follows: the \motiadd{} function uses the
  \zexist{} operator on the key {\small $\Code{``/a/b.txt"}$} (line
  {\small$6$}) to check if this path is in the file system. Since the
  context trace {\small$\alpha_0$} lacks a \zput{} event with the key
  {\small$\Code{``/a/b.txt"}$}, we have {\small
    \begin{align*}
      [\evapp{put}{``/"\,bytesDir}{()}] \vDash
      \S{exists}\Code{\ ``/a/b.txt"} \Downarrow \Code{false} \quad
      (\textsc{existsF})
    \end{align*}}\noindent
  The trace used to evaluate subsequent expressions records this event:
  {\small$\alpha_{0} \, \listconcat{} \,
    [\evapp{exists}{``/a/b.txt"}{false}]$}. The remaining evaluation of
  {\small $\Code{add\ ``/a/b.txt"\ bytesFile}$} is similar,
  yielding the following execution: %
  {\small
    \begin{align*}
      \msteptr{[\evapp{put}{``/"\,bytesDir}{()}]}{\Code{add\ ``/a/b.txt"\ bytesFile}}{[\evapp{exists}{``/a/b.txt"}{false};\, \evapp{exists}{``/a"}{false}]}{\Code{false}}
    \end{align*}}\noindent
\end{example}

\section{Type System}\label{sec:typing}

The syntax of \langname's types is shown in \autoref{fig:type-syntax}.
Our type system uses \emph{pure} refinement types to describe pure
computations and \emph{Hoare Automata Types} (HATs) to describe
effectful computations. Our pure refinement types are similar to those
of other refinement type systems~\cite{JV21}, and allow base types
(e.g., {\small\Code{int}}) to be further constrained by a logical
formula or \emph{qualifier}. Qualifiers in \langname{} are universally
quantified formulae over linear arithmetic operations
({\small$\primop$}) as well as uninterpreted functions, or
\emph{method predicates} ({\small$\mpred$}), e.g.,
{\small$\Code{isRoot}$}.  Verification conditions generated by our
type-checking algorithm can nonetheless be encoded as effectively
propositional (EPR) sentences~\cite{Ramsey1987}, which can be
efficiently handled by an off-the-shelf theorem prover such as
Z3~\cite{de2008z3}. 
We also allow function
types to be prefixed with a set of \emph{ghost variables} with base
types ({\small$\overline{x{:}b}\garr \tau$}). Ghost
variables are purely logical -- they can only appear in other
qualifiers and are implicitly instantiated when the function is
applied.

\begin{figure}[t!]
{\footnotesize
{\small\begin{flalign*}
 &\text{\textbf{Type Syntax}} &
\end{flalign*}}
\vspace{-1.8em}
{\small
\begin{alignat*}{2}
    \text{\textbf{Base Types}}& \quad & b  ::= \quad &  \Code{unit} ~|~ \Code{bool} ~|~ \Code{nat} ~|~ \Code{int} ~|~ ...
    \\ \text{\textbf{Basic Types}}& \quad & s  ::= \quad & b ~|~ s \sarr s
    \\ \text{\textbf{Value Literals}}& \quad & l ::= \quad & c ~|~ x ~|~ \primop\ \overline{l} ~|~ \mpred\ \overline{l}
\\ \text{\textbf{Qualifiers}}& \quad & \phi ::= \quad & l ~|~ \bot ~|~ \top  ~|~ \neg \phi ~|~ \phi \land \phi ~|~ \phi \lor \phi ~|~ \phi \impl \phi ~|~ \forall x{:}b.\, \phi
    \\ \text{\textbf{Refinement Types}}& \quad &t ::= \quad & \nuot{\textit{b}}{\phi} ~|~ x{:}t\sarr\tau  ~|~ x{:}b\garr t
    \\ \text{\textbf{Symbolic Finite Automata}}& \quad &A,B  ::= \quad & \evop{\effop{}}{\overline{x}}{\phi} ~|~ \evparenth{\phi} ~|~ \negA A ~|~ A \landA A ~|~ A \lorA A ~|~ A \seqA A ~|~ \nextA A ~|~ A \untilA A
    \\ \text{\textbf{Hoare Automata Types}}& \quad & \tau  ::= \quad &
    \htriple{A}{t}{A} ~|~ \tau \interty \tau
    \\\text{\textbf{Type Contexts}}& \quad &\Gamma ::= \quad  &\emptyset ~|~ x{:}t, \Gamma
  \end{alignat*}}
\vspace{-1.8em}
{\small\begin{flalign*}
 &\text{\textbf{Type Aliases}} &
\end{flalign*}}
\vspace{-1.8em}
\begin{alignat*}{5}
    \evop{\effop}{... {\sim}{\text{$v_x$}} ...}{\phi} &\doteq \evop{\effop}{... x ...}{\I{x} = v_x \land \phi}
    \quad&\finalA A &\doteq \topA \untilA A
    \quad& \globalA A &\doteq \negA \finalA \negA A
    \quad&  \lastA &\doteq \neg \nextA \topA
    \quad& b &\doteq \rawnuot{b}{\top}
  \end{alignat*}
}
\vspace{-.5cm}
\caption{\langname{} types.}
\label{fig:type-syntax}
\vspace{-.2cm}
\end{figure}

Unique to \langname{} are the HATs ascribed to a stateful
computation, which constrain the traces it may produce. HATs use
\emph{Symbolic Finite Automata} (SFAs)~\cite{Vea13,SFA-minimization,
  SFA-Transducers} to qualify traces, similar to how standard
refinement types use formulae to qualify the types of pure terms. We
adopt the symbolic version of \emph{linear temporal logic on finite
  traces}~\cite{LTLf} as the syntax of SFAs.\footnote{Our type system
  is agnostic to the syntax used to express SFAs, and our
  implementation uses the more developer-friendly syntax of Symbolic
  Regular Languages~\cite{LTLf}; this syntax is provided in
  \techreport{}. }

As shown in \autoref{fig:type-syntax}, HATs support two kinds of
atomic propositions, or \emph{symbolic effect events}:
{\small$\evop{\effop}{\overline{x}}{\phi}$} and
{\small$\evparenth{\phi}$}. A symbolic effect event
{\small$\evop{\effop}{\overline{x}}{\phi}$} describes an application
of an effectful operator {\small$\effop$} to a set of argument
variables {\small$\overline{x}$} that produces a result variable
{\small$\vnu$}. The formula {\small$\phi$} constrains the possible
instantiation of {\small$\overline{x}$} and {\small$\vnu$} in a
concrete event. The other symbolic effect event,
{\small$\evparenth{\phi}$}, is used to constrain the relationship
between pure values (e.g., ghost variables and function arguments). In
addition to the temporal next ({\small$\nextA A$}) and until operators
({\small$A \untilA A$}), the syntax of HATs includes negation
({\small$\neg A$}), intersection ({\small$A \landA A$}), union
({\small$A \lorA A$}), and concatenation ({\small$A;A$}).
\autoref{fig:type-syntax} defines notations for other useful logical
and temporal operators: implication
({\small$A \impl B \doteq \neg A \lor B$}), the eventually operator
{\small$\finalA$}, the globally operator {\small$\globalA$}, and
importantly, the last modality {\small$\lastA$}, which describes a
singleton trace, thus prohibiting a trace from including any other
effects~\cite{LTLf}.  \autoref{fig:type-syntax} also provides notations
for specifying the value of an argument {\small
  $\evop{\effop}{{\overline{x}~\sim}{\text{$v_x$}}~\overline{x}}{\phi}$}.
As is standard, we abbreviate the qualified type
{\small$\rawnuot{b}{\top}$} as {\small$b$}.
\begin{example}
  This syntax can capture rich properties over effect contexts.  For
  example, the following formulae concisely specify traces under which
  a key holds a certain value ({\small$\predA{stored}{}$}) and when a particular
  key exists in the current store ({\small$\predA{exists}{}$}): {\footnotesize
    \begin{flalign*}
      &\predA{stored}{}(\Code{key,value}) \doteq \finalA
      (\evop{put}{\df{y}{key}\,\df{z}{value}}{\top} \land \nextA \globalA
      \negA \evop{put}{\df{y}{key}\,x}{\top})
      \quad\predA{exists}{}(\Code{key}) \doteq \finalA
      \evop{put}{\df{y}{key}\,x}{\top}
    \end{flalign*}}\noindent
  \label{ex:SFA+put+get}
\end{example}

\subsection{HATs, By Example}
\label{sec:T-HAT}
Formally, a Hoare Automata Type {\small $\htriple{A}{t}{B}$} qualifies
a refinement type {\small$t$} with two SFAs: a \emph{precondition}
SFA {\small$A$} that captures the context traces in which a term can
be executed and a \emph{postcondition} SFA {\small$B$} that
describes the effect trace after the execution of the term.
Our type system also includes intersection types on HATs ({\small
  $\tau \interty \tau$}), in order to precisely specify the behaviors
of a program under different effect contexts.

\begin{example}[Built-in Effectful Operators]
  \label{ex:Typing+Delta}
  Our type system is parameterized over a (global) typing context
  $\Delta$ containing the types of built-in operators. Intuitively,
  the signatures of effectful operators in $\Delta$ captures the
  semantics the library developer ascribes to their API. Using the
  SFAs from Example~\ref{ex:SFA+put+get}, for example, the signatures
  of \zput{}, \zexist{}, and \zget{} are: {\footnotesize
\begin{flalign*}
  \Delta(\S{put}) =&\ \normalvn{k}{Path.t} \normalvn{a}{Bytes.t}
  \incrhtriple{\globalA\topA}{\Code{unit}}{(\evop{put}{\df{x}{k}\,
      \df{y}{a}}{\top}) \land \lastA}
  \\\Delta(\S{get}) =&\ \gvn{a}{Bytes.t} \normalvn{k}{Path.t}
  \incrhtriple{\predA{stored}{}(\Code{k,a})}{\nuot{Bytes.t}{\vnu =
      \Code{a}}}{(\evop{get}{\df{x}{k}}{\vnu = \Code{a}}) \land \lastA}
  \\\Delta(\S{exists}) =&\ \normalvn{k}{Path.t}
  \incrhtriple{\predA{exists}{}(\Code{k})}{\nuot{bool}{\vnu =
      \Code{true}}}{(\evop{exists}{\df{x}{k}}{\vnu = \Code{true}}) \land \lastA} \interty
  \\&\qquad\quad\quad \ \ \ \,
  \incrhtriple{\neg \predA{exists}{}(\Code{k})}{\nuot{bool}{\vnu =
      \Code{false}}}{(\evop{exists}{\df{x}{k}}{\vnu = \Code{false}}) \land \lastA}
\end{flalign*}}\noindent The postcondition SFAs of all three operators
use {\small$\lastA$}, to indicate that they append a single effect event to
the end of the effect context
({\small$\evop{\effop}{\overline{v}}{\phi} \land \lastA $}). The type
of \zput{} permits it to be used under any context ({\small
  $\globalA\topA$}). More interestingly, the built-in type of \zget{}
uses a ghost variable ({\small$\Code{a}$}) to represent the current
value of the key {\small$\Code{k}$}; its precondition stipulates that
it should only be applied to keys that have been previously
stored. Finally, the intersection type of \zexist{} describes two
possible behaviors, depending on whether the key {\small$\Code{k}$}
has been previously added to the store.
\cbnewadding{%
    Note that the pre- and post-condition SFAs in these signatures
    express different constraints on the allowed contexts in which the
    operator may execute, and the effects the operator performs.}%

\end{example}


\begin{example}[MinSet]
  \label{ex:min-set}
  Consider a developer who wants to implement an API for a set
  augmented with an additional operation that returns the minimum
  element in the set. This implementation is defined in terms of two
  other stateful libraries: a \textsf{Set} ADT that provides
  effectful operators {\small$\Code{\textbf{insert}{:}int\sarr unit}$}
  and {\small$\Code{\textbf{mem}{:}int\sarr bool}$}, and a
  \textsf{MemCell} library that provides a persistent cell with
  effectful {\small$\Code{\textbf{read}{:}unit\sarr int}$} and
  {\small$\Code{\textbf{write}{:}int\sarr unit}$} operators. One
  implementation strategy is to track the minimum element in a
  persistent cell, and maintain the representation invariant that the
  contents of this cell are always less than or equal to every element
  in the underlying set. Using SFAs, we can express this invariant as:
  {\footnotesize
    \begin{align*}
      \riA{MinSet}(\Code{el}) \doteq&\
                                     \finalA (\evop{write}{\df{x}{el}}{\top} \land \nextA \globalA \negA \evop{write}{x}{\top}) \impl
                                     \finalA \evop{insert}{\df{x}{el}}{\top} \landA \globalA \negA \evop{insert}{x}{\I{x} < \Code{el}}
      \\&\land (\globalA\neg \evop{write}{x}{\top} \impl \globalA\neg \evop{insert}{x}{\top})
    \end{align*}}\noindent
  which specifies that an element {\small$\Code{el}$} is written
  into the persistent cell only when {\small$\Code{el}$} has been inserted
  into the backing \textsf{Set} \emph{and} no element less than
  {\small$\Code{el}$} has also been inserted. The HAT ascribed to the
  function {\small$\Code{minset\_insert}$} enforces this invariant
  {\small
    \begin{align*}
      & \gvn{el}{int} \normalvn{elem}{int} \htriple{\riA{MinSet}(\Code{el})}{\Code{unit}}{\riA{MinSet}(\Code{el})}
    \end{align*}}
\end{example}

\begin{example}[LazySet]
  \label{ex:lazy-set}
  As our next example, consider a \textsf{LazySet} ADT that provides a
  lazy version of an {\small \Code{insert}} operation, delaying when
  elements are added to an underlying \textsf{Set} ADT. The ADT does
  so by returning a thunk closed over a collection of elements that
  will be added to the set when it is forced. This ADT maintains the
  representation invariant that an element is never inserted twice:
  {\small
    \begin{align*}
      \riA{LSet}(\Code{el}) \doteq
      \globalA (\evop{insert}{\df{x}{el}}{\top} \impl \nextA \neg \finalA \evop{insert}{\df{x}{el}}{\top})
    \end{align*}}\noindent
  We can specify a ``\emph{lazy}'' insert operator using the
  following HAT:
  {\small
    \begin{align*}
      \gvn{el}{int} \normalvn{elem}{int} \Code{thunk}{:}(&\Code{unit}\sarr \htriple{\riA{LSet}(\Code{el})}{\Code{unit}}{\riA{LSet}(\Code{el})}) \sarr
                                                          \Code{unit}\sarr \htriple{\riA{LSet}(\Code{el})}{\Code{unit}}{\riA{LSet}(\Code{el})}
    \end{align*}}\noindent
  This operation takes as input a (potentially new) element {\small \Code{elem}}
  to be added to the set and a thunk holding any elements
  that have not yet been inserted into the backing set, and
  returns another thunk. This HAT stipulates that both input
  thunk and output thunk preserve the representation invariant of the
  ADT:
  {\small$\Code{unit}\sarr
    \htriple{\riA{LSet}(\Code{el})}{\Code{unit}}{\riA{LSet}(\Code{el})}$}.
\end{example}



\begin{example}[DFA]
  \label{ex:mst}
  Our final example is a library built on top of a stateful graph
  library that is used to maintain the states and transitions of a
  Deterministic Finite Automaton (DFA). The underlying graph library
  exports two methods to add (\zconnect{}) and remove (\zdisconnect{})
  edges. We can specify the standard invariant that DFAs only have
  deterministic transitions: {\footnotesize
    \begin{align*}
      \riA{DFA}(\Code{n,c}) \doteq \ &\globalA
                                       \neg(\evop{connect}{\df{w}{n}\,\df{y}{c}\,n_{end}}{\top}
                                       \land \nextA(\neg
                                       \evop{disconnect}{\df{w}{n}\,\df{y}{c}\,n_{end}}{\top}
                                       \untilA \evop{connect}{\df{w}{n}\,\df{y}{c}\,n_{end}}{\top}))
    \end{align*}}\noindent
{\small$\riA{DFA}$} stipulates that a node {\small$\Code{n}$} can have at most one transition on a character {\small$\Code{c}$} ({\small
  $\evop{connect}{\df{w}{n}\,\df{y}{c}\,n_{end}}{\top}$}). Moreover, adding a new
transition from {\small$\Code{n}$} on {\small$\Code{c}$} requires that
any previous transitions on {\small$\Code{c}$} have first been removed
({\small $\nextA(\neg \evop{disconnect}{\df{w}{n}\,\df{y}{c}\,n_{end}}{\top}
  \untilA \evop{connect}{\df{w}{n}\,\df{y}{c}\,n_{end}}{\top} $}). An
{\small$\Code{add\_transition}$} method with the following signature
is thus guaranteed to preserve determinism: {\small
    \begin{align*}
    \Code{n{:}Node.t} \garr \Code{c}{:}\Code{Char.t} \garr
    \Code{n\_start}{:}\Code{Node.t} \sarr \Code{char}{:}\Code{Char.t} \sarr \Code{n\_end}{:}\Code{Node.t} \sarr
    \htriple{\riA{DFA}(\Code{n, c})}{\Code{unit}}{\riA{DFA}(\Code{n, c})}
    \end{align*}}\noindent
\end{example}

\subsection{Typing Rules}
\label{sec:type-rules}

\paragraph{Type contexts} A type context (shown in
\autoref{fig:type-syntax}) is a sequence of bindings from variables to
pure refinement types (i.e., {\small$t$}). Type contexts are not
allowed to contain HATs, as doing so breaks several structural
properties (e.g., weakening) that are used to prove type
safety.\footnote{Intuitively, since a type of the form
    $\htriple{A}{t}{B}$ describes a stateful computation, it
  cannot be duplicated or eliminated, and thus it would be unsafe to
  allow unrestricted use of ``computational'' variables, as would be
  possible if type contexts could contain such bindings.}

\begin{figure}[!t]
{\footnotesize
{\small\begin{flalign*}
 &\text{\textbf{Type Erasure}} & \fbox{$\eraserf{t} \quad \eraserf{\tau} \quad \eraserf{\Gamma}$}
\end{flalign*}}
\vspace{-1em}
\begin{alignat*}{5}
    \eraserf{\rawnuot{b}{\phi}} &\doteq b \quad&
    \eraserf{x{:}t \sarr \tau} &\doteq \eraserf{t}\sarr\eraserf{\tau} \quad&
    \eraserf{x{:}b \garr t} &\doteq \eraserf{t} \quad&
    \eraserf{\htriple{A}{t}{B}} &\doteq \eraserf{t} \quad&
    \eraserf{ \tau_1 \interty \tau_2 } &\doteq \eraserf{\tau_1}
    \\
    \eraserf{\emptyset} &\doteq \emptyset \quad&
    \eraserf{x{:}t, \Gamma} &\doteq x{:}\eraserf{t}, \eraserf{\Gamma}
\end{alignat*}
{\small
\begin{flalign*}
&\text{\textbf{Well-Formed Types}} & \fbox{$\Gamma \wellfoundedvdash t \quad \Gamma \wellfoundedvdash A \quad \Gamma \wellfoundedvdash \tau$}
\end{flalign*}
}
\begin{prooftree}
\hypo{
\parbox{50mm}{\center
  $\Gamma \wellfoundedvdash A \quad \Gamma \wellfoundedvdash B \quad \Gamma \wellfoundedvdash t$ \\
  $\forall \sigma \in \denotation{\Gamma}. \langA{\sigma(B)} \subseteq \langA{\sigma(A;\globalA\topA)}$
}
}
\infer1[\textsc{WFHoare}]{
\Gamma \wellfoundedvdash \htriple{A}{t}{B}
}
\end{prooftree}
\quad
\begin{prooftree}
\hypo{
\parbox{50mm}{\center
  $\Gamma \wellfoundedvdash \tau_1$\quad
  $\Gamma \wellfoundedvdash \tau_2$ \quad
  $\eraserf{\tau_1} = \eraserf{\tau_2}$
}
}
\infer1[\textsc{WFInter}]{
\Gamma \wellfoundedvdash \tau_1 \interty \tau_2
}
\end{prooftree}
{\small
\begin{flalign*}
&\text{\textbf{Automata Inclusion }} & \fbox{$\Gamma \vdash A \subseteq A$}\quad
&\text{\textbf{Subtyping }} & \fbox{$\Gamma \vdash t <: t \quad \Gamma \vdash \tau <: \tau$}
\end{flalign*}
}
\begin{prooftree}
\hypo{
\parbox{30mm}{\center
  $\forall \sigma \in  \denotation{\Gamma}.$
  $\langA{\sigma(A_1)} \subseteq  \langA{\sigma(A_2)} $
}
}
\infer1[\textsc{SubAutomata}]{
\Gamma \vdash A_1 \subseteq A_2
}
\end{prooftree}
\quad
\begin{prooftree}
\hypo{
\parbox{35mm}{\center
  $\Gamma \vdash A_2 \subseteq A_1$ \quad
  $\Gamma \vdash t_1 <: t_2$ \\
  $\Gamma \vdash (A_2; \globalA\topA) \landA B_1 \subseteq$ \\
  $\ \  (A_2; \globalA\topA) \landA B_2$
}
}
\infer1[\textsc{SubHoare}]{
\Gamma \vdash \htriple{A_1}{t_1}{B_1} <: \htriple{A_2}{t_2}{B_2}
}
\end{prooftree}
\quad
\ \\ \ \\ \ \\
\begin{prooftree}
\hypo{}
\infer1[\textsc{SubIntLL}]{
\Gamma \vdash \tau_1 \interty \tau_2 <: \tau_1
}
\end{prooftree}
\ \
\begin{prooftree}
\hypo{}
\infer1[\textsc{SubIntLR}]{
\Gamma \vdash \tau_1 \interty \tau_2 <: \tau_2
}
\end{prooftree}
\ \
\begin{prooftree}
\hypo{
}
\infer1[\textsc{SubIntMerge}]{
\Gamma \vdash \htriple{A_1}{t}{B} \interty \htriple{A_2}{t}{B} <: \htriple{A_1 \lor A_2}{t}{B}
}
\end{prooftree}
\ \\ \ \\ \ \\
\begin{prooftree}
\hypo{
\parbox{15mm}{\center
  $\Gamma \vdash \tau <: \tau_1 $ \quad
  $\Gamma \vdash \tau <: \tau_2 $
}
}
\infer1[\textsc{SubIntR}]{
\Gamma \vdash \tau <: \tau_1 \interty \tau_2
}
\end{prooftree}
\quad
\begin{prooftree}
\hypo{
\Gamma, x{:}\rawnuot{b}{\top} \vdash t_1 <: t_2
}
\infer1[\textsc{SubGhostR}]{
\Gamma \vdash t_1 <: x{:}b\garr t_2
}
\end{prooftree}
\quad
\begin{prooftree}
\hypo{
\parbox{30mm}{\center
  $\eraserf{\Gamma} \basicvdash v : b$ \quad
  $\Gamma \vdash t_1[x\mapsto v] <: t_2$
}
}
\infer1[\textsc{SubGhostL}]{
\Gamma \vdash x{:}b\garr t_1 <: t_2
}
\end{prooftree}
}
\caption{Selected auxiliary typing
  relations. {\small$\eraserf{\Gamma} \basicvdash e : \eraserf{t}$}
  is the standard typing judgment for basic types.}
\label{fig:aux-rules}
\end{figure}

\paragraph{Auxiliary typing relations} \autoref{fig:aux-rules} shows
selected rules from three sets of auxiliary relations used by our type
system.  The first set describes the type erasure functions
{\small$\eraserf{t}$}, {\small$\eraserf{\tau}$}, and
{\small$\eraserf{\Gamma}$}. The former two functions return basic
types by erasing all qualifiers and automata from types, while the
latter {\small$\eraserf{\Gamma}$} is the type context derived by
applying {\small$\eraserf{...}$} to all of {\small${\Gamma}$}'s
bindings. The second set describes well-formedness conditions on SFAs
({\small$\Gamma \wellfoundedvdash A$}) and types
({\small$\Gamma \wellfoundedvdash t$} and
{\small$\Gamma \wellfoundedvdash \tau$}). These rules largely ensure
that all the qualifiers appearing in a type are closed under the
current typing context {\small$\Gamma$}. The exceptions are
$\textsc{WFInter}$, which stipulates that only HATs with the same basic
type can be intersected, and $\textsc{WFHoare}$, which requires the
language accepted by the precondition SFA to be a prefix of the
postcondition automata (i.e.,
{\small$\langA{B} \subseteq \langA{A;\globalA\topA}$}, where
{\small$\globalA\topA$} denotes the SFA that accepts an arbitrary
trace) after performing a substitution consistent with the current
typing context ({\small$\sigma \in \denotation{\Gamma}$}); the
functions {\small$\langA{...}$} and {\small$\denotation{...}$} are
defined in the next subsection.

Our type system also uses a mostly-standard subtyping relation that
aligns with the denotation of the types being related.
\autoref{fig:aux-rules} highlights key subtyping rules that do not
have analogues in a standard refinement type system. The subtyping
rule for symbolic automata ($\textsc{SubAutomata}$) checks inclusion
between the languages of the automata, after performing substitutions
consistent with the current typing context. The subtyping rule for
(non-intersected) HATs ($\textsc{SubHoare}$) checks that the inclusion
is contravariant over the precondition automata and covariant over the
postcondition automata under the same context (i.e., the conjunction
of {\small$B_1$} and {\small$B_2$} with {\small$A_2;
  \globalA\topA$}). The subtyping rules for the intersection of HATs -
$\textsc{SubIntLL}$, $\textsc{SubIntLR}$, and $\textsc{SubIntR}$ - are
standard.
The $\textsc{SubIntMerge}$ rule additionally allows the precondition
automata of intersected types to be merged. Finally, the subtyping
rules for ghost variables either bind a ghost variable in the type
context ($\textsc{SubGhostR}$), or instantiate it to some concrete
value ($\textsc{SubGhostL}$).

\begin{figure}[!t]
{\footnotesize
{\small
\begin{flalign*}
 &\text{\textbf{Typing}} & \fbox{$\Gamma \vdash op : t \quad \Gamma \vdash \effop : t \quad \Gamma \vdash e : t \quad \Gamma \vdash e : \tau$}
\end{flalign*}
}
\\ \
\begin{prooftree}
\hypo{
  \parbox{12mm}{\center
    $\Gamma \wellfoundedvdash A$\quad
    $\Gamma \vdash e : t$
  }}
\infer1[\textsc{\footnotesize TEPur}]{
  \Gamma \vdash e : \htriple{A}{t}{A}
}
\end{prooftree}
\quad
\begin{prooftree}
\hypo{
\parbox{15mm}{\center
$\Gamma \wellfoundedvdash \tau_2$\quad
$\Gamma \vdash \tau_1 <: \tau_2$\quad
$\Gamma \vdash e : \tau_1$
}}
\infer1[\textsc{\footnotesize TSub}]{
\Gamma \vdash e : \tau_2
}
\end{prooftree}
\quad
\begin{prooftree}
\hypo{
\parbox{25mm}{\center
$\Gamma \wellfoundedvdash \tau_1 \interty \tau_2$\quad
$\Gamma \vdash e : \tau_1$ \quad $\Gamma \vdash e : \tau_2$
}}
\infer1[\textsc{\footnotesize TInter}]{
\Gamma \vdash e : \tau_1 \interty \tau_2
}
\end{prooftree}
\quad
\begin{prooftree}
\hypo{
\parbox{24mm}{\center
$\Gamma \wellfoundedvdash x{:}b\garr t$\quad
$\Gamma, x{:}\rawnuot{b}{\top} \vdash v :t$
}}
\infer1[\textsc{\footnotesize TGhost}]{
\Gamma \vdash v : x{:}b\garr t
}
\end{prooftree}
\\ \ \\ \ \\ \
\begin{minipage}{.155\linewidth}
\begin{tabular}{c}
  \begin{prooftree}
\hypo{
\parbox{14mm}{\center
$\Gamma \wellfoundedvdash t$\quad
$\Delta(\primop) = t$
}}
\infer1[\textsc{\footnotesize TPOp}]{
\Gamma \vdash \primop : t
}
\end{prooftree}
  \\ \\
\begin{prooftree}
\hypo{
\parbox{12mm}{\center
$\Gamma \wellfoundedvdash t$\quad
$\Delta(\effop) = t$
}}
\infer1[\textsc{\footnotesize TEOp}]{
\Gamma \vdash \effop : t
}
\end{prooftree}
\end{tabular}
\end{minipage}
\quad
\begin{prooftree}
\hypo{
\parbox{36mm}{\center
$\Gamma \wellfoundedvdash \tau$\quad
$\Gamma \vdash \primop : \overline{z_i{:}t_i}\sarr t$\quad
$\forall i. \Gamma \vdash v_i : t_i$\\
$t_x = t\overline{[z_i \mapsto v_i]}$\\
$\Gamma, x{:}t_x \vdash{} e : \tau$
}}
\infer1[\textsc{\footnotesize TPOpApp}]{
\Gamma \vdash{} \zlet{x}{\primop\ \overline{v_i}}{e} : \tau
}
\end{prooftree}
\quad
\begin{prooftree}
\hypo{
\parbox{48mm}{\center
$\Gamma \wellfoundedvdash \htriple{A}{t}{B}$\quad
$\Gamma \vdash \effop : \overline{z_i{:}t_i}\sarr \tau$\quad
$\forall i. \Gamma \vdash v_i : t_i$ \\
$\htriple{A}{t_x}{A'} = \tau\overline{[z_i \mapsto v_i]}$ \quad
$\Gamma, x{:}t_x \vdash{} e :\htriple{A'}{t}{B}$
}}
\infer1[\textsc{\footnotesize TEOpApp}]{
\Gamma \vdash{} \perform{x}{\effop{}}{\overline{v_i}}{e} : \htriple{A}{t}{B}
}
\end{prooftree}
}
\caption{Selected typing rules. All typing judgements (i.e.,
  {\small$\Gamma \vdash e : t$} and {\small$\Gamma \vdash e : \tau$})
  assume the corresponding basic type judgement
  ({\small$\eraserf{\Gamma} \basicvdash e : \eraserf{t}$} holds).
}
\label{fig:selected-typing-rules}
\end{figure}


A subset of our typing rules\footnote{The complete set of typing rules
  for \langname{} is included in the \techreport{}.} is shown
in \autoref{fig:selected-typing-rules}. Note that an stateful
computations can only be ascribed a HAT. As mentioned in
Example~\ref{ex:Typing+Delta}, our type system is parameterized over
{\small$\Delta$}, a typing context for built-in operators that
provides HATs for both pure (\textsc{TPOp}) and effectful operators
(\textsc{TEOp}). This system features the standard subsumption and
intersection rules (\textsc{TSub} and \textsc{TInter}), which enables
our type system to use fine-grained information about the effect
context when typing different control flow paths.  The rule
\textsc{TEPur} allows a pure term to be treated as a computation that
does not perform any effects.

\begin{example} The \motiadd{} function in \autoref{fig:ex-fs} has
  four possible control flow paths, depending on the effect context
  under which it is called: (1) the input path exists in the file
  system (line {\small$6$}); (2) neither the input path or its parent
  path exist (line {\small$9$}); (3) its parent path exists and is
  a directory (line {\small$13-15$}), or (4) it is not (line
  {\small$16$}).  The following four automata (recall that the
  SFA {\small$\predA{isDir}{}$} was defined in
  \autoref{sec:overview}) indicate the effect context corresponding to
  each of these scenarios:

  {\footnotesize
    \begin{flalign*}
      & A_1 \doteq \riA{FS}(\Code{p})\land
        \predA{exists}{}(\Code{path}) \\
     & A_2 \doteq
        \riA{FS}(\Code{p})\land \neg \predA{exists}{}(\Code{path}) \land
        \neg \predA{exists}{}(\Code{parent(path)})   \\
     & A_2 \doteq
        \riA{FS}(\Code{p})\land \neg \predA{exists}{}(\Code{path}) \land
        \predA{exists}{}(\Code{parent(path)}) \land \predA{isDir}{}(\Code{parent(path)})  \\
      & A_3 \doteq
        \riA{FS}(\Code{p})\land \neg \predA{exists}{}(\Code{path}) \land
        \predA{exists}{}(\Code{parent(path)}) \land \neg\predA{isDir}{}(\Code{parent(path)})
    \end{flalign*}
  }\noindent  The union of these automata is exactly the
  representation invariant ({\small$\riA{FS}(\Code{p})$}), established
  from the following subtyping relation between their intersection and the return
  type of {\small$\tauadd{}$}:
  {
    \footnotesize
    \begin{align*} \Code{p{:}}\nuot{Path.t}{\top},
      \Code{path{:}}\nuot{Path.t}{\top}, \Code{bytes{:}}\nuot{Bytes.t}{\top} \vdash \biginterty_{i = 1..4}
      \ \htriple{A_i}{\Code{bool}}{\riA{FS}(\Code{p})} <:
      \htriple{\riA{FS}(\Code{p})}{\Code{bool}}{\riA{FS}(\Code{p})}
      \tag{\textsc{SubIntMerge}}
    \end{align*}
  }\noindent Using the \textsc{TInter} and \textsc{Tsub} rules, our
  type system is able to reduce checking \motiadd{} against
  {\small$\tauadd{}$} into checking \motiadd{} against each $A_i$
  (i.e., {\small$\htriple{A_i}{\Code{bool}}{\riA{FS}(\Code{p})}$}).
  Note that \motiadd{} only adds the given path to the file system
  in the third case (line {\small$13$}), whose precondition automata
  ({\small$A_2$}) indicates that the input path does not exist in the
  file system ({\small$\neg \predA{exists}{}(\Code{path})$}),
  although its parent path does
  ({\small$\predA{exists}{}(\Code{parent(path)})$}), and is a directory
  ({\small$\predA{isDir}{}(\Code{parent(path)})$}).
  When coupled with the representation invariant that the
  parent of a path is always a directory in the file system, our
  type system is able to ensure that it is safe to perform the \zput{}
  operation.
\end{example}


\paragraph{Effectful operator application}

\textsc{TPOpApp} is the standard rule for operator application in a
typical refinement type system~\cite{JV21}.  On the other hand, rule
\textsc{TEOpApp}, the rule for effect operator application, with the
help of the subsumption rule (\textsc{TSub}), allows operators to have
function types whose ghost variables are instantiated properly;
moreover, the return type of effect operators is a non-intersected
HAT. After ensuring each argument {\small$v_i$} types against the
corresponding argument type {\small$t_i$} and substituting all
parameters ({\small$\overline{z_i}$}) with the supplied arguments
({\small$\overline{v_i}$}), the return type of an effectful operator
must be in the form {\small$\htriple{A}{t_x}{A'}$} and have exactly
the same precondition SFA ({\small$A$}) as the type of the surrounding
term. Typing the \Code{let}-body {\small$e$} is similar to
\textsc{TPOpApp}, where a new binding {\small$x{:}t_x$} is added to
the type context. Moreover, the rule replaces the precondition SFA
used to type the body of the \Code{let} with the postcondition SFA
from the return type of the effect operator {\small$A'$}, so that the
body is typed in a context reflecting the use of the effectful
operator.

\newcommand\ShiftDown[2]{\raisebox{-#1}{\upshape\scriptsize #2}}

\subsection{Type Soundness}

\begin{figure}[t!]
\footnotesize{
{\small
\begin{flalign*}
 &\text{\textbf{Trace Language}} & \fbox{$\alpha, i \models A \quad  \langA{A} \in \mathcal{P}(\alpha)$}
\end{flalign*}}
\vspace{-1em}
\begin{flalign*}
    \mathit{Tr}^\textbf{WF} &\doteq \{\alpha ~|~ \forall (\pevapp{\effop}{\overline{v_i}}{v}) \in \alpha. \emptyset \basicvdash \effop : \overline{b_i}\sarr b \land (\forall i.\emptyset \basicvdash v_i : b_i) \land \emptyset \basicvdash v : b  \}
    \quad& \langA{A} &\doteq \{\alpha \in \mathit{Tr}^\textbf{WF} ~|~ \alpha, 0 \models A \}
\end{flalign*}
\vspace{-1.5em}
\begin{flalign*}
\alpha, i &\models \evop{\effop{}}{\overline{x_j}}{\phi} \iff \alpha[i] = \pevapp{\effop}{\overline{v_j}}{v} \land \phi\overline{[x_j\mapsto v_j]}[\vnu \mapsto v]
    \quad&
\alpha, i &\models A \landA A' \iff \alpha, i \models A \land \alpha, i \models A'
\\\alpha, i &\models \evparenth{\phi} \iff \alpha[i] = \pevapp{\effop}{\overline{v_j}}{v} \land \phi
    \quad&
    \alpha, i &\models A \lorA A' \iff \alpha, i \models A \lor \alpha, i \models A'
\end{flalign*}
\vspace{-1.5em}
\begin{flalign*}
    \alpha, i &\models \nextA A \iff \alpha, i{+}1 \models A
    \quad&
    \alpha, i &\models A_1;A_2 \iff \alpha[i...\I{len(\alpha)}] = \alpha_1 \listconcat \alpha_2 \land \alpha_1 \in \langA{A_1} \land \alpha_2 \in \langA{A_2}
    \\\alpha, i &\models \neg A \iff \alpha, i \not\models A
     \quad&
     \alpha, i &\models  A \untilA A' \iff \exists j. i \leq j < \I{len}(\alpha). \alpha, j \models  A' \land \forall k. i \leq k < j \impl \alpha, k \models  A
\end{flalign*}
\vspace{-1.5em}
{\small
\begin{flalign*}
  &\text{\textbf{Type Denotation}}
  & \fbox{$\denotation{t} \in \mathcal{P}(e)
    \quad \denotation{\tau} \in \mathcal{P}(e)$}
\end{flalign*}}
\vspace{-1.5em}
\begin{align*}
  &\denotation{ \rawnuot{b}{\phi} } &&\doteq \{ e ~|~ \emptyset
                                       \basicvdash e : b \land
                                       \forall \alpha\ \alpha'\
                                       v. \; \msteptr{\alpha}{e}{\alpha'}{v} \impl \alpha' = \emptr \land \phi[\nu\mapsto v] \}
  \\ &\denotation{ x{:}t\sarr{}\tau } &&\doteq \{ e ~|~ \emptyset \basicvdash e : \eraserf{x{:}t\sarr{}\tau} \land
                                           \forall v \in \denotation{ t }.\, e\ v \in  \denotation{ \tau[x\mapsto v ] } \}
  \\ &\denotation{ x{:}b\garr{}\tau } &&\doteq \{ e ~|~ \emptyset \basicvdash e : \eraserf{\tau} \land
                                           \forall v. \emptyset \basicvdash v : b \impl e \in  \denotation{ \tau[x\mapsto v ] } \}
  \\ &\denotation{ \htriple{A}{t}{B}} &&\doteq \{e ~|~ \emptyset \basicvdash e : \eraserf{t} \land \forall \alpha\, \alpha'\, v.\; \alpha \in \langA{A} \land \msteptr{\alpha}{e}{\alpha'}{v} \impl v \in \denotation{t} \land \alpha \listconcat \alpha' \in \langA{B} \}
  \\ &\denotation{ \tau_1 \interty \tau_2 } &&\doteq \denotation{\tau_1} \cap \denotation{\tau_2}
\end{align*}
\vspace{-1.5em}
{\small
\begin{flalign*}
  &\text{\textbf{Type Context Denotation}}
  & \fbox{$\denotation{\Gamma} \in \mathcal{P}(\sigma)$}
\end{flalign*}}
\vspace{-1.2em}
\begin{flalign*}
    \denotation{ \emptyset } &\doteq \{ \emptyset \} \qquad\qquad\qquad\qquad&
    \denotation{ x{:}t, \Gamma } &\doteq \{ \sigma[x\mapsto v] ~|~ v\in \denotation{t}, \sigma \in \denotation{\Gamma[x\mapsto v]} \}
\end{flalign*}
}
\vspace{-0.4cm}
\caption{Type denotations in \langname{}}
\label{fig:type-denotation}
\vspace{-.4cm}
\end{figure}

The denotation of a SFA {\small$\langA{A}$}, shown in
\autoref{fig:type-denotation}, is the set of traces accepted or
\emph{recognized} by the automata; the denotation of refinement types
{\small$\denotation{t}$} and HATs {\small$\denotation{\tau}$} are sets
of terms; these functions are defined mutually recursively.

The language of traces ranges over the set of all \emph{well-formed
  traces} $\mathit{Tr}^\textbf{WF}$, i.e., those only containing
events ({\small$\pevapp{\effop}{\overline{v_i}}{v}$}) that are
well-typed according to the basic type system. As is
standard~\cite{LTLf}, our denotation uses an auxiliary judgement
{\small$\alpha,i \models A$} that defines when the suffix of a trace
{\small$\alpha$} starting at the {\small$i^{th}$} position is accepted
by an automaton. The denotation of an SFA $A$ is the set of
(well-formed) traces recognized by $A$, starting at index $0$.




\paragraph{Type denotations} Out type denotations use a basic typing
judgement {\small$\emptyset \basicvdash e : s$} that types an expression
{\small$e$} according to the standard typing rules of the simply-typed
lambda-calculus (STLC), erasing all qualifiers in function argument
types.  The denotation of pure refinement types is standard~\cite{JV21}. The denotation of ghost variables is similar to function
types whose parameters are restricted to base types.  The denotation
of an intersection type {\small$\tau_1 \interty \tau_2$} is the
intersection of the denotations {\small$\tau_1$} and
{\small$\tau_2$}. An expression {\small$e$} belongs to the denotation
of a HAT {\small$\htriple{A}{t}{B}$} iff every trace and value
produced by {\small$e$} is consistent with the SFA {\small$B$} and
refinement type $t$, under any effect context accepted by the SFA
{\small$A$}. Intuitively, the denotation of a HAT is naturally derived
from the language's operational semantics, as depicted by the following
correspondence: {\small
\begin{alignat*}{3}
  \text{Pure Language:}&& \ \ \mstep{e}{v} &\impl e \in
  \denotation{\rawnuot{b}{\phi}}
  \ &&\text{ where } \phi[\vnu \mapsto v] \text{ is valid} \\
  \text{\langname:}&& \ \ \msteptr{\alpha}{e}{\alpha'}{v} &\impl e
  \in \denotation{\htriple{A}{\rawnuot{b}{\phi}}{B}}
  \ &&\text{ where } \alpha \in \langA{A}, \alpha \listconcat \alpha' \in
  \langA{B}, \text{and }\phi[\vnu \mapsto v] \text{ is valid}
\end{alignat*}}\noindent
In a pure language with the simple multi-step reduction
relation {\small$\mstep{e}{v}$}, refinement types qualify the value
{\small $v$} that {\small$e$} produces. In contrast, the multi-step
reduction relation of \langname{},
{\small$\msteptr{\alpha}{e}{\alpha'}{v}$} depends on the effect
context {\small$\alpha$} and emits a trace {\small$\alpha'$} that
records the sequence of effects performed by {\small$e$} during
evaluation. Thus, HATs use precondition and postcondition automata
(i.e., {\small$A$} and {\small$B$}) to qualify the trace before and
after a term is evaluated (i.e., {\small$\alpha$} and
{\small$\alpha \listconcat \alpha'$}).

\paragraph{Type context denotation} The denotation of a type context
is a set of \emph{closing substitutions} {\small$\sigma$}, i.e., a
sequence of bindings {\small$\overline{[x_i \mapsto v_i]}$} consistent
with the type denotations of the corresponding variables in the type
context. The denotation of the empty context is a singleton set
containing only the identity substitution {\small$\emptyset$}.

\begin{definition}[Well-formed built-in operator typing context]
  \label{lemma:built-in-typing}
  The built-in operator typing context {\small$\Delta$} is well-formed
  iff the semantics of every built-in operator {\small$\theta$} is
  consistent with its type:
  {\small$\Delta(\theta) ~ = ~ \overline{x{:}b_{\I{x}}}\garr
    \overline{y{:}t_{\I{y}}}\sarr \tau \implies  \overline{\forall
      x{:}b_{\I{x}}}.\; \overline{\forall v_y \in \denotation{t_{\I{y}}}}.\;
    (\theta\, \overline{v_y}) \in \denotation{\tau\overline{[y \mapsto
        v_y]}}$}.\footnote{Recall that the semantics of an operator is
    defined by the auxiliary $e \Downarrow v$ and
    $\alpha \vDash e \Downarrow v$.}
\end{definition}




\begin{theorem}\label{theorem:fundamental}[Fundamental Theorem] Given
  a well-formed typing context for built-in operators
  {\small$\Delta$}, %
  \cbnewadding{ %
    the trace of effects produced by a well-typed term
    {\small $e$} are captured by its corresponding HAT {\small
      $\tau$}: %
  }%
  {\small$\Gamma \vdash e : \tau \implies \forall \sigma, \sigma \in
    \denotation{\Gamma} \impl \sigma(e) \in
    \denotation{\sigma(\tau)}$}.\footnote{A Coq mechanization of this
    theorem and the type soundness corollary are provided in the
    \techreport{}.}
\end{theorem}

\begin{corollary}\label{theorem:postautomata}[Type Soundness] Under
  a built-in operator type context that is well-formed, if a function
  {\small$f$} preserves the representation invariant $A$, i.e. it has
  the type: %
  {\small
    $\emptyset \vdash {\I{f}} : \overline{x{:}b_{\I{x}}}\garr
    \overline{y{:}b_{\I{y}}}\sarr \htriple{A}{t}{A}$}, then for every
  set of well-typed terms
  {\small$\emptyset \vdash \overline{v_{\I{x}} : b_{\I{x}}}$} and
  {\small$\emptyset \vdash \overline{v_{\I{y}} : b_{\I{y}}}$}, applying
  {\small $f$} to {\small $\overline{v_{\I{y}}}$} under an effect
  context $\alpha$ that is consistent with $A$ results in a context
  $\alpha\listconcat \alpha'$
  that is also consistent with $A$: {\footnotesize
    \begin{align*}
      \msteptr{\alpha}{({\I{f}}\ \overline{v_{\I{y}}})}{\alpha'}{v} \impl
      \alpha \listconcat \alpha'\in \langA{A[\overline{x \mapsto v_{\I{x}}}][\overline{y \mapsto v_{\I{y}}}]} \land v \in
      \denotation{t[\overline{x \mapsto v_{\I{x}}}][\overline{y\mapsto v_{\I{y}}}]}
    \end{align*}
  }
    %
\end{corollary}

\section{Typing Algorithm}
\label{sec:algo}

Converting our declarative type system into an efficient algorithm
requires resolving two key issues. First, we cannot directly use
existing SFA inclusion algorithms~\cite{SFA-Transducers,
  SFA-minimization} to implement the check in \textsc{SubAutomata},
because our SFAs may involve non-local variables corresponding to
function parameters and ghost variables; these variables can have
refinement types.  We deal with this wrinkle by integrating the
subtyping algorithm for pure refinement types into the existing SFA
algorithm. Second, typing the use of an effectful operator, e.g.,
\zput{} or \zexist{}, depends on the set of effects in the
precondition of the current HAT. The declarative type system can
ensure that this automaton has the right shape by applying the
\textsc{TInter} and \textsc{TSub} rules at any point in a derivation,
but an efficient implementation must perform this conversion more
intelligently. Our solution to this problem is to employ a
bidirectional type system~\cite{BidirectionalTyping} that closely
tracks a context of effects that have preceded the term being typed,
and selectively applies the subsumption rule to simplify the context
when switching between checking and synthesis modes.


\subsection{SFA Inclusion}
\label{sec:incl-algo}

Symbolic alphabets allow the character domains of SFAs to be infinite,
making them strictly more expressive than standard FAs.  This enhanced
expressivity comes at a cost, however, as the standard FA inclusion
algorithm cannot be directly applied to SFAs.  Thankfully, prior work
has shown how to reduce an inclusion check over SFAs to an inclusion
check over FAs~\cite{SFA-Transducers, SFA-minimization}.  The
high-level approach is to first construct a finite set of equivalence
classes over an SFA's infinite character domain, defined in terms of a
set of maximal satisfiable Boolean combinations of logic literals
(called \emph{minterms}) in the automaton's symbolic alphabet.  An
alphabet transformation procedure then replaces characters in the
original SFA with their corresponding equivalence classes, and
replaces the SFA's symbolic alphabet with minterms, thus allowing SFAs
to be translated into FAs. As long as the satisfiability of minterms
is decidable, so is checking inclusion between two SFAs. %
\cbnewadding{ %
  Our use of EPR formulas for qualifiers guarantees that minterm
  satisfiability is decidable.
}

\begin{algorithm}[t!]
    \Procedure{$\subqueryA(\Gamma, A, B) := $}{
    $L_{\Gamma} \leftarrow \getlits(\Gamma)$\;
    \ForEach{$\phi_{\Gamma} \in \boolcombine(L_{\Gamma})$}{
        $M \leftarrow \emptyset$\;
        \ForEach{$\effop$ where $\emptyset \basicvdash \effop : \overline{b_i}\sarr b $}{
        $L \leftarrow \getlits{}(\Gamma, A \lor B)$\;
            \ForEach{$\phi \in \boolcombine{}(L)$}{
                \If{$\Gamma, \overline{x_i{:}b_i} \not\vdash
                  \rawnuot{b}{\phi_{\Gamma} \land \phi} <:
                  \rawnuot{b}{\bot}$}{
                  $M \leftarrow M \cup \{ \evop{\effop}{\overline{x_i}}{\phi_{\Gamma} \land \phi} \}$}
            }
        }
        \If{$\langA{\mintermReplace(M, A)} \not\subseteq \langA{\mintermReplace(M, B)}$}{
            \Return{\Code{false}}\;
        }
    }
    \Return{\Code{true}}\;
    }
    \caption{Inclusion Query}
    \label{algo:inclusion}
\end{algorithm}

The pre- and post-conditions used in HATs have two features that
distinguish them from standard SFAs, however: 1) characters are drawn
from a set of distinct events and their qualifiers may range over
multiple variables, instead of a single local variable; and, 2) event
qualifiers may refer to variables in the typing context, i.e.,
function parameters and ghost variables.
\cbnewadding{ %
  Our language inclusion algorithm, shown in \autoref{algo:inclusion},
  accounts for all these differences by extending the existing SFA
  inclusion algorithm to incorporate a subtyping check between pure
  refinement types.
}
\paragraph{Candidate Minterms} The first step in checking SFA
inclusion is to construct the equivalence classes used to finitize its
character set. With HATs, these characters range over events triggered
by the use of library operations {\small
  $\pevapp{\effop}{\overline{v}}{v}$}, and return events {\small
  $\pevappret{v}$}. As each class of events is already disjoint (e.g.,
\zget{} and \zput{} events will never overlap), we only need to build
minterms that partition individual classes. The minterms of each class
of events have the form
{\small$\evop{\effop}{\overline{x_i}}{(\bigwedge \overline{l_j}) \land
    (\bigwedge \overline{\neg l_k})}$}. To build the literals used in
these minterms for an event {\small$\effop$}, we collect all the
literals {\small $\overline{l}$} used to qualify {\small$\effop$}, as
well as any literals appearing in atomic predicates
{\small$\evparenth{\phi}$}, and then construct a maximal set of
satisfiable Boolean combinations of those literals.
Notably, these literals may contain variables bound in the current
typing context.
\cbnewadding{ %
  Our algorithm divides minterms into two categories: Boolean
  combinations of literals from the typing context (line $3$),
  $\phi_{\Gamma}$, and Boolean combinations of literals appearing from
  the input automata (line $7$). The final set of candidate minterms
  are (satisfiable) conjunctions of the elements of
  $\phi_{\Gamma}$ and $\phi$ (line $9$).
}

\paragraph{Satisfiability Check} The soundness of the alphabet
transfer procedure requires that \emph{only} satisfiable minterms are
used, so that every transition in the output FA corresponds to an
actual transition in the original SFA.
To understand how this is done, observe that the typing context plays
a similar role in checking both the satisfiability of minterms and the
subtyping relation.  Thus, the algorithm reduces checking the
satisfiability of
{\small$\evop{\effop}{\overline{x_i}}{(\bigwedge \overline{l_j}) \land
    (\bigwedge \overline{\neg l_k})}$} to checking whether the
refinement type
{\small$\rawnuot{b}{(\bigwedge \overline{l_j}) \land (\bigwedge
    \overline{\neg l_k})}$} is \emph{not} a subtype of
{\small$\rawnuot{b}{\bot}$} (line $8$), that is
{\small$\Gamma, \overline{x_i{:}b_i} \not\vdash \rawnuot{b}{(\bigwedge
    \overline{l_j}) \land (\bigwedge \overline{\neg l_k})} <:
  \rawnuot{b}{\bot}$}.  Since {\small$\rawnuot{b}{\bot}$} is
uninhabited, this subtype check fails precisely when
{\small$(\bigwedge \overline{l_j}) \land (\bigwedge \overline{\neg
    l_k})$} is satisfiable under {\small$\Gamma$}.

\paragraph{Inclusion Check} Equipped with the set of satisfiable
minterms, the final inclusion check (line $10$) is a
straightforward application of the textbook inclusion check between
the FA produced by the standard alphabet transformation algorithm
(line $10$).\footnote{\cbnewadding{The details of
    alphabet transfer algorithm can be found in the \techreport{}.}} %
\cbnewadding{%
  \subqueryA{} returns $\Code{true}$ only when automata $A$ is
  included by $B$ under all instantiations of the variables in {\small
    $\Gamma$}.} %

\subsection{Bidirectional Type System}
\label{sec:bityping}

As is standard, our bidirectional type system consists of both type
synthesis ({\small$\typeinfer$}) and type checking
({\small$\typecheck$}) judgments. The bidirectional system features a
minor divergence from the declarative rules. While the declarative
system was able to use the subsumption rule (\textsc{TSub}) to freely
instantiate ghost variables, a typing algorithm requires a more
deterministic strategy. Our solution is to treat ghost variables
{\small$x{:}b \garr$} as instead being qualified by an unknown formula
which is specialized as needed, based on the information in the
current typing context. %
\cbnewadding{ %
  In order to efficiently infer the qualifiers of ghost variables, our
  algorithm only allows ghost variables to appear in SFAs. This
  restriction ensures that ghost variables are only used during SFA
  inclusion checks, and allows it to avoid the more sophisticated
  algorithms needed when ghost variables can appear in arbitrary
  refinement types~\cite{Implicit-Refinement-Type}.  This restriction
  is enforced by our algorithmic typing rules: the second line of
  \textsc{ChkEOpApp} ($\forall j. \Gamma \vdash v_j \typecheck t_j$)
  for example, only holds when no free ghost variables appear in the
  parameter types $t_j$ used by {\small$\effinfer$}
  ($\Gamma \infervdash{A} \overline{z_k{:}b_k}\garr
  \overline{y_j{:}t_j}\sarr \tau_x \effinfer \overline{z_k{:}t_k} $). }


\begin{figure}[t!]
{\footnotesize
{\small
\begin{flalign*}
 &{\text{\textbf{Type Synthesis }}} &
 \fbox{$\Gamma \vdash e \typeinfer t \quad \Gamma \vdash e \typeinfer \tau$} &&
  &{\text{\textbf{Type Check }}} &
 \fbox{$\Gamma \vdash e \typecheck t  \quad  \Gamma \vdash e \typecheck \tau$}
\end{flalign*}
}
\\ \
\begin{prooftree}
\hypo{
\parbox{50mm}{\center
$\Gamma \wellfoundedvdash x{:}b \garr \tau$\quad
  $\Gamma, x{:}\rawnuot{b}{\top} \vdash e \typecheck \tau $
}
}
\infer1[\textsc{ChkGhost}]{
\Gamma \vdash e \typecheck x{:}b \garr \tau
}
\end{prooftree}
\quad
\begin{prooftree}
\hypo{
\parbox{50mm}{\center
$\Gamma \wellfoundedvdash \tau_1 \interty \tau_2$\quad
  $\Gamma \vdash e \typecheck \tau_1$\quad
  $\Gamma \vdash e \typecheck \tau_2$
}
}
\infer1[\textsc{ChkInter}]{
\Gamma \vdash e \typecheck \tau_1 \interty \tau_2
}
\end{prooftree}
\\ \ \\ \ \\ \
\begin{prooftree}
\hypo{
\parbox{45mm}{\center
  $\Gamma \wellfoundedvdash \htriple{A_2}{t_2}{B_2}$ \quad
  $\Gamma \vdash e \typeinfer \htriple{A_1}{t_1}{B_1}$\quad
  $\Gamma \vdash A_2 \subseteq A_1$\quad
  $\Gamma \vdash t_1 <: t_2$\quad
  $\Gamma \vdash (A_2;\globalA\topA) \land B_1 \subseteq  (A_2;\globalA\topA) \land B_2$
}
}
\infer1[\textsc{ChkSub}]{
\Gamma \vdash e \typecheck \htriple{A_2}{t_2}{B_2}
}
\end{prooftree}
\quad
\begin{prooftree}
\hypo{
\parbox{60mm}{\center
  $\Gamma \wellfoundedvdash \htriple{A}{t}{B}$ \quad
  $\Delta(\effop) = \overline{z_k{:}b_k}\garr  \overline{y_j{:}t_j}\sarr \tau_x $\quad   $\forall j. \Gamma \vdash v_j \typecheck t_j$ \\
  $ \Gamma \infervdash{A}  \overline{z_k{:}b_k}\garr  \overline{y_j{:}t_j}\sarr \tau_x \effinfer \overline{z_k{:}t_k} $\\
  $\biginterty \overline{\htriple{A_i}{t_i}{A_i'}} = \tau_x\overline{[y_j \mapsto v_j]}$ \\
  $\forall i.\Gamma, \overline{z_k{:}t_k}, x{:}t_i \vdash e \typecheck
  \htriple{(A;\globalA\topA) \land A_i'}{t}{B}$
}
}
\infer1[\textsc{ChkEOpApp}]{
\Gamma \vdash \perform{x}{\effop}{\overline{v_i}}{e} \typecheck \htriple{A}{t}{B}
}
\end{prooftree}
}
\vspace{-0.5em}
\caption{Selected Bidirectional Typing Rules}\label{fig:bi-type-rules}
\end{figure}

\cbst %
\begin{figure}[t]
{\footnotesize
\begin{prooftree}
\hypo{
\parbox{50mm}{\center
  $\Gamma = \overline{x_i{:}\rawnuot{b_i}{\phi_i}}, \overline{y{:}(\overline{u{:}b_u}\garr z{:}t\sarr\tau)}$\quad
  $\forall \overline{x_i{:}b_i}. \forall \vnu{:}b. \bigwedge \phi_i \impl (\phi_1 \impl \phi_2)$
}
}
\infer1[\textsc{SubBaseAlg}]{
\Gamma \vdash \rawnuot{b}{\phi_1} <: \rawnuot{b}{\phi_2}
}
\end{prooftree}
\begin{prooftree}
\hypo{
\parbox{55mm}{\center
  $\biginterty \overline{\htriple{A_i}{t_i}{A_i'}} = \tau$ \quad $
  \Gamma'= \Gamma, \overline{y_j{:}t_j}$ \\
  $\instantiate(\Gamma, A, \bigvee A_i, \overline{z_k{:}b_k}) = \overline{z_k{:}t_k}$
}
}
\infer1[\textsc{InstAlg}]{
\Gamma \infervdash{A}  \overline{z_k{:}b_k}\garr  \overline{y_j{:}t_j}\sarr \tau \effinfer \overline{z_k{:}t_k}
}
\end{prooftree}
}
\caption{Selected Auxiliary Typing Functions}\label{fig:aux-rules-algo}
\end{figure}
\cbed %

With this minor tweak in hand, the top-level typing algorithm is
mostly a matter of bidirectionalizing the typing rules of \langname{}
by choosing appropriate modes for the assumptions of each rule. This
is largely straightforward: the type checking rule for ghost variables
(\textsc{ChkGhost}) adapts the corresponding declarative rule
(\textsc{TGhost}) in the standard way, for example. The rule for
checking effectful operations (\textsc{ChkEOpApp}) is similar,
although it adopts a fixed strategy for applying the subsumption rule:
1) instead of using \textsc{TSub} to instantiate ghost variables
directly, it now relies on an auxiliary function, $\effinfer$, to
strengthen their qualifiers based on the current context; 2) when the
result of the library operation $\effop$ is an intersection type
({\small$\biginterty \overline{\htriple{A_i}{t_i}{A_i'}}$}),
\textsc{ChkEOpApp} considers each case of the intersection, instead of
using \textsc{TSub} to focus on a single case; and 3) instead of
relying on \textsc{TSub} to align the postcondition automata
({\small$A_i'$}) of $\effop$ with the precondition of the HAT being
checked against ({\small$A$}), the rule uses the conjunction of the
automata ({\small$(A;\globalA\topA) \land A_i'$}) as the precondition
automata used to check the term that uses the result of $\effop$.
\cbnewadding{
  The implementation of $\effinfer$ relies on the \instantiate{}
  subroutine.\footnote{The full details of \instantiate{} can be found
    in the \techreport{}.} Given a typing context $\Gamma$,
  set of ghost variables $\overline{x{:}b}$, and a pair of automata
  $A$ and $A'$, \instantiate{} infers a set of qualifiers
  $\overline{\phi}$ for $\overline{x{:}b}$ sufficient enough to ensure
  that $A$ is included in $A'$, i.e.
  $\Gamma, \overline{x{:}\nuot{\textit{b}}{\phi}} \vdash A \subseteq
  A'$, or reports that none exists. \instantiate{} is an adaption of
  an existing algorithm~\cite{ZDDJ21} which infers the \emph{weakest}
  qualifiers $\phi_i$ needed for the inclusion check. %
}

\begin{theorem}\label{theorem:algo-sound}[Soundness of Algorithmic
  Typing] Given a well-formed built-in typing context
  {\small$\Delta$}, type context {\small$\Gamma$}, term {\small$e$},
  and HAT {\small$\tau$}, {\small
    $\Gamma \vdash e \typecheck \tau \implies \Gamma \vdash e :
    \tau$}.\footnote{%
    \cbnewadding{ %
      The proofs of Theorems~\ref{theorem:algo-sound}and
      \ref{theorem:algo-decidable} can be found in the \techreport{}. %
    }}
\end{theorem}
\begin{theorem}\label{theorem:algo-decidable}[Decidability of
  Algorithmic Typing]
  \cbnewadding{Type checking a term {\small $e$} against a type
  {\small $\tau$} in a typing context {\small $\Gamma$}, i.e., {\small
    $\Gamma \vdash e \typecheck \tau$}, is decidable.}
\end{theorem} %

\section{Implementation and Evaluation}
\label{sec:evaluation}

We have implemented a tool based on the above approach, called
\name{}, that targets ADTs implemented in terms of other stateful
libraries. \name{} consists of approximately 12K lines of OCaml and
uses Z3~\cite{de2008z3} as its backend solver for both SMT and FA
inclusion queries.\footnote{Our
  \techreport{} provides a docker image
  that contains the source code of \name{} and all our benchmarks.}

\name{} takes as an input the implementation of an ADT in OCaml, enhanced
signatures of ADT operations that include representation invariants expressed
as HATs,  and specifications of the
supporting libraries as HATs (e.g., the signatures in
Example~\ref{ex:Typing+Delta}).
The typing context used by \name{} includes signatures for a number of
(pure) OCaml primitives, including the pure operators listed in
\autoref{fig:term-syntax}. \name{} also includes a set of predefined method
predicates (i.e., {\small$mp$} in \autoref{fig:type-syntax}) that
allow qualifiers to capture non-trivial datatype properties. For
example, the method predicates {\small$\Code{isDir}(\Code{x})$} and
{\small$\Code{isDel}(\Code{y})$} from the motivating example in
\autoref{sec:overview} encode that {\small$\Code{x}$} holds the
contents of a directory, and that {\small$\Code{y}$} is marked as
deleted, respectively. The semantics of method predicates are defined
via a set of lemmas in FOL, in order to enable automated verification;
e.g., the axiom {\small$\forall \Code{x}.\Code{isDir}(\Code{x}) \impl
  \neg\Code{isDel}(\Code{x})$} encodes that a set of bytes cannot
simultaneously be an active directory and marked as deleted.

\begin{table}[]
  \renewcommand{\arraystretch}{0.8}
  \caption{\small Experimental results of using \name{} to verify
    representation invariants of ADTs. Benchmarks from prior works are
    annotated with their source:
    Okasaki~\cite{okasaki1999purely}($^{\dagger}$),
    Hanoi~\cite{MP+20}($^{\star}$). %
    \cbcontent{%
      We have rewritten functional implementations of these benchmarks into stateful
      versions by assuming there is a hidden mutable datatype instance that ADT methods use to perform their operations. %
    } %
  The backing stateful libraries are
  drawn from a verified OCaml library~\cite{vocal}. }
\vspace*{-.15in}
\footnotesize
\setlength{\tabcolsep}{2.5pt}
\begin{tabular}{cc|ccc|c||cc|ccc|cc}
  \toprule
 ADT &  Library &  \#Method &  \#Ghost &  s$_{I}\ \ $  &  t$_{\text{total}}$ (s) &  \#Branch &  \#App &  \#SAT &  \#FA$_\subseteq$ &  avg. s$_{\text{FA}}$ &  t$_{\text{SAT}}$ (s) &  t$_{\text{FA}_\subseteq}$ (s)\\
\midrule
\multirow{2}{*}{\textsf{Stack}$^{\dagger}$} & \textsf{LinkedList} & 7 & 0 & 4 & 5.94 & 4 & 7 & 297 & 7 & 98 & 1.63 & 0.06 \\
 & \textsf{KVStore} & 7 & 1 & 9 & 17.88 & 4 & 10 & 874 & 17 & 226 & 5.78 & 0.23 \\
\midrule
\multirow{2}{*}{\textsf{Queue}$^{\dagger}$} & \textsf{LinkedList} & 6 & 0 & 4 & 4.88 & 4 & 9 & 190 & 4 & 96 & 1.08 & 0.06 \\
 & \textsf{Graph} & 6 & 1 & 24 & 28.24 & 5 & 12 & 1212 & 14 & 525 & 8.19 & 0.86 \\
\midrule
\multirow{2}{*}{\textsf{Set}$^{\star}$} & \textsf{Tree} & 5 & 0 & 12 & 27.12 & 5 & 12 & 1589 & 11 & 531 & 9.34 & 0.53 \\
 & \textsf{KVStore} & 3 & 1 & 9 & 2.39 & 3 & 5 & 245 & 6 & 160 & 1.51 & 0.06 \\
\midrule
\multirow{2}{*}{\textsf{Heap}$^{\star}$} & \textsf{Tree} & 7 & 0 & 12 & 25.71 & 5 & 12 & 1589 & 11 & 531 & 9.33 & 0.52 \\
 & \textsf{LinkedList} & 6 & 0 & 4 & 8.84 & 4 & 8 & 497 & 8 & 118 & 3.03 & 0.08 \\
\midrule
\multirow{2}{*}{\textsf{MinSet}} & \textsf{Set} & 4 & 1 & 28 & 10.55 & 3 & 6 & 612 & 17 & 294 & 4.19 & 1.58 \\
 & \textsf{KVStore} & 4 & 1 & 25 & 32.84 & 5 & 8 & 2227 & 23 & 519 & 10.93 & 9.34 \\
\midrule
\multirow{3}{*}{\textsf{LazySet}} & \textsf{Tree} & 6 & 0 & 12 & 27.10 & 5 & 12 & 1589 & 11 & 531 & 9.33 & 0.57 \\
 & \textsf{Set} & 4 & 1 & 9 & 1.42 & 2 & 3 & 101 & 4 & 106 & 0.57 & 0.04 \\
 & \textsf{KVStore} & 5 & 1 & 9 & 3.10 & 3 & 5 & 245 & 6 & 160 & 1.49 & 0.06 \\
\midrule
\multirow{2}{*}{\textsf{FileSystem}} & \textsf{Tree} & 6 & 1 & 20 & 58.80 & 3 & 8 & 2085 & 17 & 652 & 14.15 & 2.21 \\
 & \textsf{KVStore} & 4 & 1 & 17 & 157.27 & 4 & 10 & 8144 & 43 & 481 & 56.64 & 16.54 \\
\midrule
\multirow{2}{*}{\textsf{DFA}} & \textsf{KVStore} & 5 & 2 & 18 & 42.62 & 3 & 3 & 3604 & 25 & 228 & 18.84 & 1.60 \\
 & \textsf{Graph} & 5 & 2 & 11 & 78.44 & 4 & 4 & 3625 & 27 & 225 & 25.53 & 3.42 \\
\midrule
\multirow{2}{*}{\textsf{ConnectedGraph}} & \textsf{Set} & 5 & 2 & 9 & 80.46 & 4 & 4 & 3889 & 50 & 357 & 27.39 & 12.80 \\
 & \textsf{Graph} & 5 & 1 & 20 & 176.89 & 4 & 3 & 1349 & 17 & 360 & 19.19 & 53.90 \\
\bottomrule
\end{tabular}

\label{tab:evaluation}
\end{table}

\cbst %
\begin{table}[]
  \renewcommand{\arraystretch}{0.8}
  \caption{\small \cbcontent{Details of the properties used in
      the benchmarks shown in \autoref{tab:evaluation}. Representation
      invariants of client ADTs are expressed in
      terms of their interactions with the backing stateful library.} } %
  \vspace*{-.15in} 
  \footnotesize \setlength{\tabcolsep}{2.5pt}
\begin{tabular}{c|c|c|c}
  \toprule
  \multirow{2}{*}{ Client ADT} &  Representation & Underlying & \multirow{2}{*}{ Policy governing
                                                      interactions with
                                                      underlying API} \\
                               &  Invariant & Library & \\
  \midrule
  \multirow{2}{*}{\textsf{Stack}} & \multirow{2}{*}{LIFO-property} & \textsf{LinkedList} &  The addresses that store elements are unique \\
          & & \textsf{KVStore} & Not a circular linked list \\
  \midrule
  \multirow{2}{*}{\textsf{Queue}} & \multirow{2}{*}{FIFO-property} & \textsf{LinkedList} & Not a circular linked list \\
          & & \textsf{Graph} & No self-loops in graph, degree of nodes in graph are at most $1$ \\
  \midrule
  \multirow{2}{*}{\textsf{Set}} & \multirow{1}{*}{Unique} &
                                                                \textsf{Tree}
                                                              & The
                                                                underlying
                                                                tree
                                                                is a binary search tree \\
          & \multirow{1}{*}{elements}& \textsf{KVStore} & Every key is associated with a distinct value \\
  \midrule
  \multirow{3}{*}{\textsf{Heap}} & \multirow{3}{*}{Min-heap property}  & \multirow{2}{*}{\textsf{Tree}} & The value of the parent node in the tree \\
          & & & is less than the value of its child nodes \\
          & & \textsf{LinkedList} & Not a circular linked list; the elements are sorted \\
  \midrule
  \multirow{4}{*}{\textsf{MinSet}} & & \multirow{2}{*}{\textsf{Set}} & The cached element is less than all inserted elements \\
          & \multirow{1}{*}{Uniqueness and} & & in the set and has
                                                been inserted into the set  \\
          & \multirow{1}{*}{minimality} & \multirow{2}{*}{\textsf{KVStore}} & The cached element is less than all put elements \\
          & & & in the store and has been put before \\
  \midrule
  \multirow{3}{*}{\textsf{LazySet}} & \multirow{1}{*}{Uniqueness} & \textsf{Tree} & The
                                                                underlying
                                                                tree
                                                                is a binary search tree \\
          &\multirow{2}{*}{of elements} & \textsf{Set} & An element
                                                         has never
                                                         been inserted twice \\
          & & \textsf{KVStore} & Every key is associated with a distinct value \\
  \midrule
  \multirow{4}{*}{\textsf{FileSystem}} & &
                                           \multirow{2}{*}{\textsf{Tree}} & The Parent node stores a path that is prefix of its child nodes; \\
          & \multirow{1}{*}{Policy of Unix-like} & & a non-root parent node stores directories \\
          & \multirow{1}{*}{filesystem paths} & \multirow{2}{*}{\textsf{KVStore}} & Any non-root path stored as a key in the key-value store must \\
          & & & have its parent stored as a non-deleted directory in the store \\
  \midrule
  \multirow{4}{*}{\textsf{DFA}} & \multirow{1}{*}{Determinism}  & \multirow{2}{*}{\textsf{KVStore}} & All stored transitions are represented as tuples (start, char, end); \\
          &  \multirow{2}{*}{of Transitions} & & starting state has at most have one transition for a character \\
          & & \multirow{2}{*}{\textsf{Graph}} & Two nodes
                                                (which represent states in FA) can have at most one edge, \\
          & & & which is labeled with a character \\
  \midrule
  \multirow{3}{*}{\textsf{ConnectedGraph}} & \multirow{3}{*}{Connectivity} & \multirow{2}{*}{\textsf{Set}} & The set stores unique pairs (fst, snd); only elements that have\\
          & & &  both in and out edges are treated as a value node in the graph \\
          & & \textsf{Graph} & 	All nodes in the graph are connected \\
  \bottomrule
\end{tabular}

\label{tab:evaluation-prop}
\vspace*{-.15in}
\end{table} %
\cbed %

\cbnewadding{ %
Our experimental evaluation addresses three research questions:
\begin{itemize}
\item[\textbf{Q1}:] Is \name{} \emph{effective}? Can it verify a
  diverse range of ADTs that are implemented using a variety of
  backing stateful libraries?
\item[\textbf{Q2}:] Is \name{} \emph{expressive}? Are HATs able to
  capture a rich set of representation invariants?
\item[\textbf{Q3}:] Is \name{} \emph{efficient}? Is it able verify
  clients of stateful APIs in a reasonable amount of time?
\end{itemize} %
}

We have evaluated \name{} on a corpus of complex and realistic ADTs
drawn from a variety of sources (shown in the title of
\autoref{tab:evaluation}), including general datatypes such as stacks,
queues, sets, heaps, graphs, as well as custom data structures
including the Unix-style file system from \autoref{sec:intro} and the
deterministic finite automata described in \autoref{sec:typing}
(\textbf{Q1}). \autoref{tab:evaluation} presents the
results of this evaluation. Each datatype includes multiple
implementations using different underlying libraries (Library column),
each of which provides a distinct set of operations. Our backing APIs
include linked lists, a persistent key-value store, sets, trees, and
graphs, each of which induces a different encoding of the target ADT's
representation invariant.

\autoref{tab:evaluation} divides the results of our experiments into
two categories, separated by the double bars; the first gives
comprehensive results about analyzing the entire datatype, while the
second reports information about the most complex method of each ADT
implementation.\footnote{Our \techreport{} includes a full
  table listing the corresponding information for each ADT method.}
The first group of columns describes broad characteristics of each
ADT, including its number of methods (\#Method), the number of ghost
variables in its representation invariant (\#Ghost), and the size of
the formula encoding the representation invariant (s$_I$), after it is
desugared into a symbolic regular expression. The next column reports
the total time needed to verify all the method of each ADT
(t$_{\text{total}}$). \name{} performs quite reasonably on all of our
benchmarks, taking between 1.42 to 176.89 seconds for each ADT. As
expected, more complicated ADT implementations (i.e., those with
larger values in the \#Method, \#Ghost, and s$_I$ columns), take
longer to verify (\textbf{Q3}).

\cbnewadding{ %
  Details about the representation invariants used in our benchmarks
  are shown in \autoref{tab:evaluation-prop} (\textbf{Q2}). The stack
  and queue ADTs require a low-level guarantee that the underlying
  linked list implementation is not circular. The set ADTs maintain
  the invariant that there are no duplicate elements. The connected
  graph benchmark requires that every pair of nodes is connected,
  necessitating the need for ghost variables. As shown in
  \autoref{tab:evaluation}, all of these properties, as well as those
  discussed in previous sections, can be expressed using LTL$_f$ using
  no more than $2$ ghost variables (\#Ghost) and $28$ literals
  (s$_I$). %
}


The first group of columns in the second half of \autoref{tab:evaluation} list
features relevant to the complexity of the most demanding method in
the ADT. These methods feature between {\small$3$} and {\small$5$}
branches (\#Branch) and {\small$3$} to {\small$12$} uses of built-in
operators (\#App). The last two groups of columns present type
checking results for these methods. The \#SAT and \#FA$_\subseteq$
columns list the number of SMT queries and finite automata (FA)
inclusion checks needed to type check the method. The next column
(avg. s$_{\text{FA}}$) gives the average number of transitions in the
finite state automata (FA) after the alphabet transformation described
in \autoref{sec:algo}. These numbers are roughly proportional to code
complexity (column \#Branch and \#App) and invariant complexity
(column \#Ghost and s$_I$) --- intuitively, programs that have more
uses of operators and larger specifications, require more queries and
produce more complicated FAs. The last two columns report verification
time for the method (\textbf{Q3}). These times are
typically dominated by the SFA inclusion checks (ranging from .63 to
73.09 seconds) that result from minterm satisfiability checks during
the alphabet transformation phase (t$_{\text{SAT}}$) and FA inclusion
checks (t$_{\text{FA}_\subseteq}$). Unsurprisingly, generating more
queries (both \#SMT and \#FA$_\subseteq$) result in more time spent
checking minterm satisfiability and FA inclusion. Our results also
indicate that FA inclusion checks are not particularly sensitive to
the size of the FA involved; rather, the cost of these checks
corresponds more closely to the deep semantics of the representation
invariant, in which the choice of the underlying library is an
important factor.  Taking the \textsf{FileSystem} benchmark as an
example, both \textsf{Tree} and \textsf{KVStore} implementations lead
to similar sizes for the invariant, in term of \#Ghost. s$_I$, and FA
(avg.s$_{\text{FA}}$). However, the use of the former library results
in a significantly shorter verification time since the relevant
property captured by the invariant ``\emph{every child has its parent
  in the file system}'' is naturally supported by the \textsf{Tree}
library; the only remaining verification task in this case for the
client to handle involves ensuring these parents are indeed
directories in the file system. In contrast, the \textsf{KVStore}
provides no such guarantees on how its elements are related, requiring
substantially more verification effort to ensure the invariant is
preserved.

\section{Related Work}\label{sec:related}

\textit{Representation Invariant Inference.}  \citet{MP+20} develop a
data-driven CEGIS-based inference approach, while ~\citet{MP+07}
develop a solution derived from examining testcases. Solvers used to
infer predicates that satisfy a set of Constrained Horn Clauses
(CHCs)~\cite{HR18,ZMJ18,EN+18} can also be thought of as broadly
addressing similar goals. Our focus in this paper is orthogonal to
these efforts, however, as it is centered on the automated
verification of user-specified representation invariants expressed in terms of SFAs.

\textit{Effect Verification.} There has been significant prior work
that considers the type-based verification of functional programs
extended with support for (structured) effectful computation.
Ynot~\cite{NM+08} and F*~\cite{FStar} are two well-known examples
whose type specifications allow fine-grained effect tracking of
stateful functional programs.  These systems allow writing
specifications that leverage type abstractions like the Hoare
monad~\cite{NMB08} or the Dijkstra
monad~\cite{Dijkstra-Monads-for-Free, MAA+19,SW+13, Steel, SteelCore}.  For example,
specifications using the Hoare monad involve writing type-level
assertions in the form of pre- and post-conditions over the heaps that
exist before and after a stateful computation executes.  The
Dijkstra monad generalizes this idea by allowing specifications to be
written in the style of a (weakest precondition) predicate transformer
semantics, thus making it more amenable for improved automation and VC
generation.  While our goals are broadly similar to these other
efforts, our setup (and consequently, our approach) is fundamentally
different.  Because we assume that the implementation of effectful
operations are hidden behind an opaque interface, and that the
specifications of these operations are not tailored for any specific
client application, our verification methodology must necessarily
reason about stateful library actions, indirectly in terms of the
\emph{history} of calls (and returns) to library methods made by the
client, rather than in terms of predicates over the concrete
representation of the state.
One
important way that our  significantly different representation choice
impacts our verification methodology is that unlike e.g., the Dijkstra
monad that uses a predicate transformer semantics to establish a
relation between pre- and post-states of an effectful action, our
typing of interaction history in terms of HATs allows us to simply
reuse a standard subtying relation (suitably adapted), instead.
Consequently, the implementation of our typing algorithm is amenable
to a significant degree of automation, on par with other refinement
type systems like Liquid Haskell~\cite{LiquidHaskell}.

\cbnewadding{Monadic-based techniques~\cite{WD+17, Tlon-Embedding,
    Choice-Trees, Interaction-Trees-VST, Interaction-Trees} to reason
  about effectful programs have also been well-studied, especially in
  the context of interactive (mechanized) verification. One notable
  instance are interaction trees (ITrees)~\cite{Interaction-Trees}, a
  coinductive variant of the free monad that also provide an
  abstraction of an (infinite) sequence of visible events (i.e.,
  traces) produced by effectful programs. Analogous to HATs, context
  and effect traces can be expressed as pre- and post-
  ITrees~\cite{Interaction-Trees-VST}. In contrast to ITrees, HATs use
  of SFAs provides a decidable inclusion check, which enables fully
  automated verification. Furthermore, because SFAs denote a regular
  trace language, they enable the use of LTL$_f$, an arguably more
  readable and lighter-weight specification language.}

\textit{Type and Effect Systems.} Type and effect systems
comprise a broad range of techniques that provide state guarantees
about both \emph{what} values a program computes and \emph{how} it
computes them. These systems have been developed for a number of
different applications: static exception checking in Java~\cite{Java},
ensuring effects are safely handled in the languages with algebraic
effects~\cite{Eff}, and reasoning about memory
accesses~\cite{EffectSystem}, in order to ensure, e.g., programs can
be safely parallelized~\cite{ABHI+08}. Notably, a majority of these
systems are agnostic to the order in which these effects are produced:
the effect system in Java, for example, only tracks whether or not a
computation may raise an exception at run-time.

In contrast, \emph{sequential} effect systems~\cite{Tate+13} provide
finer grained information about the order in which a computation's
effects may occur.
More closely related to this work are type and effect systems that
target \emph{temporal} properties on the sequences of effects a
program may \emph{produce}. For example, \citet{SS+04} present a type
and effect system for reasoning about the shape of histories (i.e.,
finite traces) of events embedded in a program, with a focus on
security-related properties expressed as regular expressions augmented
with fixed points. Their specification language is not rich enough to
capture data-dependent properties: the histories of a conditional
expression must include the histories of both branches, even when its
condition is a variable that is provably false. \citet{KT+14} present
a type and effect system that additionally supports verification
properties of infinite traces, specified as B\"{u}chi automata. Their
system features refinement types and targets a language similar to
\langname{}. Their effect specifications also include a second
component describing the sets of \emph{infinite} traces a
(non-terminating) program may produce, enabling it to reason about
liveness properties. Unlike \langname{}, however, events in their
system are not allowed to refer to program variables. This restriction
was later lifted by \citet{NUKT+18}, whose results support value-dependent
predicates in effect specifications written in a first-order fixpoint
logic. To solve the resulting verification conditions, they introduce
a novel proof system for this logic. More recently,
\citet{Temporal-Verification} consider how to support richer control
flow structures, e.g., delimited continuations, in such an effect
system. To the best of our knowledge, all of these effect
systems lack an analogue to the precondition automaton of HATs, and
instead have to rely on the arguments of a dependently typed function
to constrain its set of output traces, as opposed to constructing and
managing the context of previously seen events and their results.

While our focus has been on the development of a practical type system
for the purposes of verification, there have also been several works
that provide foundational characterizations of sequential effect
systems. Most of these works adopt a semantics-oriented approach and
seek to develop categorical semantics for effectful
languages~\cite{Tate+13, Katsumata+14,MOP+15}.  More closely related
is the recent work of Gordon~\cite{Quantales}, which defines a generic
model of sequential effects systems in terms of a parameterized
system. The key idea is to represent effects as a \emph{quantale}, a
join semi-lattice equipped with a top element and whose underlying set
is also a monoid. The proposed type system is parameterized over an
arbitrary effect quantale; this system is proven to be safe for any
appropriate instantiation of this parameter. 

\textit{Type-Based Protocol Enforcement.} Because HATs express a
history of the interaction between a client and an stateful library,
they naturally serve as a form of protocol specification on client
behavior \cbnewadding{(e.g., session types~\cite{Hon93,Vasconcelos2009})}, although the kinds of protocols we consider in this work are
limited to those that are relevant to defining a representation
invariant, i.e., those that capture constraints on library
methods sufficient to ensure a desired client invariant. In this
sense, HATs play a similar role as typestate in object-oriented
languages~\cite{deline2004typestates, bodden2012clara, SNSAT+11},
which augments the interface of objects with coarse-grained
information about the state of an object, in order to constrain its
set of available methods.  Typestate specifications differ in obvious
ways from HATs, most notably with respect to the level of granularity
that is expressible; in particular, HATs are designed to capture
fine-grained interactions between libraries and clients that are
necessary to specify useful representation invariants.


\section{Conclusion}
\label{sec:conclusion}

This paper explores the integration of SFAs within a refinement type
system as a means to specify and verify client-specific representation
invariants of functional programs that interact with stateful
libraries.  Our key innovation is the manifestation of SFAs as HATs in
the type system, which allows the specification of context traces
(preconditions) as well as computation traces (postconditions) that
allow us to characterize the space of allowed client/library
interactions and thus enable us to prove that these interactions do
not violate provided representation invariants.  HATs integrate
cleanly within a refinement type system, enjoy pleasant decidability
properties, and are amenable to a high-degree of automation.

\cbnewadding{There are several interesting directions for future
  work. While the focus of this paper is the verification of
  representation invariants, there is no instrinsic requirement that
  HATs use the same automaton in their pre- and postcondition. We intend to explore more
  general verification problems under this relaxation.} Another potential direction is extending our
current formalization to support richer automata classes (e.g.,
symbolic visibly pushdown~\cite{DA14} or B\"{u}chi automata), in order
to support a larger class of safety properties, \cbnewadding{e.g.,
  re-entrant properties of locks, quantitative properties, and
  properties over infinite traces) or more general categories of
  effects, e.g., control effects~\cite{Eff})}. %


\section*{Acknowledgements}

We thank the anonymous reviewers for their detailed comments and
suggestions, and Ashish Mishra, Julia Belyakova, Patrick Lafontaine 
Yongwei Yuan (Purdue), John Wiegley (Kadena), and Scott Moore (Galois)
for numerous stimulating discussions.  This material is based upon
work supported by the Defense Advanced Research Projects Agency
(DARPA) and Naval Information Warfare Center Pacific (NIWC Pacific)
under N6600 1-22-C-4027 and the National Science Foundation under
CCF-SHF 2321680.

\bibliography{bibliography}


\ifdefined\originalmode
\else
\newpage
\appendix
\section{Operational Semantics}\label{sec:tech:semantics}

The auxiliary big-step reduction rules for effect operators and the
small-step operational semantics of our core language are shown in
\autoref{fig:semantics}.

\begin{figure}[h!]
{\footnotesize
{\small\begin{flalign*}
 &\text{\textbf{Auxiliary Operational Semantics }} & \fbox{$\alpha \vDash \effop\ \overline{v} \Downarrow v$}
\end{flalign*}}
    \begin{prooftree}
      \hypo{ (\pevapp{put}{k\,v}{()}) \not\in \alpha'}
      \infer1[\textsc{get}]{
        \alpha \listconcat [\pevapp{put}{k\,v}{()}] \listconcat \alpha' \vDash \S{get}\,k \Downarrow v
      }
    \end{prooftree}
    \quad
    \begin{prooftree}
      \hypo{ (\pevapp{put}{k\,v}{()}) \not\in \alpha}
      \infer1[\textsc{existsF}]{
        \alpha \vDash \S{exists}\,k \Downarrow \Code{false}
      }
    \end{prooftree}
    \quad
    \begin{prooftree}
      \hypo{ (\pevapp{put}{k\,v}{()}) \in \alpha}
      \infer1[\textsc{existsT}]{
        \alpha \vDash \S{exists}\,k \Downarrow \Code{true}
      }
    \end{prooftree}
\\ \ \\ \ \\
    \begin{prooftree}
      \hypo{}
      \infer1[\textsc{put}]{
        \alpha \vDash \S{put}\, k\, v \Downarrow ()
      }
    \end{prooftree}
    \quad
\begin{prooftree}
      \hypo{ (\pevapp{insert}{n}{()}) \in \alpha}
      \infer1[\textsc{memT}]{
        \alpha \vDash \S{mem}\,v \Downarrow \Code{true}
      }
\end{prooftree}
\quad
\begin{prooftree}
      \hypo{ \pevapp{insert}{n}{()} \not\in \alpha }
      \infer1[\textsc{memF}]{
        \alpha \vDash \S{mem}\,v \Downarrow \Code{false}
      }
\end{prooftree}
\quad
\begin{prooftree}
      \hypo{ }
      \infer1[\textsc{insert}]{
        \alpha  \vDash \S{insert}\,v \Downarrow \Code{()}
      }
    \end{prooftree}
    {\small
      \begin{flalign*}
        &\text{\textbf{Operational Semantics }} & \fbox{$\steptr{\alpha}{e}{\alpha}{e}$}
      \end{flalign*}}
    \begin{prooftree}
      \hypo{\primop\ \overline{v} \Downarrow v_x }
      \infer1[\textsc{\small StPureOp}]{
        \steptr{\alpha}{\zlet{x}{\primop\ \overline{v}}{e}}{\emptr}{e[x\mapsto v_x]}
      }
    \end{prooftree}
    \
    \begin{prooftree}
      \hypo{ \alpha \vDash \effop\, \overline{v} \Downarrow v_x }
      \infer1[\textsc{\small StEffOp}]{
        \steptr{\alpha}{\perform{x}{\effop{}}{\overline{v}}{e}}{[\pevapp{\effop{}}{\overline{v}}{v_x}]}{e[x\mapsto v_x]}
      }
    \end{prooftree}
    \\ \ \\ \ \\
    \begin{prooftree}
      \hypo{\steptr{\alpha}{e_1}{\alpha'}{e_1'}}
      \infer1[\textsc{StLetE1}]{
        \steptr{\alpha}{(\zlet{y}{e_1}{e_2})}{\alpha'}{(\zlet{y}{e_1'}{e_2})}
      }
    \end{prooftree}
    \quad
    \begin{prooftree}
      \hypo{}
      \infer1[\textsc{StLetE2}]{
        \steptr{\alpha}{\zlet{y}{v}{e}}{\emptr}{e[y\mapsto v]}
      }
    \end{prooftree}
    \\ \ \\ \ \\
    \begin{prooftree}
      \hypo{}
      \infer1[\textsc{StLetAppLam}]{
        \steptr{\alpha}{(\zlet{y}{\zlam{x}{t}{e_1}\ v_x}{e_2})}{\emptr}{(\zlet{y}{e_1[x\mapsto v_x]}{e_2})}
      }
    \end{prooftree}
    \\ \ \\ \ \\
    \begin{prooftree}
      \hypo{}
      \infer1[\textsc{StLetAppFix}]{
        \steptr{\alpha}{\zlet{y}{\zfix{f}{t}{x}{t_x}{e_1}\ v_x}{e_2}}{\emptr}{
          \zlet{y}{(\zlam{f}{t}{e_1[x\mapsto v_x]}) \ (\zfix{f}{t}{x}{t_x}{e_1})}{e_2}}
      }
    \end{prooftree}
    \\ \ \\ \ \\
    \begin{prooftree}
      \hypo{}
      \infer1[\textsc{StMatch}]{
        \steptr{\alpha}{\match{d_i \ \overline{v_j}} \overline{d_i\ \overline{y_j} \to e_i}}{\emptr}{e_i\overline{[y_j \mapsto v_j]}
        }}
    \end{prooftree}
  }
  \caption{Operational Semantics}
  \label{fig:semantics}
\end{figure}

\newpage
\section{Basic Typing Rules}\label{sec:tech:basic-typing}

The basic typing rules of our core language and qualifiers are shown
in \autoref{fig:basic-type-rules} and
\autoref{fig:basic-qualifier-type-rules}. We use an auxiliary function
$\S{Ty}$ to provide a basic type for the primitives of our language,
e.g., constants, built-in operators, and data constructors.

\begin{figure}[h!]
 {\footnotesize
 {\small
\begin{flalign*}
 &\text{\textbf{Basic Typing }} & \fbox{$\Gamma \basicvdash e : s$}
\end{flalign*}}
\\ \
\begin{prooftree}
\hypo{}
\infer1[\textsc{BtConst}]{
\Gamma \basicvdash c : \S{Ty}(c)
}
\end{prooftree}
\quad
\begin{prooftree}
\hypo{ \Gamma(x) = s }
\infer1[\textsc{BtVar}]{
\Gamma \basicvdash x : s
}
\end{prooftree}
\quad
\begin{prooftree}
\hypo{\Gamma, x{:}\eraserf{t_1} \basicvdash e : s_2}
\infer1[\textsc{BtFun}]{
\Gamma \basicvdash \lambda x{:}t_1.e : \eraserf{t_1}{\sarr}s_2
}
\end{prooftree}
\\ \ \\ \ \\
\begin{prooftree}
\hypo{\Gamma, f{:}\eraserf{t_1{\sarr}t_2} \basicvdash \lambda x{:}t_1.e : \eraserf{t_1{\sarr}t_2}}
\infer1[\textsc{BtFix}]{
\Gamma \basicvdash \zfix{f}{t_1{\sarr}t_2}{x}{t_1}{e}
: \eraserf{t_1{\sarr}t_2}
}
\end{prooftree}
\quad
\begin{prooftree}
\hypo{\S{Ty}(\primop) = \overline{s_i}{\sarr}s_x \quad \forall i. \Gamma \basicvdash v_i : s_i \quad \Gamma, x{:}s_x \basicvdash e : s}
\infer1[\textsc{BtPureOp}]{
\Gamma \basicvdash \zlet{x}{\primop\, \overline{v_i}}{e} : s
}
\end{prooftree}
\\ \ \\ \ \\
\begin{prooftree}
\hypo{\Gamma \basicvdash v_1 : s_2 {\sarr} s_x \quad \Gamma \basicvdash v_2 : s_2 \quad \Gamma, x{:}s_x \basicvdash e : s}
\infer1[\textsc{BtApp}]{
\Gamma \basicvdash \zlet{x}{v_1\,v_2}{e} : s
}
\end{prooftree}
\quad
\begin{prooftree}
\hypo{\S{Ty}(\effop) = \overline{s_i}{\sarr}s_x \quad \forall i. \Gamma \basicvdash v_i : s_i \quad \Gamma, x{:}s_x \basicvdash e : s}
\infer1[\textsc{BtEffOp}]{
\Gamma \basicvdash \zlet{x}{\effop\, \overline{v_i}}{e} : s
}
\end{prooftree}
\\ \ \\ \ \\
\begin{prooftree}
\hypo{\emptyset \basicvdash e_1 : s_x \quad \Gamma, x{:}s_x \basicvdash e_2 : s}
\infer1[\textsc{BtLetE}]{
\Gamma \basicvdash \zlet{x}{e_1}{e_2} : s
}
\end{prooftree}
\quad
\begin{prooftree}
\hypo{\Gamma \basicvdash v : s_v \quad
\forall i, \S{Ty}(d_i) = \overline{s_j} {\sarr} s_v \quad
\Gamma, \overline{y_j{:}s_j} \basicvdash e_i : s}
\infer1[\textsc{BtMatch}]{
\Gamma \basicvdash \match{v} \overline{d_i\,\overline{y_j} \to e_i} : s
}
\end{prooftree}
}
\caption{Basic Typing Rules}
    \label{fig:basic-type-rules}
\end{figure}

\begin{figure}
 {\footnotesize
 {\small
\begin{flalign*}
 &\text{\textbf{Basic Qualifier Typing }} & \fbox{$\Gamma \basicvdash l: s  \quad \Gamma \basicvdash \phi: s$}
\end{flalign*}}
\\ \
\begin{prooftree}
\hypo{\S{Ty}(c) = s }
\infer1[\textsc{BtLitConst}]{
\Gamma \basicvdash c : s
}
\end{prooftree}
\quad
\begin{prooftree}
\hypo{ \Gamma(x) = s }
\infer1[\textsc{BtLitVar}]{
\Gamma \basicvdash x : s
}
\end{prooftree}
\quad
\begin{prooftree}
\hypo{ }
\infer1[\textsc{BtTop}]{
\Gamma \basicvdash \top : \Code{bool}
}
\end{prooftree}
\quad
\begin{prooftree}
\hypo{ }
\infer1[\textsc{BtBot}]{
\Gamma \basicvdash \bot : \Code{bool}
}
\end{prooftree}
\\ \ \\ \ \\
\begin{prooftree}
\hypo{\S{Ty}(\primop) = \overline{s_i}{\sarr}s \quad \forall i. \Gamma \basicvdash l_i : s_i }
\infer1[\textsc{BtLitOp}]{
\Gamma \basicvdash \primop\,\overline{l_i} : s
}
\end{prooftree}
\quad
\begin{prooftree}
\hypo{\S{Ty}(mp) = \overline{s_i}{\sarr}\Code{bool} \quad \forall i. \Gamma \basicvdash l_i : s_i }
\infer1[\textsc{BtLitMp}]{
\Gamma \basicvdash mp\,\overline{l_i} : s
}
\end{prooftree}
\quad
\begin{prooftree}
\hypo{
\Gamma \basicvdash \phi : \Code{bool} }
\infer1[\textsc{BtNeg}]{
\Gamma \basicvdash \neg \phi : \Code{bool}
}
\end{prooftree}
\\ \ \\ \ \\
\begin{prooftree}
\hypo{
\Gamma \basicvdash \phi_1 : \Code{bool} \quad
\Gamma \basicvdash \phi_2 : \Code{bool} }
\infer1[\textsc{BtAnd}]{
\Gamma \basicvdash \phi_1 \land \phi_2 : \Code{bool}
}
\end{prooftree}
\quad
\begin{prooftree}
\hypo{
\Gamma \basicvdash \phi_1 : \Code{bool} \quad
\Gamma \basicvdash \phi_2 : \Code{bool} }
\infer1[\textsc{BtOr}]{
\Gamma \basicvdash \phi_1 \lor \phi_2 : \Code{bool}
}
\end{prooftree}
\quad
\begin{prooftree}
\hypo{
\Gamma, x{:}b \basicvdash \phi : \Code{bool} }
\infer1[\textsc{BtForall}]{
\Gamma \basicvdash \forall x{:}b. \phi : \Code{bool}
}
\end{prooftree}
}
\caption{Basic Qualifier Typing Rules}
    \label{fig:basic-qualifier-type-rules}
\end{figure}

\newpage
\section{Declarative Typing Rules}\label{sec:tech:typing}

The full set of rules for our auxiliary typing relations are shown in
\autoref{fig:aux-rules-full}, and the full set of declarative typing
rules are shown in \autoref{fig:full-typing-rules}. We elide the basic
typing relation ({\small$\emptyset \basicvdash e : s$}) in the
premises of the rules in \autoref{fig:full-typing-rules}; which
assume all terms have a basic type.

\begin{figure}[!ht]
{\footnotesize
{\small
\begin{flalign*}
 &\text{\textbf{Well-Formed Types }} & 
 \fbox{$\Gamma \wellfoundedvdash A \quad \Gamma \wellfoundedvdash t  \quad \Gamma \wellfoundedvdash \tau$}
\end{flalign*}}
 \\ \
\begin{prooftree}
\hypo{
\eraserf{\Gamma}, \vnu{:}b \basicvdash \phi : \Code{bool}
}
\infer1[\textsc{WfBase}]{
\Gamma \wellfoundedvdash \rawnuot{b}{\phi}
}
\end{prooftree}
\quad
\begin{prooftree}
\hypo{
\Gamma \wellfoundedvdash t_x
}
\hypo{
\Gamma, x{:}\eraserf{t_x} \wellfoundedvdash t
}
\infer2[\textsc{WfArr}]{
\Gamma \wellfoundedvdash x{:}t_x \sarr t
}
\end{prooftree}
\quad
\begin{prooftree}
\hypo{
\Gamma, x{:}b \wellfoundedvdash t
}
\infer1[\textsc{WfGhost}]{
\Gamma \wellfoundedvdash x{:}b \garr t
}
\end{prooftree}
 \\ \ \\ \ \\
\begin{prooftree}
\hypo{
\S{Ty}(\effop) = \overline{x_i{:}b_i}\sarr b \quad  \eraserf{\Gamma}, \overline{x_i{:}b_i}, \vnu{:}b \basicvdash \phi : \Code{bool}
}
\infer1[\textsc{WfOpEvent}]{
\Gamma \wellfoundedvdash \evop{\effop{}}{\overline{x_i}}{\phi}
}
\end{prooftree}
\quad
\begin{prooftree}
\hypo{
\eraserf{\Gamma} \basicvdash \phi : \Code{bool}
}
\infer1[\textsc{WfTestEvent}]{
\Gamma \wellfoundedvdash \evparenth{\phi}
}
\end{prooftree}
 \\ \ \\ \ \\
\begin{prooftree}
\hypo{\Gamma \wellfoundedvdash A}
\infer1[\textsc{WfNeg}]{
\Gamma \wellfoundedvdash \neg A
}
\end{prooftree}
\quad
\begin{prooftree}
\hypo{
\Gamma \wellfoundedvdash A_1 \quad \Gamma \wellfoundedvdash A_2
}
\infer1[\textsc{WfAnd}]{
\Gamma \wellfoundedvdash A_1 \land A_2
}
\end{prooftree}
\quad
\begin{prooftree}
\hypo{
\Gamma \wellfoundedvdash A_1 \quad \Gamma \wellfoundedvdash A_2
}
\infer1[\textsc{WfOr}]{
\Gamma \wellfoundedvdash A_1 \lor A_2
}
\end{prooftree}
 \\ \ \\ \ \\
 \begin{prooftree}
\hypo{
\Gamma \wellfoundedvdash A_1 \quad \Gamma \wellfoundedvdash A_2
}
\infer1[\textsc{WfConcat}]{
\Gamma \wellfoundedvdash A_1;A_2
}
\end{prooftree}
\quad
\begin{prooftree}
\hypo{
\Gamma \wellfoundedvdash A
}
\infer1[\textsc{WfNext}]{
\Gamma \wellfoundedvdash \nextA A
}
\end{prooftree}
\quad
\begin{prooftree}
\hypo{
\Gamma \wellfoundedvdash A_1 \quad \Gamma \wellfoundedvdash A_2
}
\infer1[\textsc{WfUntil}]{
\Gamma \wellfoundedvdash A_1 \untilA A_2
}
\end{prooftree}
\ \\ \ \\ \ \\
\begin{prooftree}
\hypo{
\parbox{50mm}{\center
  $\Gamma \wellfoundedvdash A \quad \Gamma \wellfoundedvdash B \quad \Gamma \wellfoundedvdash t$ \\
  $\forall \sigma \in \denotation{\Gamma}. \langA{\sigma(B)} \subseteq \langA{\sigma(A;\globalA\topA)}$
}
}
\infer1[\textsc{WFHoare}]{
\Gamma \wellfoundedvdash \htriple{A}{t}{B}
}
\end{prooftree}
\quad
\begin{prooftree}
\hypo{
\parbox{50mm}{\center
  $\Gamma \wellfoundedvdash \tau_1$\quad
  $\Gamma \wellfoundedvdash \tau_2$ \quad
  $\eraserf{\tau_1} = \eraserf{\tau_2}$
}
}
\infer1[\textsc{WFInter}]{
\Gamma \wellfoundedvdash \tau_1 \interty \tau_2
}
\end{prooftree}
 \\ \
{\small
\begin{flalign*}
&\text{\textbf{Automata Inclusion }} & \fbox{$\Gamma \vdash A \subseteq A$}\quad
&\text{\textbf{Subtyping }} & \fbox{$\Gamma \vdash t <: t \quad \Gamma \vdash \tau <: \tau$}
\end{flalign*}
}
\begin{prooftree}
\hypo{
\parbox{30mm}{\center
  $\forall \sigma \in  \denotation{\Gamma}.$
  $\langA{\sigma(A_1)} \subseteq  \langA{\sigma(A_2)} $
}
}
\infer1[\textsc{SubAutomata}]{
\Gamma \vdash A_1 \subseteq A_2
}
\end{prooftree}
\quad
\begin{prooftree}
\hypo{
\parbox{35mm}{\center
  $\Gamma \vdash A_2 \subseteq A_1$ \quad
  $\Gamma \vdash t_1 <: t_2$ \\
  $\Gamma \vdash (A_2; \globalA\topA) \landA B_1 \subseteq$ \\
  $\ \  (A_2; \globalA\topA) \landA B_2$
}
}
\infer1[\textsc{SubHoare}]{
\Gamma \vdash \htriple{A_1}{t_1}{B_1} <: \htriple{A_2}{t_2}{B_2}
}
\end{prooftree}
\quad
\ \\ \ \\ \ \\
\begin{prooftree}
\hypo{}
\infer1[\textsc{SubIntLL}]{
\Gamma \vdash \tau_1 \interty \tau_2 <: \tau_1
}
\end{prooftree}
\ \
\begin{prooftree}
\hypo{}
\infer1[\textsc{SubIntLR}]{
\Gamma \vdash \tau_1 \interty \tau_2 <: \tau_2
}
\end{prooftree}
\ \
\begin{prooftree}
\hypo{
}
\infer1[\textsc{SubIntMerge}]{
\Gamma \vdash \htriple{A_1}{t}{B} \interty \htriple{A_2}{t}{B} <: \htriple{A_1 \lor A_2}{t}{B}
}
\end{prooftree}
\ \\ \ \\ \ \\
\begin{prooftree}
\hypo{
\parbox{15mm}{\center
  $\Gamma \vdash \tau <: \tau_1 $ \quad
  $\Gamma \vdash \tau <: \tau_2 $
}
}
\infer1[\textsc{SubIntR}]{
\Gamma \vdash \tau <: \tau_1 \interty \tau_2
}
\end{prooftree}
\quad
\begin{prooftree}
\hypo{
\Gamma, x{:}\rawnuot{b}{\top} \vdash t_1 <: t_2
}
\infer1[\textsc{SubGhostR}]{
\Gamma \vdash t_1 <: x{:}b\garr t_2
}
\end{prooftree}
\quad
\begin{prooftree}
\hypo{
\parbox{30mm}{\center
  $\exists v. \eraserf{\Gamma} \basicvdash v : b$ \quad
  $\Gamma \vdash t_1[x\mapsto v] <: t_2$
}
}
\infer1[\textsc{SubGhostL}]{
\Gamma \vdash x{:}b\garr t_1 <: t_2
}
\end{prooftree}
}
\caption{Full auxiliary typing rules.}\label{fig:aux-rules-full}
\end{figure}

\begin{figure}[!ht]
{\footnotesize
{\small
\begin{flalign*}
 &\text{\textbf{Typing}} & \fbox{$\Gamma \vdash op : t \quad \Gamma \vdash \effop : t \quad \Gamma \vdash e : t \quad \Gamma \vdash e : \tau$}
\end{flalign*}
}
\\ \
\begin{prooftree}
\hypo{
\parbox{25mm}{\center
$\Gamma \wellfoundedvdash \rawnuot{b}{\vnu = x}$\quad
$(x, \rawnuot{b}{\phi}) \in \Gamma$
}}
\infer1[\textsc{\footnotesize TBaseVar}]{
\Gamma \vdash x : \rawnuot{b}{\vnu = x}
}
\end{prooftree}
\quad
\begin{prooftree}
\hypo{
\parbox{20mm}{\center
$\Gamma \wellfoundedvdash z{:}t\sarr\tau$\quad
$(x, z{:}t\sarr\tau) \in \Gamma$
}}
\infer1[\textsc{\footnotesize TFuncVar}]{
\Gamma \vdash x : z{:}t\sarr\tau
}
\end{prooftree}
\quad
\begin{prooftree}
\hypo{
\parbox{25mm}{\center
$\Gamma \wellfoundedvdash \rawnuot{b}{\vnu = c}$\quad
$\emptyset \basicvdash c : b$
}}
\infer1[\textsc{\footnotesize TConst}]{
\Gamma \vdash c : \rawnuot{b}{\vnu = c}
}
\end{prooftree}
\\ \ \\ \ \\ \
\begin{prooftree}
\hypo{
\parbox{14mm}{\center
$\Gamma \wellfoundedvdash t$\quad
$\Delta(\primop) = t$
}}
\infer1[\textsc{\footnotesize TPOp}]{
\Gamma \vdash \primop : t
}
\end{prooftree}
\quad
\begin{prooftree}
\hypo{
\parbox{12mm}{\center
$\Gamma \wellfoundedvdash t$\quad
$\Delta(\effop) = t$
}}
\infer1[\textsc{\footnotesize TEOp}]{
\Gamma \vdash \effop : t
}
\end{prooftree}
\quad
\begin{prooftree}
\hypo{
\parbox{27mm}{\center
$\Gamma \wellfoundedvdash t_2$\quad
$\Gamma \vdash t_1 <: t_2$\quad
$\Gamma \vdash op : t_1$
}}
\infer1[\textsc{\footnotesize TSubPOp}]{
\Gamma \vdash op : t_2
}
\end{prooftree}
\quad
\begin{prooftree}
\hypo{
\parbox{27mm}{\center
$\Gamma \wellfoundedvdash t_2$\quad
$\Gamma \vdash t_1 <: t_2$\quad
$\Gamma \vdash \effop : t_1$
}}
\infer1[\textsc{\footnotesize TSubEOp}]{
\Gamma \vdash \effop : t_2
}
\end{prooftree}
\\ \ \\ \ \\ \
\begin{prooftree}
\hypo{
\parbox{27mm}{\center
$\Gamma \wellfoundedvdash \tau_2$\quad
$\Gamma \vdash \tau_1 <: \tau_2$\quad
$\Gamma \vdash e : \tau_1$
}}
\infer1[\textsc{\footnotesize TSub}]{
\Gamma \vdash e : \tau_2
}
\end{prooftree}
\quad
\begin{prooftree}
\hypo{
\parbox{27mm}{\center
$\Gamma \wellfoundedvdash t_2$\quad
$\Gamma \vdash t_1 <: t_2$\quad
$\Gamma \vdash e : t_1$
}}
\infer1[\textsc{\footnotesize TSubPP}]{
\Gamma \vdash e : t_2
}
\end{prooftree}
\quad
\begin{prooftree}
\hypo{
  \parbox{12mm}{\center
    $\Gamma \wellfoundedvdash A$\quad
    $\Gamma \vdash e : t$
  }}
\infer1[\textsc{\footnotesize TEPur}]{
  \Gamma \vdash e : \htriple{A}{t}{A}
}
\end{prooftree}
\\ \ \\ \ \\ \
\begin{prooftree}
\hypo{
\parbox{25mm}{\center
$\Gamma \wellfoundedvdash \tau_1 \interty \tau_2$\quad
$\Gamma \vdash e : \tau_1$ \quad $\Gamma \vdash e : \tau_2$
}}
\infer1[\textsc{\footnotesize TInter}]{
\Gamma \vdash e : \tau_1 \interty \tau_2
}
\end{prooftree}
\quad
\begin{prooftree}
\hypo{
\parbox{24mm}{\center
$\Gamma \wellfoundedvdash x{:}b\garr t$\quad
$\Gamma, x{:}\rawnuot{b}{\top} \vdash v : t$
}}
\infer1[\textsc{\footnotesize TGhost}]{
\Gamma \vdash v : x{:}b\garr t
}
\end{prooftree}
\quad
\begin{prooftree}
\hypo{
\parbox{24mm}{\center
$\Gamma \wellfoundedvdash x{:}t_x\sarr \tau$\quad
$\Gamma, x{:}t_x \effvdash{} e : \tau$
}
}
\infer1[\textsc{\footnotesize TFun}]{
\Gamma \vdash \zlam{x}{t_x}{e} : x{:}t_x\sarr \tau
}
\end{prooftree}
\\ \ \\ \ \\ \
\begin{prooftree}
\hypo{
\parbox{38mm}{\center
$\Gamma \wellfoundedvdash \tau$\quad
$\Gamma \vdash \primop : \overline{z_i{:}t_i}\sarr t$\quad
$\forall i. \Gamma \vdash v_i : t_i$\\
$t_x = t\overline{[z_i \mapsto v_i]}$\\
$\Gamma, x{:}t_x \vdash{} e : \tau$
}}
\infer1[\textsc{\footnotesize TPOpApp}]{
\Gamma \vdash{} \zlet{x}{\primop\ \overline{v_i}}{e} : \tau
}
\end{prooftree}
\quad
\begin{prooftree}
\hypo{
\parbox{48mm}{\center
$\Gamma \wellfoundedvdash \htriple{A}{t}{B}$\quad
$\Gamma \vdash \effop : \overline{z_i{:}t_i}\sarr \tau$\quad
$\forall i. \Gamma \vdash v_i : t_i$ \\
$\htriple{A}{t_x}{A'} = \tau\overline{[z_i \mapsto v_i]}$ \quad
$\Gamma, x{:}t_x \vdash{} e :\htriple{A'}{t}{B}$
}}
\infer1[\textsc{\footnotesize TEOpApp}]{
\Gamma \vdash{} \perform{x}{\effop{}}{\overline{v_i}}{e} : \htriple{A}{t}{B}
}
\end{prooftree}
\\ \ \\ \ \\ \
\begin{prooftree}
\hypo{
\parbox{48mm}{\center
$\Gamma \wellfoundedvdash \htriple{A}{t}{B}$\quad
$\Gamma \vdash v_1 : y{:}t_2\sarr \tau$\quad
$\Gamma \vdash v_2 : t_2$ \\
$\htriple{A}{t_x}{A'} = \tau\overline{[y \mapsto v_2]}$ \quad
$\Gamma, x{:}t_x \vdash{} e :\htriple{A'}{t}{B}$
}}
\infer1[\textsc{\footnotesize TApp}]{
\Gamma \vdash{} \zlet{x}{v_1\ v_2}{e} : \htriple{A}{t}{B}
}
\end{prooftree}
\quad
\begin{prooftree}
\hypo{
\parbox{40mm}{\center
$\Gamma \wellfoundedvdash \htriple{A}{t}{B}$\quad
$\Gamma \vdash e_x : \htriple{A}{t_x}{A'}$ \quad
$\Gamma, x{:}t_x \vdash e : \htriple{A'}{t}{B}$
}}
\infer1[\textsc{\footnotesize TLetE}]{
\Gamma \effvdash{} \zlet{x}{e_x}{e} : \htriple{A}{t}{B}
}
\end{prooftree}
\\ \ \\ \ \\ \
\begin{prooftree}
\hypo{
\parbox{70mm}{\center
$\Gamma \wellfoundedvdash \tau$\quad
$\forall i. \Delta(d_i) = \overline{{y_{j}}{:}t_{j}}\sarr\rawnuot{b}{\phi_i}$ \\
$\Gamma, \overline{{y_{j}}{:}t_{j}}, z{:}\rawnuot{b}{\vnu = v \land \phi_i} \vdash e_i : \tau$ \quad \text{ where $z$ is fresh}
}}
\infer1[\textsc{\footnotesize TMatch}]{
\Gamma \effvdash{} (\match{v} \overline{d_i\, \overline{y_{j}} \to e_i}) :
\tau }
\end{prooftree}
\quad
\begin{prooftree}
\hypo{
\parbox{24mm}{\center
$\Gamma \wellfoundedvdash t$\quad
$\Gamma, f{:}t \effvdash{} \zlam{x}{t_x}{e} : t$
}
}
\infer1[\textsc{\footnotesize TFix}]{
\Gamma \vdash \zfix{f}{t}{x}{t_x}{e} : t
}
\end{prooftree}
}
\caption{Type rules.}\label{fig:full-typing-rules}
\end{figure}

\newpage
\section{Typing Algorithm}\label{sec:tech:algo}

In this section, we provide the details of our typing algorithm. The
full set of bi-directional typing rules for this algorithm are shown
in \autoref{fig:full-bi-type-rules}; again, we again omit the basic
typing relation ({\small$\emptyset \basicvdash e : s$}) in the
precondition of these rules. The detail of alphabet transformation
($\mintermReplace$) and SFA instantiation ($\instantiate$) procedures
are provided in \autoref{algo:alphabet} and
\autoref{algo:instantiate}, respectively.

\begin{figure}[h!]
{\footnotesize
{\small
\begin{flalign*}
 &{\text{\textbf{Type Synthesis }}} &
 \fbox{$\Gamma \vdash e \typeinfer t \quad \Gamma \vdash e \typeinfer \tau$} &&
  &{\text{\textbf{Type Check }}} &
 \fbox{$\Gamma \vdash e \typecheck t  \quad  \Gamma \vdash e \typecheck \tau$}
\end{flalign*}
}
\\ \
\begin{prooftree}
\hypo{
\emptyset \wellfoundedvdash \rawnuot{b}{\vnu = c}
}
\infer1[\textsc{SynConst}]{
\Gamma \infervdash{A} c \typeinfer \rawnuot{b}{\vnu = c}
}
\end{prooftree}
\quad
\begin{prooftree}
\hypo{\Gamma \wellfoundedvdash \rawnuot{b}{\vnu = x}\quad (x, \rawnuot{b}{\phi}) \in \Gamma}
\infer1[\textsc{SynBaseVar}]{
\Gamma \infervdash{A} x \typeinfer \rawnuot{b}{\vnu = x}
}
\end{prooftree}
\\ \ \\ \ \\ \
\begin{prooftree}
\hypo{
\parbox{30mm}{\center
$\Gamma \wellfoundedvdash \overline{y{:}b}\garr z{:}t\sarr\tau$ \quad
$(x, \overline{y{:}b}\garr z{:}t\sarr\tau) \in \Gamma$
}
}
\infer1[\textsc{SynFuncVar}]{
\Gamma \infervdash{A} x \typeinfer \overline{y{:}b}\garr z{:}t\sarr\tau
}
\end{prooftree}
\quad
\begin{prooftree}
\hypo{
\parbox{25mm}{\center
$\Gamma \wellfoundedvdash x{:}b \garr \tau$\quad
  $\Gamma, x{:}\rawnuot{b}{\top} \vdash e \typecheck \tau $
}
}
\infer1[\textsc{ChkGhost}]{
\Gamma \vdash e \typecheck x{:}b \garr \tau
}
\end{prooftree}
\quad
\begin{prooftree}
\hypo{
\parbox{25mm}{\center
$\Gamma \wellfoundedvdash \tau_1 \interty \tau_2$\quad
  $\Gamma \vdash e \typecheck \tau_1$\quad
  $\Gamma \vdash e \typecheck \tau_2$
}
}
\infer1[\textsc{ChkInter}]{
\Gamma \vdash e \typecheck \tau_1 \interty \tau_2
}
\end{prooftree}
\\ \ \\ \ \\ \
\begin{prooftree}
\hypo{
\parbox{30mm}{\center
  $\Gamma \wellfoundedvdash \htriple{A}{t}{B}$ \\
  $\Gamma \vdash v_x \typeinfer t_x$\\
  $\Gamma, x{:}t_x \vdash e \typecheck \htriple{A}{t}{B}$
}
}
\infer1[\textsc{ChkLetV}]{
\Gamma \vdash \zlet{x}{v_x}{e} \typecheck \htriple{A}{t}{B}
}
\end{prooftree}
\quad
\begin{prooftree}
\hypo{
\parbox{50mm}{\center
  $\Gamma \wellfoundedvdash \htriple{A}{t}{B}$ \\
  $\Gamma \vdash e_x \typeinfer \biginterty \overline{\htriple{A_i}{t_i}{A_i'}}$ \quad $\Gamma \vdash A \subseteq \bigvee A_i $ \\
  $\forall i.\Gamma, x{:}t_i \vdash e \typecheck \htriple{(A;\globalA\topA) \land A_i'}{t}{B}$
}
}
\infer1[\textsc{ChkLetE}]{
\Gamma \vdash \zlet{x}{e_x}{e} \typecheck \htriple{A}{t}{B}
}
\end{prooftree}
\\ \ \\ \ \\ \
\begin{prooftree}
\hypo{
\parbox{35mm}{\center
  $\Gamma \wellfoundedvdash \htriple{A}{t}{B}$ \\
  $\Delta(\primop) = \overline{y_j{:}t_j}\sarr t_x $\\
  $\forall j. \Gamma \vdash v_j \typecheck t_j$ \\
  $\Gamma, x{:}t_x \vdash e \typecheck
  \htriple{A}{t}{B}$
}
}
\infer1[\textsc{ChkPOpApp}]{
\Gamma \vdash \perform{x}{\effop}{\overline{v_i}}{e} \typecheck \htriple{A}{t}{B}
}
\end{prooftree}
\quad
\begin{prooftree}
\hypo{
\parbox{60mm}{\center
  $\Gamma \wellfoundedvdash \htriple{A}{t}{B}$ \quad
  $\Delta(\effop) = \overline{z_k{:}b_k}\garr  \overline{y_j{:}t_j}\sarr \tau_x $\quad   $\forall j. \Gamma \vdash v_j \typecheck t_j$ \\
  $ \Gamma \infervdash{A}  \overline{z_k{:}b_k}\garr  \overline{y_j{:}t_j}\sarr \tau_x \effinfer \overline{z_k{:}t_k} $\\
  $\biginterty \overline{\htriple{A_i}{t_i}{A_i'}} = \tau_x\overline{[y_j \mapsto v_j]}$ \\
  $\forall i.\Gamma, \overline{z_k{:}t_k}, x{:}t_i \vdash e \typecheck
  \htriple{(A;\globalA\topA) \land A_i'}{t}{B}$
}
}
\infer1[\textsc{ChkEOpApp}]{
\Gamma \vdash \perform{x}{\effop}{\overline{v_i}}{e} \typecheck \htriple{A}{t}{B}
}
\end{prooftree}
\\ \ \\ \ \\ \
\begin{prooftree}
\hypo{
\parbox{55mm}{\center
  $\Gamma \wellfoundedvdash \htriple{A}{t}{B}$ \quad
  $\Gamma \vdash v_1 \typeinfer \overline{z_k{:}b_k}\garr  y{:}t_y \sarr \tau_x $\\
  $\Gamma \vdash v_2 \typecheck t_y$\\
 $ \Gamma \infervdash{A}  \overline{z_k{:}b_k}\garr  y{:}t_y \sarr \tau_x \effinfer \overline{z_k{:}t_k} $\\
   $\biginterty \overline{\htriple{A_i}{t_i}{A_i'}} = \tau_x[y \mapsto v_y]$ \\
  $\forall i.\Gamma, \overline{z_k{:}t_k}, x{:}t_i \vdash e \typecheck
  \htriple{(A;\globalA\topA) \land A_i'}{t}{B}$
}
}
\infer1[\textsc{ChkApp}]{
  \Gamma \vdash \zlet{x}{v_1\ v_2}{e} \typecheck \htriple{A}{t}{B}
}
\end{prooftree}
\quad
\begin{prooftree}
\hypo{
\parbox{40mm}{\center
  $\Gamma \wellfoundedvdash x{:}t_x' \sarr \tau$ \quad
  $\Gamma \vdash t_x' <: t_x$ \\
  $\Gamma, x{:}t_x' \vdash e \typecheck \tau$
}
}
\infer1[\textsc{ChkFunc}]{
\Gamma \vdash \zlam{x}{t_x}{e} \typecheck x{:}t_x' \sarr \tau
}
\end{prooftree}
\\ \ \\ \ \\ \
\begin{prooftree}
\hypo{
\parbox{35mm}{\center
  $\Gamma \wellfoundedvdash t_f'$ \quad
  $\Gamma \vdash t_f' <: t_f$ \quad
  $\Gamma, f{:}t_f' \vdash \zlam{x}{t_x}{e} \typecheck t_f'$
}
}
\infer1[\textsc{ChkFix}]{
\Gamma \vdash \zfix{f}{t_f}{x}{t_x}{e} \typecheck t_f'
}
\end{prooftree}
\quad
\begin{prooftree}
\hypo{
\parbox{70mm}{\center
  $\forall i. (d_i, \overline{{y_{j}}{:}t_{j}}\sarr\rawnuot{b}{\phi_i}) \in \Delta$ \\
  $\Gamma, \overline{{y_{j}}{:}t_{j}}, z{:}\rawnuot{b}{\vnu = v \land \phi_i} \vdash e_i \typecheck \tau$ \quad \text{ where $z$ is fresh}
}
}
\infer1[\textsc{ChkMatch}]{
\Gamma \vdash \match{v} \overline{d_i\, \overline{y_{j}} \to e_i} \typecheck \tau
}
\end{prooftree}
\\ \ \\ \ \\ \
\begin{prooftree}
\hypo{
\parbox{45mm}{\center
  $\Gamma \wellfoundedvdash \htriple{A_2}{t_2}{B_2}$ \quad
  $\Gamma \vdash e \typeinfer \htriple{A_1}{t_1}{B_1}$\quad
  $\Gamma \vdash A_2 <: A_1$\quad
  $\Gamma \vdash t_1 <: t_2$\quad
  $\Gamma \vdash (A_2;\globalA\topA) \land (A_2;\globalA\topA) \land B_1 \subseteq  B_2$
}
}
\infer1[\textsc{ChkSub}]{
\Gamma \vdash e \typecheck \htriple{A_2}{t_2}{B_2}
}
\end{prooftree}
\quad
\begin{prooftree}
\hypo{
\parbox{35mm}{\center
  $\Gamma \wellfoundedvdash t$ \quad
  $\Gamma \vdash v \typeinfer t'$\quad
  $\Gamma \vdash t' <: t$
}
}
\infer1[\textsc{ChkSubV}]{
\Gamma \vdash v \typecheck t
}
\end{prooftree}
}
\caption{Bidirectional Typing Rules}\label{fig:full-bi-type-rules}
\end{figure}


\begin{algorithm}[h!]
    \Procedure{$\mintermReplace(M, A) := $}{
    \Match{$A$}{
    \lCase{$\emptyset$ or $\epsilon$ or $\anyA$}{\Return{$A$}}
    \Case{$\evop{\effop}{\overline{x_i}}{\phi}$ or $\evparenth{\phi}$}{
        $M' \leftarrow \{ \evop{\effop}{\overline{x_i}}{\phi'} \in M ~|~  \phi' \implies \phi\}$\;
        \Return{$\bigvee_{m \in M'} \Code{Hash}(m)$}
    }
    \lCase{$A_1;A_2$}{\Return{$\mintermReplace(M, A_1); \mintermReplace(M, A_2)$}}
    \lCase{$A_1 \lorA A_2$}{\Return{$\mintermReplace(M, A_1) \lorA \mintermReplace(M, A_2)$}}
    \lCase{$A_1 \landA A_2$}{\Return{$\mintermReplace(M, A_1) \landA \mintermReplace(M, A_2)$}}
    \lCase{$A_1 \setminus A_2$}{\Return{$\mintermReplace(M, A_1) \setminus \mintermReplace(M, A_2)$}}
    \lCase{$A^*$}{\Return{$\mintermReplace(M, A_1)^*$}}
    }
    }
    \caption{Alphabet Transformation}
    \label{algo:alphabet}
\end{algorithm}

\paragraph{Alphabet Transformation} The alphabet transformation
function $\mintermReplace$ translates a SFA into a FA that shares the
same regular operators. Notably, the alphabet transformation for
effect events gathers all minterms that can imply the qualifier of the
effect event (line $17$); moreover, the effect event
$\evop{\effop}{\overline{x_i}}{\phi}$ requires the gathered minterms
have the same operator ({\small$\effop$}).

\begin{algorithm}[t!]
    \Procedure{$\instantiate(\Gamma, A, A', \overline{x_i{:}b_i}) := $}{
    \lIf{$\Gamma, \overline{x_i{:}\rawnuot{b_i}{\top}} \not\vdash A \subseteq A'$}{
        \Return{$\bot$}
    }
    $L \leftarrow \getlits(\Gamma, A \lorA B)$\;
    $M^? \leftarrow \bigcup_i \boolcombine{}(L \cup x_i)$\;
    $M^+ \leftarrow \emptyset$; $M^- \leftarrow \emptyset$; $\phi\leftarrow \top$\;
    \While{$\exists m \in M^?. m \impl \phi $}{
        $M^? \leftarrow M^? \setminus \{m\}$\;
        $\phi'\leftarrow \synthesizer(M^+ \cup M^?, M^- \cup \{m\})$\;
        \uIf{$\Gamma, \overline{x_i{:}\rawnuot{b_i}{\top}}, z{:}\nuot{unit}{\phi'} \not\vdash A \subseteq A'$}{
        $M^- \leftarrow M^- \cup \{m\}$\;
        $\phi \leftarrow \phi'$\;
        }
        \lElse{
        $M^+ \leftarrow M^+ \cup \{m\}$
        }
    }
    \ForEach{$x_i{:}b_i$}{
        $M \leftarrow \{ m \in \boolcombine{}(L_{\Gamma} \cup x_i) ~|~ m \impl \phi \}$\;
        $t_i \leftarrow \rawnuot{b_i}{\bigvee_{m \in M} m[x_i \mapsto \vnu]}$\;
    }
    \Return{$\overline{x_i{:}t_i}$}\;
    }
    \caption{Ghost Variable Inference}
    \label{algo:instantiate}
  \end{algorithm}

  \paragraph{Ghost Variable Instantiatation} As we show in
  \autoref{algo:instantiate}, the abductive synthesizer \instantiate{}
  adapts an existing algorithm~\cite{ZDDJ21} that synthesizes a
  \emph{maximal} assignment of unknown qualifier $\phi_i$ for the
  given ghost variables $x_i{:}b_i$. These inferred qualifiers are
  strong enough to prove the given inclusion relation between two
  input automata under the given type context (i.e.,
  $\Gamma \vdash A \subseteq A'$). Note that the ghost variables only
  make sense when the inclusion relation holds with all unknown
  formulae $\phi_i$ are maximally weak (i.e., true) (line $2$).
  Similar to the minterm algorithm in \autoref{algo:alphabet}, the
  abductive algorithm gathers all literals that appear in the pre- and
  post-condition automata and type context (line $3$) that will be
  used to generate candidate minterms (feature vectors, in the
  literature of abductive synthesis~\cite{ZDDJ21}), to build the
  hypothesis space for synthesis.  The function \instantiate{} uses a
  counterexample-guided refinement loop (line $6$ - $12$) adapted from
  the existing algorithm~\cite{ZDDJ21} manipulating three sets of
  minterms: unknown minterms {\small$M^?$}, positive minterms
  {\small$M^+$}, and negative minterms {\small$M^-$}. Thanks to the
  multi-abduction approach from the existing algorithm~\cite{ZDDJ21},
  inferring a formula $\phi$ from the union of all minterms induces
  all the assignment of unknown qualifiers $\overline{\phi_i}$ at one
  shot.  The auxiliary synthesizer \synthesizer{} (line $8$) returns a
  formula that can be implied by given positive minterm and unknown
  minterms but cannot be implied by any given negative
  minterms. Following the existing algorithm~\cite{ZDDJ21}, function
  \synthesizer{} is implemented as a Decision Tree (DT) that
  classifies these two sets of minterms. On the other hand, the
  inclusion query on line $9$ decides if unknown minterms should be
  positive or negative; it also guarantees the soundness of the
  algorithm, that is, a synthesized type is always consistent with the
  given automaton $A$ and $A'$. Finally, the function \instantiate{}
  returns refinement type $\rawnuot{b_i}{\phi_i}$ for each ghost
  variable $x_i$, where $\phi_i$ is the union of all local minterms
  that can imply $\phi$ (line $14$ - $15$).

\newpage
\section{Proofs}\label{sec:tech:proof}

\paragraph{Type Soundness}
The Coq formalization of our core language, typing rules, and the proof of \autoref{theorem:fundamental} and Corollary~\ref{theorem:postautomata} are available on Zenodo

\href{https://doi.org/10.5281/zenodo.10806686}{https://doi.org/10.5281/zenodo.10806686}

\paragraph{Soundness of Algorithmic Typing} We present the proof for \autoref{theorem:algo-sound} from \autoref{sec:algo}. The proof requires the following lemmas about the $\subqueryA$ subroutine and the instantiation function $\instantiate$.

\begin{lemma}\label{lemma:subtyping-sound}[Soundness of Inclusion Algorithm] Given type context {\small$\Gamma$}, automaton $A$ and $A'$,
{\small\begin{align*}
    \subqueryA(\Gamma, A, A') \impl \Gamma \vdash A \subseteq A'
\end{align*}}
\end{lemma}

\begin{lemma}\label{lemma:infer-sound}[Soundness of Instantiation Function] Given a type context {\small$\Gamma$}, ghost variables $\overline{x_i{:}b_i}$, automaton $A$ and $A'$,
{\small\begin{align*}
    \instantiate(\Gamma, A, A', \overline{x_i{:}b_i}) \implies  \Gamma, \overline{x_i{:}\rawnuot{b_i}{\phi_i}} \vdash A \subseteq A'
\end{align*}}
\end{lemma}

The soundness of \subqueryA establishes that it is safe to treat the subtyping relation in our typing algorithm as in declarative type system. On the other hand, the soundness of instantiation function induces the following lemma.

\begin{lemma}\label{lemma:ghost}[Instantiation Implies Precondition Automata Inclusion] Given a type context {\small$\Gamma$}, term $e$, consider the following the refinement type that has ghost variables $\overline{z_k{:}b_k}\garr  \overline{y_j{:}t_j}\sarr \biginterty \overline{\htriple{A_i}{t_i}{A_i'}}$,
{\small\begin{align*}
    \Gamma \infervdash{A} \overline{z_k{:}b_k}\garr  \overline{y_j{:}t_j}\sarr \overline{\htriple{A_i}{t_i}{A_i'}} \effinfer \overline{z_k{:}t_k} \implies &  \Gamma, \overline{z_k{:}t_k} \vdash A \subseteq \bigvee A_i \land
    \\ &\Gamma, \overline{z_k{:}t_k} \vdash \overline{z_k{:}b_k}\garr  \overline{y_j{:}t_j}\sarr \overline{\htriple{A_i}{t_i}{A_i'}} <: \overline{y_j{:}t_j}\sarr \overline{\htriple{A_i}{t_i}{A_i'}}
\end{align*}}
\end{lemma}

We need three auxiliary lemmas about declarative typing, which
establish that we can drop unused binding from the type context,
well-typed substitution preserves typing judgement, and
well-formedness implies subtyping.

\begin{lemma}\label{lemma:strengthening}[Strengthening] Given a type context {\small$\Gamma$}, term {\small$e$} and HAT {\small$\tau$}, for any free variable {\small$x$} and pure refinement type {\small$t$}:
{\small$
    \Gamma, x{:}t \vdash e : \tau \land \Gamma \wellfoundedvdash \tau \impl \Gamma \vdash e : \tau$}
\end{lemma}

\begin{lemma}\label{lemma:subst-typing}[Substitution Preserves Typing] Given a type context {\small$\Gamma$}, term {\small$e$} and HAT {\small$\tau$}, for any free variable {\small$x$} and pure refinement type {\small$t$}:
{\small$
    \Gamma, x{:}t \vdash e : \tau \land \Gamma \vdash v : t \impl \Gamma \vdash e : \tau[x\mapsto v]$}
\end{lemma}

\begin{lemma}\label{lemma:wf-sub}[Well-formedness implies Subtyping] Given a type context {\small$\Gamma$}, term {\small$e$} and HAT {\small$\htriple{A}{t}{B}$}, we have
{\small$
    \Gamma \wellfoundedvdash \htriple{A}{t}{B} \impl \Gamma \vdash \htriple{A}{t}{B} <: \htriple{A \land A' }{t}{B \land (A';\globalA\topA) }$}.
\end{lemma}
\begin{proof}
Consider the denotation of HAT $\htriple{A}{t}{B}$ and  $\htriple{A \land A' }{t}{B \land (A';\globalA\topA) }$:
{\footnotesize\begin{align*}
    \denotation{ \htriple{A}{t}{B}} &= \{e ~|~ \emptyset \basicvdash e : \eraserf{t} \land \forall \alpha\, \alpha'\, v.\; \alpha \in \langA{A} \land \msteptr{\alpha}{e}{\alpha'}{v} \impl v \in \denotation{t} \land \alpha \listconcat \alpha' \in \langA{B} \}
    \\ \denotation{ \htriple{A \land A' }{t}{B \land (A';\globalA\topA) }} &= \{e ~|~ \emptyset \basicvdash e : \eraserf{t} \land \forall \alpha\, \alpha'\, v.\; \alpha \in \langA{A \land A'} \land \msteptr{\alpha}{e}{\alpha'}{v} \impl
    \\&\qquad\qquad  v \in \denotation{t} \land \alpha \listconcat \alpha' \in \langA{B \land (A';\globalA\topA)} \}
\end{align*}}\noindent
Consider a trace $\alpha \in \langA{A \land A'}$, it also can be recognized by automata $A$, thus for any $\msteptr{\alpha}{e}{\alpha'}{v}$, we have $\alpha \listconcat \alpha' \in \langA{B}$. Moreover, the trace $\alpha \listconcat \alpha'$ has a prefix $\alpha$, thus is naturally recognized by automata $(A \land A');\globalA\topA$. Thus, we know $\alpha \listconcat \alpha' \in \langA{(A \land A');\globalA\topA \land B}$.
Note that the rule \textsc{WFHoare} requires that precondition is always the ``prefix'' of the postcondition automata, thus $\Gamma \wellfoundedvdash \htriple{A}{t}{B}$ implies $(A;\globalA\topA) \subseteq B$, which means that $\langA{(A \land A');\globalA\topA \land B} = \langA{A';\globalA\topA \land B}$. Thus the term $e$ is in the denotation of type $\htriple{A \land A' }{t}{B \land (A';\globalA\topA) }$ when it is in the denotation of type $\htriple{A}{t}{B}$.
\end{proof}

Now we can prove the soundness theorem of our typing algorithm with respect to our declarative type system. As the type synthesis rules are defined mutually recursively, we simultaneously prove both are correct. The following theorem states a stronger inductive invariant that is sufficient to prove the soundness of the typing algorithm.
\begin{theorem}\label{theorem:algo-ind}[Soundness of the type synthesis and type check algorithm] For all type context
  $\Gamma$, term $e$, pure refinement type $t$ and HAT $\tau$,
{\small\begin{align*}
  &\Gamma \vdash e \typecheck \tau \implies \Gamma \vdash e : \tau \\
  &\Gamma \vdash e \typecheck t \implies \Gamma \vdash e : t \\
  &\Gamma \vdash e \typeinfer \tau \implies \Gamma \vdash e : \tau \\
  &\Gamma \vdash e \typeinfer t \implies \Gamma \vdash e : t
\end{align*}}
\end{theorem}
\begin{proof} We proceed by induction over the mutual recursive structure of type synthesis ($\typeinfer$) and type check ($\typecheck$) rules. In the cases for; synthesis and checking rules of rule \textsc{SynConst}, \textsc{SynBaseVar}, \textsc{SynFuncVar}, \textsc{ChkGhost}, \textsc{ChkInter}, \textsc{ChkFunc}, \textsc{ChkFix}, \textsc{ChkMatch}, the typing rules in \autoref{fig:full-typing-rules} align exactly with these rules. There are seven remaining cases that are about let-bindings (\textsc{ChkLetV} and \textsc{ChkLetE}), application (\textsc{ChkApp}, \textsc{ChkPOpApp}, and \textsc{ChkEOpApp}), and subtyping (\textsc{ChkSubV} and \textsc{ChkSub}). The rule \textsc{ChkLetV}, \textsc{ChkPOpApp}, and \textsc{ChkSubV} are less interesting, which are similar with standard refinement typing algorithm\cite{JV21} and simply drop the pre- and post-condition automata from their effectful version. Consequently, we show the proof of three interesting cases (\textsc{ChkSub}, \textsc{ChkLetE}, \textsc{ChkEOpApp}) below, where the case \textsc{ChkApp} is almost the same as \textsc{ChkEOpApp}.

\begin{enumerate}[label=Case]

\item \textsc{ChkSub}:
\ \\ \ \\
{\footnotesize
\begin{prooftree}
\hypo{
\parbox{50mm}{\center
  $\Gamma \wellfoundedvdash \htriple{A_2}{t_2}{B_2}$ \quad
  $\Gamma \vdash e \typeinfer \htriple{A_1}{t_1}{B_1}$\quad
  $\Gamma \vdash A_2 <: A_1$\quad
  $\Gamma \vdash t_1 <: t_2$\quad
  $\Gamma \vdash (A_2;\globalA\topA) \land B_1 \subseteq  (A_2;\globalA\topA) \land B_2$
}
}
\infer1[\textsc{ChkSub}]{
\Gamma \vdash e \typecheck \htriple{A_2}{t_2}{B_2}
}
\end{prooftree}
}
\ \\ \ \\
This rule connects the relation between the type synthesis and type check judgement. From the induction hypothesis and the precondition of \textsc{ChkSub}, we know
    {\footnotesize
    \begin{alignat}{2}
    \setcounter{equation}{0}
        &\Gamma \vdash A_2 <: A_1
        \ \ &\ \
        &\text{from precondition}\\
        &\Gamma \vdash t_1 <: t_2
        \ \ &\ \
        &\text{from precondition}\\
        &\Gamma \vdash (A_2;\globalA\topA) \land B_1 \subseteq  (A_2;\globalA\topA) \land B_2
        \ \ &\ \
        &\text{from precondition}\\
        &\Gamma \vdash \htriple{A_1}{t_1}{B_1} <: \htriple{A_2}{t_2}{B_2}
        \ \ &\ \
        &\text{via rule \textsc{SubHoare} and (1) (2) (3)}\\
        &\Gamma \vdash e : \htriple{A_1}{t_1}{B_1}
        \ \ &\ \
        &\text{inductive hypothesis}\\
        &\Gamma \vdash e : \htriple{A_2}{t_2}{B_2}
        \ \ &\ \
        &\text{via rule \textsc{TSub} and (4) (5)}
    \end{alignat}}
    Then the proof immediate holds in this case.
\ \\
\item \textsc{ChkLetE}:
\ \\ \ \\
{\footnotesize
\begin{prooftree}
\hypo{
\parbox{50mm}{\center
  $\Gamma \wellfoundedvdash \htriple{A}{t}{B}$ \\
  $\Gamma \vdash e_x \typeinfer \biginterty \overline{\htriple{A_i}{t_i}{A_i'}}$ \quad $\Gamma \vdash A \subseteq \bigvee A_i $ \\
  $\forall i.\Gamma, x{:}t_i \vdash e \typecheck \htriple{(A;\globalA\topA) \land A_i'}{t}{B}$
}
}
\infer1[\textsc{ChkLetE}]{
\Gamma \vdash \zlet{x}{e_x}{e} \typecheck \htriple{A}{t}{B}
}
\end{prooftree}
}
\ \\ \ \\
This rule can be treated as a combination of subtyping of intersection types (e.g., \textsc{SubIntLL}) and \textsc{TLet}. From the induction hypothesis and the precondition of \textsc{ChkApp}, we know
    {\footnotesize\begin{alignat}{2}
    \setcounter{equation}{0}
        &\Gamma \vdash e_x : \biginterty \htriple{A_i}{t}{A_i'}
        \ \ &\ \
        &\text{inductive hypothesis}\\
        &\forall i. \Gamma \vdash e_x : \overline{\htriple{A_i}{t_i}{A_i'}}
        \ \ &\ \
        &\text{via rule \textsc{TSub}, rule \textsc{SubIntLL} and (1)}\\
        &\forall i. \Gamma \vdash e_x : \overline{\htriple{A \land A_i}{t_i}{(A;\globalA\topA) \land A_i'}}
        \ \ &\ \
        &\text{via Lemma~\ref{lemma:wf-sub}, precondition, and (2)}\\
        &\forall i.\Gamma, x{:}t_i \vdash e : \htriple{(A;\globalA\topA) \land A_i'}{t}{B}
        \ \ &\ \
        &\text{inductive hypothesis}\\
        &\forall i.\Gamma \vdash \zlet{x}{e_x}{e} : \htriple{A \land A_i'}{t}{B}
        \ \ &\ \
        &\text{rule \textsc{TLet} and (3) (4)}\\
        &\Gamma \vdash \zlet{x}{e_x}{e} : \biginterty \htriple{A \land A_i}{t}{B}
        \ \ &\ \
        &\text{rule \textsc{TInter} and (5)}\\
        &\Gamma \vdash \zlet{x}{e_x}{e} : \htriple{A \land \bigvee A_i}{t}{B}
        \ \ &\ \
        &\text{via rule \textsc{TSub}, rule \textsc{SubIntMerge} and (6)}\\
        &\Gamma \vdash A \subseteq \bigvee A_i
        \ \ &\ \
        &\text{precondition}\\
        &\Gamma \vdash \zlet{x}{e_x}{e} : \htriple{A}{t}{B}
        \ \ &\ \
        &\text{via rule \textsc{TSub}, rule \textsc{SubHoare} and (7) (8)}
    \end{alignat}}\noindent
    which is exactly what we needed to prove for this case.
\ \\
\item \textsc{ChkEOpApp}:
\ \\ \ \\
{\footnotesize
\begin{prooftree}
\hypo{
\parbox{60mm}{\center
  $\Gamma \wellfoundedvdash \htriple{A}{t}{B}$ \quad
  $\Delta(\effop) = \overline{z_k{:}b_k}\garr  \overline{y_j{:}t_j}\sarr \tau_x $\quad   $\forall j. \Gamma \vdash v_j \typecheck t_j$ \\
  $ \Gamma \infervdash{A}  \overline{z_k{:}b_k}\garr  \overline{y_j{:}t_j}\sarr \tau_x \effinfer \overline{z_k{:}t_k} $\\
  $\biginterty \overline{\htriple{A_i}{t_i}{A_i'}} = \tau_x\overline{[y_j \mapsto v_j]}$ \\
  $\forall i.\Gamma, \overline{z_k{:}t_k}, x{:}t_i \vdash e \typecheck
  \htriple{(A;\globalA\topA) \land A_i'}{t}{B}$
}
}
\infer1[\textsc{ChkEOpApp}]{
\Gamma \vdash \perform{x}{\effop}{\overline{v_i}}{e} \typecheck \htriple{A}{t}{B}
}
\end{prooftree}
}
\ \\ \ \\
This rule can be treated as a combination of \textsc{TSub} and \textsc{TEffOp}. From the induction hypothesis and the precondition of \textsc{ChkEffOp}, we know
    {\footnotesize\begin{alignat}{2}
    \setcounter{equation}{0}
        &\Delta(\effop) = \overline{z_k{:}b_k}\garr  \overline{y_j{:}t_j}\sarr \tau_x
        \ \ &\ \
        &\text{precondition}\\
        &\Gamma \vdash \effop : \overline{z_k{:}b_k}\garr  \overline{y_j{:}t_j}\sarr \tau_x
        \ \ &\ \
        &\text{via rule \textsc{TEOp} and (1)}\\
        &\forall j. \Gamma \vdash v_j \typecheck t_j
        \ \ &\ \
        &\text{precondition}\\
        &\Gamma \vdash \effop : \overline{z_k{:}b_k}\garr  \overline{y_j{:}t_j}\sarr \tau_x\overline{[y_j \mapsto v_j]}
        \ \ &\ \
        &\text{Lemma~\ref{lemma:subst-typing} and (2) (3)}\\
        &\Gamma \vdash \effop : \overline{z_k{:}b_k}\garr  \overline{y_j{:}t_j}\sarr \biginterty \overline{\htriple{A_i}{t_i}{A_i'}}
        \ \ &\ \
        &\text{precondition and (4)}\\
        &\Gamma \infervdash{A}  \overline{z_k{:}b_k}\garr  \overline{y_j{:}t_j}\sarr \tau_x \effinfer \overline{z_k{:}t_k}
         \ \ &\ \
        &\text{precondition}\\
        &\Gamma,  \overline{z_k{:}t_k} \vdash A \subseteq \bigvee A_i
         \ \ &\ \
        &\text{via Lemma~\ref{lemma:ghost}}\\
        &\Gamma, \overline{z_k{:}t_k}\vdash \effop : \overline{y_j{:}t_j}\sarr \biginterty \overline{\htriple{A_i}{t_i}{A_i'}}
         \ \ &\ \
        &\text{via Lemma~\ref{lemma:ghost} and rule \textsc{TSub}}
    \end{alignat}
    }\noindent
    Now the operator $\effop$ has the return type $\biginterty \overline{\htriple{A_i}{t_i}{A_i'}}$. The rest of proof it almost the same as the proof in the case \textsc{ChkLetE}:
    {\footnotesize\begin{alignat}{2}
        &\forall i. \Gamma, \overline{z_k{:}t_k} \vdash \effop : \overline{\htriple{A_i}{t_i}{A_i'}}
        \ \ &\ \
        &\text{via rule \textsc{TSub}, rule \textsc{SubIntLL} and (8)}\\
        &\forall i. \Gamma, \overline{z_k{:}t_k}  \vdash \effop : \overline{\htriple{A \land A_i}{t_i}{(A;\globalA\topA) \land A_i'}}
        \ \ &\ \
        &\text{via Lemma~\ref{lemma:wf-sub}, precondition, and (9)}\\
        &\forall i.\Gamma, \overline{z_k{:}t_k}, x{:}t_i \vdash e : \htriple{(A;\globalA\topA) \land A_i'}{t}{B}
        \ \ &\ \
        &\text{inductive hypothesis}\\
        &\forall i.\Gamma, \overline{z_k{:}t_k} \vdash  \perform{x}{\effop}{\overline{v_i}}{e}  : \htriple{A \land A_i'}{t}{B}
        \ \ &\ \
        &\text{rule \textsc{TEOpApp} and (10) (11)}\\
        &\Gamma, \overline{z_k{:}t_k} \vdash \perform{x}{\effop}{\overline{v_i}}{e} : \biginterty \htriple{A \land A_i}{t}{B}
        \ \ &\ \
        &\text{rule \textsc{TInter} and (12)}\\
        &\Gamma, \overline{z_k{:}t_k} \vdash  \perform{x}{\effop}{\overline{v_i}}{e} : \htriple{A \land \bigvee A_i}{t}{B}
        \ \ &\ \
        &\text{via rule \textsc{TSub}, rule \textsc{SubIntMerge} and (13)}\\
        &\Gamma,\overline{z_k{:}t_k} \vdash  \perform{x}{\effop}{\overline{v_i}}{e} : \htriple{A}{t}{B}
        \ \ &\ \
        &\text{via rule \textsc{TSub}, rule \textsc{SubHoare} and (7) (14)}\\
        &\Gamma \vdash  \perform{x}{\effop}{\overline{v_i}}{e} : \htriple{A}{t}{B}
        \ \ &\ \
        &\text{via Lemma~\ref{lemma:strengthening},  precondition and (15)}
    \end{alignat}}\noindent
    which are valid.
\end{enumerate}
\end{proof}

\paragraph{Decidability of Algorithmic Typing} We present the proof for \autoref{theorem:algo-decidable} from \autoref{sec:algo}. The proof requires the following lemmas about the $\subqueryA$ subroutine and the instantiation function $\instantiate$.

\begin{lemma}\label{lemma:sub-base-deciable}[Decidability of Base Subtyping] the $\Gamma \vdash \rawnuot{b}{\phi_1} <: \rawnuot{b}{\phi_2}$ is decidable.
\end{lemma}
\begin{proof} As we mentioned in Section 5.1, similar to standard refinement type system, the subtyping check will be encoded as a query in EPR, which is in the decidable fragment of FOL. Thus, the base subtyping check is decidable.
\end{proof}

\begin{lemma}\label{lemma:subqueryA-deciable}[Decidability of Inclusion Algorithm] the $\subqueryA$ is decidable.
\end{lemma}
\begin{proof} The gathered literals are finite, thus the induced boolean combination is also finite. Then the loop in $\subqueryA$ (line $3$ in \autoref{algo:inclusion}) has a finite bound. According to Lemma~\ref{lemma:sub-base-deciable}, $\subqueryA$ is decidable.
\end{proof}

\begin{lemma}\label{lemma:instantiate-deciable}[Decidability of Instantiation Algorithm] the $\instantiate$ is decidable.
\end{lemma}
\begin{proof} The gathered literals are finite, thus the induced boolean combination is also finite. Then the loop in $\instantiate$ (line $6$ in \autoref{algo:inclusion}) has a finite bound. According to Lemma~\ref{lemma:sub-base-deciable}, $\instantiate$ is decidable.
\end{proof}

\begin{lemma}\label{lemma:well-formedness-deciable}[Decidability of Well-Formedness] For all type context $\Gamma$, type $t$, and $\tau$, the $\Gamma \wellfoundedvdash t$ and $\Gamma \wellfoundedvdash \tau$ is decidable.
\end{lemma}
\begin{proof} According to the well-formedness rules shown in \autoref{fig:aux-rules-full}, these rules are inductively defined and structural decreasing over the types ($t_1$ or $\tau_1$). Thus, they always terminate.
\end{proof}

\begin{lemma}\label{lemma:subtyping-deciable}[Decidability of Subtyping] For all type context $\Gamma$, type $t_1$, $t_2$, $\tau_1$, and $\tau_2$ , the $\Gamma \vdash t_1 <: t_2$ and $\Gamma \vdash \tau_1 <: \tau_2$ is decidable.
\end{lemma}
\begin{proof} According to the subtyping rules shown in \autoref{fig:aux-rules-full}, Lemma~\autoref{lemma:sub-base-deciable} and Lemma~\autoref{lemma:subqueryA-deciable}, these rules are inductively defined and structural decreasing over the types ($t_1$ or $\tau_1$). Thus, they always terminate.
\end{proof}

\begin{lemma}\label{lemma:inst-deciable}[Decidability of Instantiation] For all type context $\Gamma$ and automata $A$,

\noindent  $\Gamma \infervdash{A}  \overline{z_k{:}b_k}\garr  \overline{y_j{:}t_j}\sarr \tau_x \effinfer \overline{z_k{:}t_k}$ is decidable.
\end{lemma}
\begin{proof} According to the rule shown in \autoref{fig:aux-rules-full} and Lemma~\ref{lemma:instantiate-deciable}, it always terminates.
\end{proof}

Now we can prove the decidability theorem of our typing algorithm. As the type synthesis rules are defined mutually recursively, we simultaneously prove both are decidable. The following theorem states a stronger inductive invariant that is sufficient to prove the decidability of the typing algorithm.
\begin{theorem}[Decidability of
  Algorithmic Typing]
  Type checking a term {\small $e$} against a type
  {\small $\tau$} in a typing context {\small $\Gamma$}, all {\small
    $\Gamma \vdash e \typecheck \tau$}, {\small $\Gamma \vdash v \typecheck t$}, {\small $\Gamma \vdash e \typeinfer \tau$}, and {\small
    $\Gamma \vdash v \typeinfer t$} is decidable.
\end{theorem}
\begin{proof} According to Lemma~\ref{lemma:well-formedness-deciable}, Lemma~\ref{lemma:subtyping-deciable}, and Lemma~\ref{lemma:inst-deciable}, the auxiliary functions (well-formedness, subtyping, instantiation) always terminate.
Thus, the key of the decidability proof is finding a ranking function over the inputs of typing algorithm, which is based on the size of terms ($\sizeE$) and types ($\sizeT$):
\begin{align*}
    \sizeE(c) \ &\doteq 1
    \\\sizeE(x) \ &\doteq 1
    \\\sizeE(d(\overline{v_i})) \ &\doteq 1 + \Sigma_i \sizeE(v_i)
    \\\sizeE(\zlam{x}{t}{e}) \ &\doteq 1 + \sizeE(e)
    \\\sizeE(\zfix{f}{t}{x}{t}{e}) \ &\doteq 2 + \sizeE(e)
    \\\sizeE(\zlet{x}{\effop\ \overline{v_i}}{e}) \ &\doteq 1 + \sizeE(e) + \Sigma_i \sizeE(v_i)
    \\\sizeE(\zlet{x}{\primop\ \overline{v_1}}{e}) \ &\doteq 1 + \sizeE(e) + \Sigma_i \sizeE(v_i)
    \\\sizeE(\zlet{x}{v_1\ v_2}{e}) \ &\doteq 1 + \sizeE(e) + \sizeE(v_1) + \sizeE(v_2)
    \\\sizeE(\zlet{x}{e_1}{e_2}) \ &\doteq 1 + \sizeE(e_1) + \sizeE(e_2)
    \\\sizeE(\match{v}{\overline{d\ \overline{y} \sarr {e_i}}}) \ &\doteq 1 + \sizeE v + \Sigma_i \sizeE(e_i)
    \\\sizeT(\rawnuot{b}{\phi}) \ &\doteq 1
    \\\sizeT(x{:}t\sarr\tau) \ &\doteq 1 + \sizeT(t) + \sizeT(\tau)
    \\\sizeT(x{:}b\garr\tau) \ &\doteq 1 + \sizeT(\tau)
    \\\sizeT(\htriple{A}{t}{A}) \ &\doteq 1 + \sizeT(t)
    \\\sizeT(\tau_1 \interty \tau_2) \ &\doteq 1 + \sizeT(\tau_1) + \sizeT(\tau_2)
\end{align*}\noindent
With the help of $\sizeE$, we define a lexicographic ranking function as a string contains two natural numbers (a pair of natural numbers).
\begin{align*}
    \rankE(\Gamma \vdash e \typecheck \tau) \ &\doteq (\sizeE(e), \sizeT(\tau))
    \\\rankE(\Gamma \vdash e \typeinfer \tau) \ &\doteq (\sizeE(e), 0)
\end{align*}\noindent
where one rank is less than another iff
\begin{align*}
    (a_1, b_1) < (a_2, b_2) \iff a_1 < b_1 \lor (a_1 = b_1 \land b_1 < b_2)
\end{align*}\noindent
We proceed by induction over the mutual recursive structure of type synthesis ($\typeinfer$) and type check ($\typecheck$) rules. In the cases for; synthesis rule \textsc{SynConst}, \textsc{SynBaseVar}, and \textsc{SynFuncVar} don't have recursive call, thus always terminates. Consequently, we try to prove that the rank current call is greater than recursive calls in the rest of cases below.
\begin{enumerate}[label=Case]

\item \textsc{ChkGhost}:
\ \\ \ \\
{\footnotesize
\begin{prooftree}
\hypo{
\parbox{25mm}{\center
$\Gamma \wellfoundedvdash x{:}b \garr \tau$\quad
  $\Gamma, x{:}\rawnuot{b}{\top} \vdash e \typecheck \tau $
}
}
\infer1[\textsc{ChkGhost}]{
\Gamma \vdash e \typecheck x{:}b \garr \tau
}
\end{prooftree}
}
\ \\ \ \\
We the rank current call is
\begin{align*}
    \rankE(\Gamma \vdash e \typecheck x{:}b \garr \tau) &= (\sizeE(e), 1 + \sizeT(\tau))
\end{align*}\noindent
which is greater than recursive call:
\begin{align*}
    \rankE(\Gamma, x{:}\rawnuot{b}{\top} \vdash e \typecheck \tau) &= (\sizeE(e), \sizeT(\tau))
\end{align*}\noindent

\item \textsc{ChkInter}:
\ \\ \ \\
{\footnotesize
\begin{prooftree}
\hypo{
\parbox{25mm}{\center
$\Gamma \wellfoundedvdash \tau_1 \interty \tau_2$\quad
  $\Gamma \vdash e \typecheck \tau_1$\quad
  $\Gamma \vdash e \typecheck \tau_2$
}
}
\infer1[\textsc{ChkInter}]{
\Gamma \vdash e \typecheck \tau_1 \interty \tau_2
}
\end{prooftree}
}
\ \\ \ \\
We the rank current call is
\begin{align*}
    \rankE(\Gamma \vdash e \typecheck \tau_1 \interty \tau_2) &= (\sizeE(e), 1 + \sizeT(\tau_1) + \sizeT(\tau_2))
\end{align*}\noindent
which is greater than recursive call:
\begin{align*}
    \rankE(\Gamma \vdash e \typecheck \tau_1) &= (\sizeE(e), \sizeT(\tau_1))
    \\\rankE(\Gamma \vdash e \typecheck \tau_2) &= (\sizeE(e), \sizeT(\tau_2))
\end{align*}\noindent

\item \textsc{ChkLetV}:
\ \\ \ \\
{\footnotesize
\begin{prooftree}
\hypo{
\parbox{30mm}{\center
  $\Gamma \wellfoundedvdash \htriple{A}{t}{B}$ \\
  $\Gamma \vdash v_x \typeinfer t_x$\\
  $\Gamma, x{:}t_x \vdash e \typecheck \htriple{A}{t}{B}$
}
}
\infer1[\textsc{ChkLetV}]{
\Gamma \vdash \zlet{x}{v_x}{e} \typecheck \htriple{A}{t}{B}
}
\end{prooftree}
}
\ \\ \ \\
We the rank current call is
\begin{align*}
    \rankE(\Gamma \vdash \zlet{x}{v_x}{e} \typecheck \htriple{A}{t}{B}) &= (1 + \sizeE(v_x) +\sizeE(e), 1 + \sizeT(t))
\end{align*}\noindent
which is greater than recursive calls:
\begin{align*}
    \rankE(\Gamma \vdash v_x \typeinfer t_x) &= (\sizeE(v_x), 0)
    \\\rankE(\Gamma, x{:}t_x \vdash e \typecheck \htriple{A}{t}{B}) &= (\sizeE(e), 1 + \sizeT(t))
\end{align*}\noindent

\item \textsc{ChkLetE}:
\ \\ \ \\
{\footnotesize
\begin{prooftree}
\hypo{
\parbox{50mm}{\center
  $\Gamma \wellfoundedvdash \htriple{A}{t}{B}$ \\
  $\Gamma \vdash e_x \typeinfer \biginterty \overline{\htriple{A_i}{t_i}{A_i'}}$ \quad $\Gamma \vdash A \subseteq \bigvee A_i $ \\
  $\forall i.\Gamma, x{:}t_i \vdash e \typecheck \htriple{(A;\globalA\topA) \land A_i'}{t}{B}$
}
}
\infer1[\textsc{ChkLetE}]{
\Gamma \vdash \zlet{x}{e_x}{e} \typecheck \htriple{A}{t}{B}
}
\end{prooftree}
}
\ \\ \ \\
We the rank current call is
\begin{align*}
    \rankE(\Gamma \vdash \zlet{x}{e_x}{e} \typecheck \htriple{A}{t}{B}) &= (1 + \sizeE(e_x) +\sizeE(e), 1 + \sizeT(t))
\end{align*}\noindent
which is greater than recursive calls:
\begin{align*}
    \rankE(\Gamma \vdash e_x \typeinfer \biginterty \overline{\htriple{A_i}{t_i}{A_i'}}) &= (\sizeE(e_x), 0)
    \\\rankE(\Gamma, x{:}t_i \vdash e \typecheck \htriple{(A;\globalA\topA) \land A_i'}{t}{B}) &= (\sizeE(e), 1 + \sizeT(t))
\end{align*}\noindent

\item \textsc{ChkPOpApp}:
\ \\ \ \\
{\footnotesize
\begin{prooftree}
\hypo{
\parbox{35mm}{\center
  $\Gamma \wellfoundedvdash \htriple{A}{t}{B}$ \\
  $\Delta(\primop) = \overline{y_j{:}t_j}\sarr t_x $\\
  $\forall j. \Gamma \vdash v_j \typecheck t_j$ \\
  $\Gamma, x{:}t_x \vdash e \typecheck
  \htriple{A}{t}{B}$
}
}
\infer1[\textsc{ChkPOpApp}]{
\Gamma \vdash \perform{x}{\effop}{\overline{v_i}}{e} \typecheck \htriple{A}{t}{B}
}
\end{prooftree}
}
\ \\ \ \\
We the rank current call is
\begin{align*}
    \rankE(\Gamma \vdash \perform{x}{\effop}{\overline{v_i}}{e} \typecheck \htriple{A}{t}{B}) &= (1 + \sizeE(e) + \Sigma \sizeE(v_i), 1 + \sizeT(t))
\end{align*}\noindent
which is greater than recursive calls:
\begin{align*}
    \rankE(\Gamma \vdash v_j \typecheck t_j) &= (\sizeE(v_j), \sizeT(t_j))
    \\\rankE(\Gamma, x{:}t_i \vdash e \typecheck \htriple{(A;\globalA\topA) \land A_i'}{t}{B}) &= (\sizeE(e), 1 + \sizeT(t))
\end{align*}\noindent

\item \textsc{ChkEOpApp}:
\ \\ \ \\
{\footnotesize
\begin{prooftree}
\hypo{
\parbox{60mm}{\center
  $\Gamma \wellfoundedvdash \htriple{A}{t}{B}$ \quad
  $\Delta(\effop) = \overline{z_k{:}b_k}\garr  \overline{y_j{:}t_j}\sarr \tau_x $\quad   $\forall j. \Gamma \vdash v_j \typecheck t_j$ \\
  $ \Gamma \infervdash{A}  \overline{z_k{:}b_k}\garr  \overline{y_j{:}t_j}\sarr \tau_x \effinfer \overline{z_k{:}t_k} $\\
  $\biginterty \overline{\htriple{A_i}{t_i}{A_i'}} = \tau_x\overline{[y_j \mapsto v_j]}$ \\
  $\forall i.\Gamma, \overline{z_k{:}t_k}, x{:}t_i \vdash e \typecheck
  \htriple{(A;\globalA\topA) \land A_i'}{t}{B}$
}
}
\infer1[\textsc{ChkEOpApp}]{
\Gamma \vdash \perform{x}{\effop}{\overline{v_i}}{e} \typecheck \htriple{A}{t}{B}
}
\end{prooftree}
}
\ \\ \ \\
We the rank current call is
\begin{align*}
    \rankE(\Gamma \vdash \perform{x}{\effop}{\overline{v_i}}{e} \typecheck \htriple{A}{t}{B}) &= (1 + \sizeE(e) + \Sigma \sizeE(v_i), 1 + \sizeT(t))
\end{align*}\noindent
which is greater than recursive calls:
\begin{align*}
    \rankE(\Gamma \vdash v_j \typecheck t_j) &= (\sizeE(v_j), \sizeT(t_j))
    \\\rankE(\Gamma, \overline{z_k{:}t_k}, x{:}t_i \vdash e \typecheck
  \htriple{(A;\globalA\topA) \land A_i'}{t}{B}) &= (\sizeE(e), 1 + \sizeT(t))
\end{align*}\noindent

\item \textsc{ChkApp}:
\ \\ \ \\
{\footnotesize
\begin{prooftree}
\hypo{
\parbox{55mm}{\center
  $\Gamma \wellfoundedvdash \htriple{A}{t}{B}$ \quad
  $\Gamma \vdash v_1 \typeinfer \overline{z_k{:}b_k}\garr  y{:}t_y \sarr \tau_x $\\
  $\Gamma \vdash v_2 \typecheck t_y$\\
 $ \Gamma \infervdash{A}  \overline{z_k{:}b_k}\garr  y{:}t_y \sarr \tau_x \effinfer \overline{z_k{:}t_k} $\\
   $\biginterty \overline{\htriple{A_i}{t_i}{A_i'}} = \tau_x[y \mapsto v_y]$ \\
  $\forall i.\Gamma, \overline{z_k{:}t_k}, x{:}t_i \vdash e \typecheck
  \htriple{(A;\globalA\topA) \land A_i'}{t}{B}$
}
}
\infer1[\textsc{ChkApp}]{
  \Gamma \vdash \zlet{x}{v_1\ v_2}{e} \typecheck \htriple{A}{t}{B}
}
\end{prooftree}
}
\ \\ \ \\
We the rank current call is
{\small
\begin{align*}
    \rankE(\Gamma \vdash \zlet{x}{v_1\ v_2}{e} \typecheck \htriple{A}{t}{B}) &= (1 + \sizeE(e) + \sizeE(v_1) + \sizeE(v_2), 1 + \sizeT(t))
\end{align*}}\noindent
which is greater than recursive calls:
\begin{align*}
    \rankE(\Gamma \vdash v_1 \typeinfer \overline{z_k{:}b_k}\garr  y{:}t_y \sarr \tau_x) &= (\sizeE(v_1), 0)
    \\\rankE(\Gamma \vdash v_2 \typecheck t_y) &= (\sizeE(v_2), \sizeT(t_y))
    \\\rankE(\Gamma, \overline{z_k{:}t_k}, x{:}t_i \vdash e \typecheck
  \htriple{(A;\globalA\topA) \land A_i'}{t}{B}) &= (\sizeE(e), 1 + \sizeT(t))
\end{align*}\noindent

\item \textsc{ChkFunc}:
\ \\ \ \\
{\footnotesize
\begin{prooftree}
\hypo{
\parbox{40mm}{\center
  $\Gamma \wellfoundedvdash x{:}t_x' \sarr \tau$ \quad
  $\Gamma \vdash t_x' <: t_x$ \\
  $\Gamma, x{:}t_x' \vdash e \typecheck \tau$
}
}
\infer1[\textsc{ChkFunc}]{
\Gamma \vdash \zlam{x}{t_x}{e} \typecheck x{:}t_x' \sarr \tau
}
\end{prooftree}
}
\ \\ \ \\
We the rank current call is
\begin{align*}
    \rankE(\Gamma \vdash \zlam{x}{t_x}{e} \typecheck x{:}t_x' \sarr \tau) &= (1 + \sizeE(e) , 1 + \sizeT(t_x') + \sizeT(\tau))
\end{align*}\noindent
which is greater than recursive calls:
\begin{align*}
    \rankE(\Gamma, x{:}t_x' \vdash e \typecheck \tau) &= (\sizeE(e), \sizeT(\tau))
\end{align*}\noindent

\item \textsc{ChkFix}:
\ \\ \ \\
{\footnotesize
\begin{prooftree}
\hypo{
\parbox{35mm}{\center
  $\Gamma \wellfoundedvdash t_f'$ \quad
  $\Gamma \vdash t_f' <: t_f$ \quad
  $\Gamma, f{:}t_f' \vdash \zlam{x}{t_x}{e} \typecheck t_f'$
}
}
\infer1[\textsc{ChkFix}]{
\Gamma \vdash \zfix{f}{t_f}{x}{t_x}{e} \typecheck t_f'
}
\end{prooftree}
}
\ \\ \ \\
We the rank current call is
\begin{align*}
    \rankE(\Gamma \vdash \zfix{f}{t_f}{x}{t_x}{e} \typecheck t_f') &= (2 + \sizeE(e) , \sizeT(t_f'))
\end{align*}\noindent
which is greater than recursive calls:
\begin{align*}
    \rankE(\Gamma, f{:}t_f' \vdash \zlam{x}{t_x}{e} \typecheck t_f') &= (1 + \sizeE(e), \sizeT(t_f'))
\end{align*}\noindent

\item \textsc{ChkMatch}:
\ \\ \ \\
{\footnotesize
\begin{prooftree}
\hypo{
\parbox{70mm}{\center
  $\forall i. (d_i, \overline{{y_{j}}{:}t_{j}}\sarr\rawnuot{b}{\phi_i}) \in \Delta$ \\
  $\Gamma, \overline{{y_{j}}{:}t_{j}}, z{:}\rawnuot{b}{\vnu = v \land \phi_i} \vdash e_i \typecheck \tau$ \quad \text{ where $z$ is fresh}
}
}
\infer1[\textsc{ChkMatch}]{
\Gamma \vdash \match{v} \overline{d_i\, \overline{y_{j}} \to e_i} \typecheck \tau
}
\end{prooftree}
}
\ \\ \ \\
We the rank current call is
\begin{align*}
    \rankE(\Gamma \vdash \match{v} \overline{d_i\, \overline{y_{j}} \to e_i} \typecheck \tau) &= (1 + \sizeE(v) + \Sigma \sizeE(e_i) , \sizeT(\tau))
\end{align*}\noindent
which is greater than recursive calls:
\begin{align*}
    \rankE(\Gamma, \overline{{y_{j}}{:}t_{j}}, z{:}\rawnuot{b}{\vnu = v \land \phi_i} \vdash e_i \typecheck \tau) &= (\sizeE(e_i), \sizeT(\tau))
\end{align*}\noindent

\item \textsc{ChkSub}:
\ \\ \ \\
{\footnotesize
\begin{prooftree}
\hypo{
\parbox{45mm}{\center
  $\Gamma \wellfoundedvdash \htriple{A_2}{t_2}{B_2}$ \quad
  $\Gamma \vdash e \typeinfer \htriple{A_1}{t_1}{B_1}$\quad
  $\Gamma \vdash A_2 <: A_1$\quad
  $\Gamma \vdash t_1 <: t_2$\quad
  $\Gamma \vdash (A_2;\globalA\topA) \land (A_2;\globalA\topA) \land B_1 \subseteq  B_2$
}
}
\infer1[\textsc{ChkSub}]{
\Gamma \vdash e \typecheck \htriple{A_2}{t_2}{B_2}
}
\end{prooftree}
}
\ \\ \ \\
We the rank current call is
\begin{align*}
    \rankE(\Gamma \vdash e \typecheck \htriple{A_2}{t_2}{B_2}) &= (\sizeE(e) , 1 + \sizeT(t_2))
\end{align*}\noindent
which is greater than recursive calls:
\begin{align*}
    \rankE(\Gamma \vdash e \typeinfer \htriple{A_1}{t_1}{B_1}) &= (\sizeE(e), 0)
\end{align*}\noindent

\item \textsc{ChkSubV}:
\ \\ \ \\
{\footnotesize
\begin{prooftree}
\hypo{
\parbox{35mm}{\center
  $\Gamma \wellfoundedvdash t$ \quad
  $\Gamma \vdash v \typeinfer t'$\quad
  $\Gamma \vdash t' <: t$
}
}
\infer1[\textsc{ChkSubV}]{
\Gamma \vdash v \typecheck t
}
\end{prooftree}
}
\ \\ \ \\
We the rank current call is
\begin{align*}
    \rankE(\Gamma \vdash v \typecheck t) &= (\sizeE(v) , \sizeT(t))
\end{align*}\noindent
which is greater than recursive calls:
\begin{align*}
    \rankE(\Gamma \vdash v \typeinfer t') &= (\sizeE(v),  0)
\end{align*}\noindent

\end{enumerate}
Thus, the typing algorithm is decidable.
\end{proof}

\newpage
\section{Evaluation Details}\label{sec:tech:evaluation}

\autoref{tab:evaluation-long-1} and \autoref{tab:evaluation-long-2}
list the information for individual ADT methods from
\autoref{tab:evaluation}.  All our benchmarks suite, and corresponding Coq proofs are available on Zenodo \href{https://doi.org/10.5281/zenodo.10806686}{https://doi.org/10.5281/zenodo.10806686}

\begin{table}[h!]
  \renewcommand{\arraystretch}{0.8}
  \caption{\small Full table of \autoref{tab:evaluation}, first part. }
\vspace*{-.05in}
\footnotesize
\setlength{\tabcolsep}{2.5pt}

\begin{tabular}{cc|cc|c||cc|ccc|cc}
\toprule
Datatype & Library & \#Ghost & s$_{I}$ & Method & \#Branch & \#App & \#SAT & \#Inc & avg. s$_{A}$ & t$_{\text{SAT}}$ (s) & t$_{\text{Inc}}$ (s)\\
\midrule
\multirow{14}{*}{\textsf{Stack}} & \textsf{LinkedList} & 0 & 4 & \textsf{is\_empty} & 3 & 2 & 60 & 3 & 59 & 0.29 & 0.03 \\
 & \textsf{LinkedList} & 0 & 4 & \textsf{empty} & 2 & 3 & 68 & 2 & 47 & 0.37 & 0.01 \\
 & \textsf{LinkedList} & 0 & 4 & \textsf{concat\_aux} & 4 & 7 & 297 & 7 & 98 & 1.63 & 0.06 \\
 & \textsf{LinkedList} & 0 & 4 & \textsf{concat} & 2 & 3 & 34 & 2 & 57 & 0.16 & 0.01 \\
 & \textsf{LinkedList} & 0 & 4 & \textsf{cons} & 3 & 7 & 203 & 5 & 112 & 1.18 & 0.05 \\
 & \textsf{LinkedList} & 0 & 4 & \textsf{head} & 2 & 2 & 65 & 3 & 55 & 0.36 & 0.03 \\
 & \textsf{LinkedList} & 0 & 4 & \textsf{tail} & 2 & 2 & 159 & 4 & 69 & 0.79 & 0.04 \\
 & \textsf{KVStore} & 1 & 9 & \textsf{is\_empty} & 2 & 1 & 93 & 4 & 79 & 0.55 & 0.04 \\
 & \textsf{KVStore} & 1 & 9 & \textsf{empty} & 3 & 5 & 213 & 6 & 176 & 1.29 & 0.06 \\
 & \textsf{KVStore} & 1 & 9 & \textsf{concat\_aux} & 3 & 6 & 709 & 16 & 167 & 3.84 & 0.19 \\
 & \textsf{KVStore} & 1 & 9 & \textsf{concat} & 2 & 3 & 100 & 4 & 89 & 0.58 & 0.03 \\
 & \textsf{KVStore} & 1 & 9 & \textsf{cons} & 4 & 10 & 874 & 17 & 226 & 5.78 & 0.23 \\
 & \textsf{KVStore} & 1 & 9 & \textsf{head} & 2 & 2 & 183 & 5 & 98 & 0.97 & 0.06 \\
 & \textsf{KVStore} & 1 & 9 & \textsf{tail} & 2 & 2 & 396 & 8 & 125 & 1.79 & 0.12 \\
\midrule
\multirow{8}{*}{\textsf{Set}} & \textsf{Tree} & 0 & 12 & \textsf{mem} & 3 & 5 & 115 & 4 & 212 & 0.69 & 0.05 \\
 & \textsf{Tree} & 0 & 12 & \textsf{insert} & 2 & 5 & 85 & 3 & 215 & 0.51 & 0.04 \\
 & \textsf{Tree} & 0 & 12 & \textsf{empty} & 1 & 0 & 17 & 1 & 111 & 0.11 & 0.01 \\
 & \textsf{Tree} & 0 & 12 & \textsf{insert\_aux} & 5 & 12 & 1589 & 11 & 531 & 9.34 & 0.53 \\
 & \textsf{Tree} & 0 & 12 & \textsf{mem\_aux} & 5 & 10 & 1469 & 11 & 490 & 8.48 & 0.41 \\
 & \textsf{KVStore} & 1 & 9 & \textsf{insert} & 3 & 5 & 245 & 6 & 160 & 1.51 & 0.06 \\
 & \textsf{KVStore} & 1 & 9 & \textsf{empty} & 1 & 0 & 7 & 1 & 42 & 0.04 & 0.01 \\
 & \textsf{KVStore} & 1 & 9 & \textsf{mem} & 1 & 1 & 97 & 4 & 73 & 0.60 & 0.03 \\
\midrule
\multirow{12}{*}{\textsf{Queue}} & \textsf{LinkedList} & 0 & 4 & \textsf{snoc} & 2 & 6 & 171 & 4 & 110 & 0.84 & 0.03 \\
 & \textsf{LinkedList} & 0 & 4 & \textsf{head} & 2 & 2 & 65 & 3 & 55 & 0.36 & 0.03 \\
 & \textsf{LinkedList} & 0 & 4 & \textsf{append} & 4 & 9 & 190 & 4 & 96 & 1.08 & 0.06 \\
 & \textsf{LinkedList} & 0 & 4 & \textsf{empty} & 2 & 3 & 68 & 2 & 47 & 0.37 & 0.01 \\
 & \textsf{LinkedList} & 0 & 4 & \textsf{is\_empty} & 3 & 2 & 60 & 3 & 59 & 0.29 & 0.03 \\
 & \textsf{LinkedList} & 0 & 4 & \textsf{tail} & 2 & 2 & 167 & 4 & 93 & 0.84 & 0.04 \\
 & \textsf{Graph} & 1 & 24 & \textsf{snoc} & 3 & 9 & 984 & 12 & 491 & 6.60 & 0.20 \\
 & \textsf{Graph} & 1 & 24 & \textsf{head} & 2 & 2 & 291 & 5 & 251 & 1.57 & 0.24 \\
 & \textsf{Graph} & 1 & 24 & \textsf{append} & 5 & 12 & 1212 & 14 & 525 & 8.19 & 0.86 \\
 & \textsf{Graph} & 1 & 24 & \textsf{empty} & 2 & 3 & 202 & 4 & 318 & 1.27 & 0.08 \\
 & \textsf{Graph} & 1 & 24 & \textsf{is\_empty} & 2 & 1 & 165 & 4 & 234 & 1.03 & 0.10 \\
 & \textsf{Graph} & 1 & 24 & \textsf{tail} & 3 & 4 & 980 & 12 & 409 & 6.58 & 0.59 \\
\midrule
\multirow{8}{*}{\textsf{MinSet}} & \textsf{Set} & 1 & 28 & \textsf{minset\_mem} & 1 & 1 & 199 & 6 & 241 & 1.43 & 0.24 \\
 & \textsf{Set} & 1 & 28 & \textsf{minimum} & 2 & 2 & 181 & 5 & 251 & 0.89 & 0.37 \\
 & \textsf{Set} & 1 & 28 & \textsf{minset\_insert} & 3 & 6 & 612 & 17 & 294 & 4.19 & 1.58 \\
 & \textsf{Set} & 1 & 28 & \textsf{minset\_singleton} & 2 & 3 & 120 & 4 & 259 & 0.80 & 0.13 \\
 & \textsf{KVStore} & 1 & 25 & \textsf{minset\_mem} & 1 & 1 & 207 & 6 & 223 & 1.49 & 0.31 \\
 & \textsf{KVStore} & 1 & 25 & \textsf{minimum} & 2 & 2 & 191 & 5 & 234 & 0.95 & 0.36 \\
 & \textsf{KVStore} & 1 & 25 & \textsf{minset\_insert} & 5 & 8 & 2227 & 23 & 519 & 10.93 & 9.34 \\
 & \textsf{KVStore} & 1 & 25 & \textsf{minset\_singleton} & 3 & 6 & 235 & 5 & 408 & 1.59 & 0.32 \\
\midrule
\multirow{15}{*}{\textsf{LazySet}} & \textsf{Tree} & 0 & 12 & \textsf{mem\_aux} & 5 & 10 & 1469 & 11 & 490 & 8.37 & 0.41 \\
 & \textsf{Tree} & 0 & 12 & \textsf{lazy\_insert} & 2 & 6 & 85 & 3 & 301 & 0.50 & 0.04 \\
 & \textsf{Tree} & 0 & 12 & \textsf{force} & 1 & 1 & 17 & 1 & 171 & 0.11 & 0.01 \\
 & \textsf{Tree} & 0 & 12 & \textsf{insert\_aux} & 5 & 12 & 1589 & 11 & 531 & 9.33 & 0.57 \\
 & \textsf{Tree} & 0 & 12 & \textsf{lazy\_mem} & 3 & 6 & 115 & 4 & 298 & 0.68 & 0.07 \\
 & \textsf{Tree} & 0 & 12 & \textsf{new\_thunk} & 1 & 0 & 17 & 1 & 111 & 0.11 & 0.01 \\
 & \textsf{Set} & 1 & 9 & \textsf{lazy\_mem} & 1 & 2 & 98 & 4 & 105 & 0.59 & 0.04 \\
 & \textsf{Set} & 1 & 9 & \textsf{force} & 1 & 1 & 7 & 1 & 66 & 0.03 & 0.01 \\
 & \textsf{Set} & 1 & 9 & \textsf{lazy\_insert} & 2 & 3 & 101 & 4 & 106 & 0.57 & 0.04 \\
 & \textsf{Set} & 1 & 9 & \textsf{new\_thunk} & 1 & 0 & 7 & 1 & 42 & 0.03 & 0.01 \\
 & \textsf{KVStore} & 1 & 9 & \textsf{lazy\_mem} & 1 & 2 & 98 & 4 & 105 & 0.59 & 0.04 \\
 & \textsf{KVStore} & 1 & 9 & \textsf{insert\_aux} & 3 & 5 & 245 & 6 & 160 & 1.49 & 0.06 \\
 & \textsf{KVStore} & 1 & 9 & \textsf{force} & 1 & 1 & 7 & 1 & 66 & 0.03 & 0.01 \\
 & \textsf{KVStore} & 1 & 9 & \textsf{lazy\_insert} & 2 & 3 & 101 & 4 & 121 & 0.58 & 0.04 \\
 & \textsf{KVStore} & 1 & 9 & \textsf{new\_thunk} & 1 & 0 & 7 & 1 & 42 & 0.03 & 0.01 \\
\bottomrule
\end{tabular}

\label{tab:evaluation-long-1}
\vspace*{-.15in}
\end{table}

\begin{table}[]
  \renewcommand{\arraystretch}{0.8}
  \caption{\small Full table of \autoref{tab:evaluation}, second part. }
\vspace*{-.05in}
\footnotesize
\setlength{\tabcolsep}{2.5pt}

\begin{tabular}{cc|cc|c||cc|ccc|cc}
\toprule
Datatype & Library & \#Ghost & s$_{I}$ & Method & \#Branch & \#App & \#SAT & \#Inc & avg. s$_{A}$ & t$_{\text{SAT}}$ (s) & t$_{\text{Inc}}$ (s)\\
\midrule
\multirow{13}{*}{\textsf{Heap}} & \textsf{Tree} & 0 & 12 & \textsf{minimum} & 2 & 3 & 82 & 3 & 207 & 0.49 & 0.03 \\
 & \textsf{Tree} & 0 & 12 & \textsf{insert} & 2 & 5 & 85 & 3 & 215 & 0.50 & 0.04 \\
 & \textsf{Tree} & 0 & 12 & \textsf{contains\_aux} & 5 & 10 & 871 & 11 & 311 & 4.87 & 0.17 \\
 & \textsf{Tree} & 0 & 12 & \textsf{contains} & 3 & 5 & 115 & 4 & 212 & 0.68 & 0.06 \\
 & \textsf{Tree} & 0 & 12 & \textsf{empty} & 1 & 0 & 17 & 1 & 111 & 0.11 & 0.01 \\
 & \textsf{Tree} & 0 & 12 & \textsf{insert\_aux} & 5 & 12 & 1589 & 11 & 531 & 9.33 & 0.52 \\
 & \textsf{Tree} & 0 & 12 & \textsf{minimum\_aux} & 2 & 3 & 408 & 5 & 334 & 2.11 & 0.09 \\
 & \textsf{LinkedList} & 0 & 4 & \textsf{insert\_aux} & 4 & 8 & 497 & 8 & 118 & 3.03 & 0.08 \\
 & \textsf{LinkedList} & 0 & 4 & \textsf{empty} & 1 & 0 & 11 & 1 & 19 & 0.07 & 0.01 \\
 & \textsf{LinkedList} & 0 & 4 & \textsf{contains} & 2 & 4 & 65 & 3 & 72 & 0.36 & 0.02 \\
 & \textsf{LinkedList} & 0 & 4 & \textsf{contains\_aux} & 3 & 5 & 359 & 6 & 129 & 1.88 & 0.05 \\
 & \textsf{LinkedList} & 0 & 4 & \textsf{insert} & 2 & 6 & 137 & 5 & 93 & 0.88 & 0.05 \\
 & \textsf{LinkedList} & 0 & 4 & \textsf{minimum} & 2 & 2 & 63 & 3 & 65 & 0.36 & 0.03 \\
\midrule
\multirow{10}{*}{\textsf{FileSystem}} & \textsf{Tree} & 1 & 20 & \textsf{deleteChildren} & 3 & 7 & 1055 & 13 & 470 & 6.39 & 1.79 \\
 & \textsf{Tree} & 1 & 20 & \textsf{del\_path\_in\_dir} & 2 & 5 & 452 & 5 & 284 & 1.76 & 0.17 \\
 & \textsf{Tree} & 1 & 20 & \textsf{add} & 3 & 8 & 2085 & 17 & 652 & 14.15 & 2.21 \\
 & \textsf{Tree} & 1 & 20 & \textsf{add\_path\_in\_dir} & 2 & 5 & 837 & 9 & 424 & 4.90 & 7.25 \\
 & \textsf{Tree} & 1 & 20 & \textsf{delete} & 4 & 16 & 1064 & 12 & 520 & 6.95 & 0.76 \\
 & \textsf{Tree} & 1 & 20 & \textsf{init} & 1 & 5 & 431 & 7 & 323 & 3.22 & 0.49 \\
 & \textsf{KVStore} & 1 & 17 & \textsf{deleteChildren} & 3 & 7 & 2244 & 19 & 392 & 14.00 & 1.38 \\
 & \textsf{KVStore} & 1 & 17 & \textsf{add} & 4 & 10 & 8144 & 43 & 481 & 56.64 & 16.54 \\
 & \textsf{KVStore} & 1 & 17 & \textsf{delete} & 4 & 20 & 5658 & 10 & 234 & 7.91 & 1.49 \\
 & \textsf{KVStore} & 1 & 17 & \textsf{init} & 1 & 3 & 449 & 5 & 166 & 3.45 & 0.08 \\
\midrule
\multirow{10}{*}{\textsf{DFA}} & \textsf{KVStore} & 2 & 18 & \textsf{is\_node} & 1 & 0 & 19 & 1 & 79 & 0.11 & 0.01 \\
 & \textsf{KVStore} & 2 & 18 & \textsf{add\_transition} & 2 & 2 & 1417 & 14 & 176 & 9.68 & 0.45 \\
 & \textsf{KVStore} & 2 & 18 & \textsf{add\_node} & 1 & 0 & 19 & 1 & 79 & 0.11 & 0.01 \\
 & \textsf{KVStore} & 2 & 18 & \textsf{del\_transition} & 2 & 2 & 704 & 8 & 152 & 4.72 & 0.32 \\
 & \textsf{KVStore} & 2 & 18 & \textsf{is\_transition} & 3 & 3 & 3604 & 25 & 228 & 18.84 & 1.60 \\
 & \textsf{Graph} & 2 & 11 & \textsf{is\_node} & 1 & 1 & 115 & 4 & 95 & 0.74 & 0.04 \\
 & \textsf{Graph} & 2 & 11 & \textsf{add\_transition} & 4 & 4 & 2420 & 27 & 192 & 16.73 & 2.12 \\
 & \textsf{Graph} & 2 & 11 & \textsf{add\_node} & 2 & 2 & 118 & 4 & 96 & 0.71 & 0.04 \\
 & \textsf{Graph} & 2 & 11 & \textsf{del\_transition} & 4 & 4 & 3625 & 27 & 225 & 25.53 & 3.42 \\
 & \textsf{Graph} & 2 & 11 & \textsf{is\_transition} & 4 & 3 & 3622 & 27 & 208 & 25.86 & 2.72 \\
\midrule
\multirow{10}{*}{\textsf{ConnectedGraph}} & \textsf{Set} & 2 & 9 & \textsf{is\_transition} & 2 & 1 & 1503 & 30 & 77 & 10.03 & 0.25 \\
 & \textsf{Set} & 2 & 9 & \textsf{add\_transition} & 4 & 4 & 3889 & 50 & 357 & 27.39 & 12.80 \\
 & \textsf{Set} & 2 & 9 & \textsf{add\_node} & 3 & 3 & 2332 & 35 & 314 & 15.80 & 1.24 \\
 & \textsf{Set} & 2 & 9 & \textsf{is\_node} & 3 & 2 & 1441 & 30 & 197 & 9.59 & 0.45 \\
 & \textsf{Set} & 2 & 9 & \textsf{singleton} & 2 & 2 & 362 & 10 & 160 & 2.30 & 0.09 \\
 & \textsf{Graph} & 1 & 20 & \textsf{is\_transition} & 4 & 3 & 1349 & 17 & 360 & 19.19 & 53.90 \\
 & \textsf{Graph} & 1 & 20 & \textsf{add\_transition} & 4 & 4 & 1351 & 17 & 361 & 19.23 & 52.38 \\
 & \textsf{Graph} & 1 & 20 & \textsf{add\_node} & 3 & 4 & 666 & 12 & 300 & 14.40 & 14.34 \\
 & \textsf{Graph} & 1 & 20 & \textsf{is\_node} & 1 & 1 & 139 & 4 & 237 & 0.90 & 0.34 \\
 & \textsf{Graph} & 1 & 20 & \textsf{singleton} & 2 & 3 & 202 & 4 & 255 & 1.28 & 0.47 \\
\bottomrule
\end{tabular}

\label{tab:evaluation-long-2}
\vspace*{-.15in}
\end{table}

\fi



\end{document}